\documentclass[preprint,12pt]{elsarticle}

\usepackage{amssymb}

\usepackage{lineno}
\usepackage{pdflscape}

\usepackage{color}

\journal{--}

\begin{document}

\begin{frontmatter}



\title{Exploring the dynamics of protest against National Register of Citizens \& Citizenship Amendment Act through online social media: the Indian experience}

 \author[]{Souvik Roy, Milan Mukherjee,  Priyadarsini Sinha, Sukanta Das, Subhasis Bandopadhyay, Abhik Mukherjee}


\address{Indian Institute of Engineering Science and Technology, Shibpur\\
Howrah, West Bengal, India – 711103}

\begin{abstract}
The generic fluidity observed in the nature of political protest movements across the world during the last decade weigh heavily with the presence of social media. As such, there is a possibility to study the contemporary movements with an interdisciplinary approach combining computational analytics with social science perspectives. The present study has put efforts to understand such dynamics in the context of the ongoing nationwide movement in India opposing the NRC-CAA enactment. The transformative nature of individual discontent into collective mobilization, especially with a reflective intervention in social media across a sensitive region of the nation state, is presented here with a combination of qualitative (fieldwork) and quantitative (computing) techniques. The study is augmented further by the primary data generation coupled with real-time application of analytical approaches.
\end{abstract}


\end{frontmatter}

\section{Introduction}

During the last decade, social media has provided instrumental means of communication for moulding the nature of political protests in the recent past\cite{Jost78,Sandra11,Pablo15}, such as, mass demonstrations in Moldova \cite{Faris10,IJoC1246,LYSENKO2012341,Pippidi}, Turkish protest movement in June, 2013 \cite{Tucker2016BigDS}, Egyptian revolution of 2011 \cite{Starbird12,IJoC1246}, Ukrainian protests of 2014 \cite{Onuch14}, Occupy Wall Street protests \cite{Langer18}, mass mobilization in Spain in May 2011 \cite{Sandra11}, protest in Hong Kong in 2014-15 \cite{Parker} etc.. John T. Jost et. al. \cite{Jost78} have summarized evidence from a variety of studies of protest movements in the United States, Spain, Turkey and Ukraine and presented the following findings about the usage of social media to facilitate political protest - (a) News about the coordination of protest activities (such as transportation, turnout, police presence, violence etc.) is spread quickly and seamlessly through social media channels; (b) Social media transmits emotional and motivational messages about protest (i.e. normativity, social justice and deprivation); and (c) The structure of on-line social network also plays an important role in the success and failure of protest movements. However, the research on political participation has also long emphasized the debate of `raise of political awareness' and `feel good politics with hollow consequences' to understand the meaningful and void effect of social media on political protest \cite{Jost78,Sandra11,Pablo15,Gladwell}.

Computer scientist have mainly examined the diffusion process of protests using the recruitment patterns in the social media (i.e. Facebook, Twitter etc.) network in the context of mass mobilization \cite{Jost78,Sandra11,Pablo15,Bond12,Parker,Shirky,Gladwell,Morozov11}. According to \cite{Pablo15}, higher density of ties are present at the core of the network, where participants are on average more active in posting messages and retweets. The information has flown largely from the core to the periphery where users are significantly less active on a per capita basis but who contribute as many messages at the aggregate level. The most interesting fact is that removing the lowest five periphery zones results in a dip of slightly more than 50\% in reach. According to \cite{Sandra11}, the vast majority of contagion chains die soon, with only a very small fraction reaching global dimensions. This result is also supported by other findings ~\cite{Bakshy11,1Haewoon,Sun2009GesundheitMC}.

However, long before the presence of organizing tool internet, sociologists \cite{Jenkins,Granovetter,michael2005a,Schelling,Marwell,Michael,Granovetter83,Watts5766,Watts,Marwell88,Michael91,Roger93,Centola07,David09} have also discussed the role of critical mass and their importance in resource mobilization to understand the success of social movements. This includes historical empirical results, such as, revolt in Paris commune (1871) \cite{Roger91}, 1960’s civil right struggles in the US \cite{McAdam86}, demonstrations in East Germany prior to the fall of the Berlin wall \cite{lohmann_1994,Opp93} etc.

In a different research effort, computer scientists have also analysed the informational and motivational content of social media posts during the mass mobilization period of political protest \cite{Tucker2016BigDS,Onuch14,Langer18}. Again, in this context, social scientists have widely discussed about social psychological factors -- moral outrage, social identification and group efficacy
\cite{Barbalet98,John12,Kawakami95,Stefan,Nicole11,Wakslak07,Zomeren,Smith15,McGarty14,Klandomans97,Kelly96,Drury09,Mazzoni15,Martijn12}   
. However, there is a lack of understanding about political protest by aggregating both the social media and field experience together which is a limitation identified in the reviewed literature on social media analysis.

In this backdrop, this article studies the dynamics of protest movement against the policy of countrywide National Register of Citizens (NRC) \& Citizenship Amendment Act (CAA) in India by analyzing social media as well as data collected from field work. This article covers the first six months since the Central Govt that came to power in June 2019 announced that the National Registry of Citizens (NRC) will be prepared in lines with a similar exercise conducted in Assam \cite{news6}, a state located in the North Eastern zone of India. The exercise in Assam culminated in declaration of a sizable population as de-voters and forcible detention of people in camps pending foreigner tribunal cases \cite{news5}. Hence many peace-loving people across the country got scared when the newly elected Govt  announced this NRC as their pan-India policy. In fact, the Govt subsequently passed a legislation to amend the existing Citizenship Act (CAA) along communal lines \cite{news3}, widely seen as a precursor to the nationwide implementation of NRC. The gripping fear, with clear evidence from Assam as indication of what lies ahead, resulted in widespread resent that burst into protest throughout the country. The protests had an interesting feature of spontaneity and is fast becoming the new movement cult or norms,  with sizeable  presence  in the social media fronts. Two modes of social media interactions have been considered, the deliberations within a group formed with a purpose and individuals trying to reach out to their followers with some purpose. Based on availability, Facebook and Twitter are respectively chosen for such purpose.  

In this work, different social media (such as, Facebook and Twitter) activities are monitored over the first six months of the Contra-NRC \& CAA movement. Group activities and related dynamics is explored from contra NRC \& CAA Facebook groups that evolved during this time period. The proliferation of such groups, growth of their membership and sharing of their categorized contents among people with different social identities have been considered. Spatial, temporal and informational dynamics of individual responses to the contra NRC \& CAA topics as reflected through Twitter accounts have also been scrutinized to understand the movement.










Somewhat extensive field work has also been conducted in the bordering districts of West Bengal and Assam. These districts share border with independent sovereign of Bangladesh and are considered to be the hotspot of historical migration since the partition of Bengal along communal lines. This study attempts to close the gap between the ground   reality and the findings of social media activity through such field work. The work investigates to what extent the contra NRC \& CAA movement has acquired the characteristics of mass movement and the associated representation of societal dynamics.

In detail, this paper is organized as follows:

\begin{itemize}
\item In Section~\ref{section1}, the dynamics of the creation of the contra NRC \& CAA $123$ Facebook pages and $79$ Facebook groups are depicted.

\item The growth in membership of the newly created $11$ contra NRC \& CAA Facebook groups containing total $1,98,463$ members are described in Section~\ref{section2}.

\item  To understand the structure of the Facebook groups from the perspective of mass movement, the number of common members in West Bengal based $37$ contra NRC \& CAA Facebook groups are analyzed in Section~\ref{section3}.

\item In Section~\ref{section4}, the contagion chain dynamics and religious participation within it is depicted for contra NRC \& CAA $14$ viral Facebook posts containing total $11,642$ Facebook profile users.

\item The informational and motivational content of $3200$ Facebook posts during the mobilization period of NRC \& CAA protest and $500$ Facebook posts during spread of COVID-19 period are analyzed in Section~\ref{section5}.

\item In Section~\ref{section6}, the dynamics of growth in number of tweets, location and structure of twitter network are analyzed for $4,90,978$ tweets during the mobilization period of contra NRC \& CAA protest.

\item The topic wise responses and expressiveness of the responders in the field interviews are analyzed in Section~\ref{section7}.

\item Finally, Section~\ref{section8} concludes the paper with critical comparison of different signature behaviour between social media and ground reality.

\end{itemize}


\section{Creation of groups and pages in social media}
\label{section1}

This section studies the growth of Facebook pages and Facebook groups in numbers during mobilization period of contra NRC \& CAA protest. To do so, we have collected dataset from random but purposive $123$ Facebook pages and $79$ Facebook groups during the period of mid-September, $2019$ to mid-January, $2020$. To facilitate this study, we have used following notations in the analysis. 

\begin{itemize}
\item $N$($k$) indicates total number of Facebook pages/groups present at instant $k$.
Total time of the study is divided into slots of $3$-days interval. 

\item $N$($n$) depicts the total number of pages/groups at the end of our time window, i.e. January $15$, 2020. Here, $k$ runs from $1$ to $n$. Therefore, $N$($n$) = $123$ (resp. $79$) for Facebook pages (resp. Facebook groups). When we consider both pages and groups together, then $N$($n$) = $202$.
\end{itemize}

\subsection{Results and Observations}
\begin{figure}[!htbp]
\begin{tabular}{cc}
\includegraphics[width=70mm]{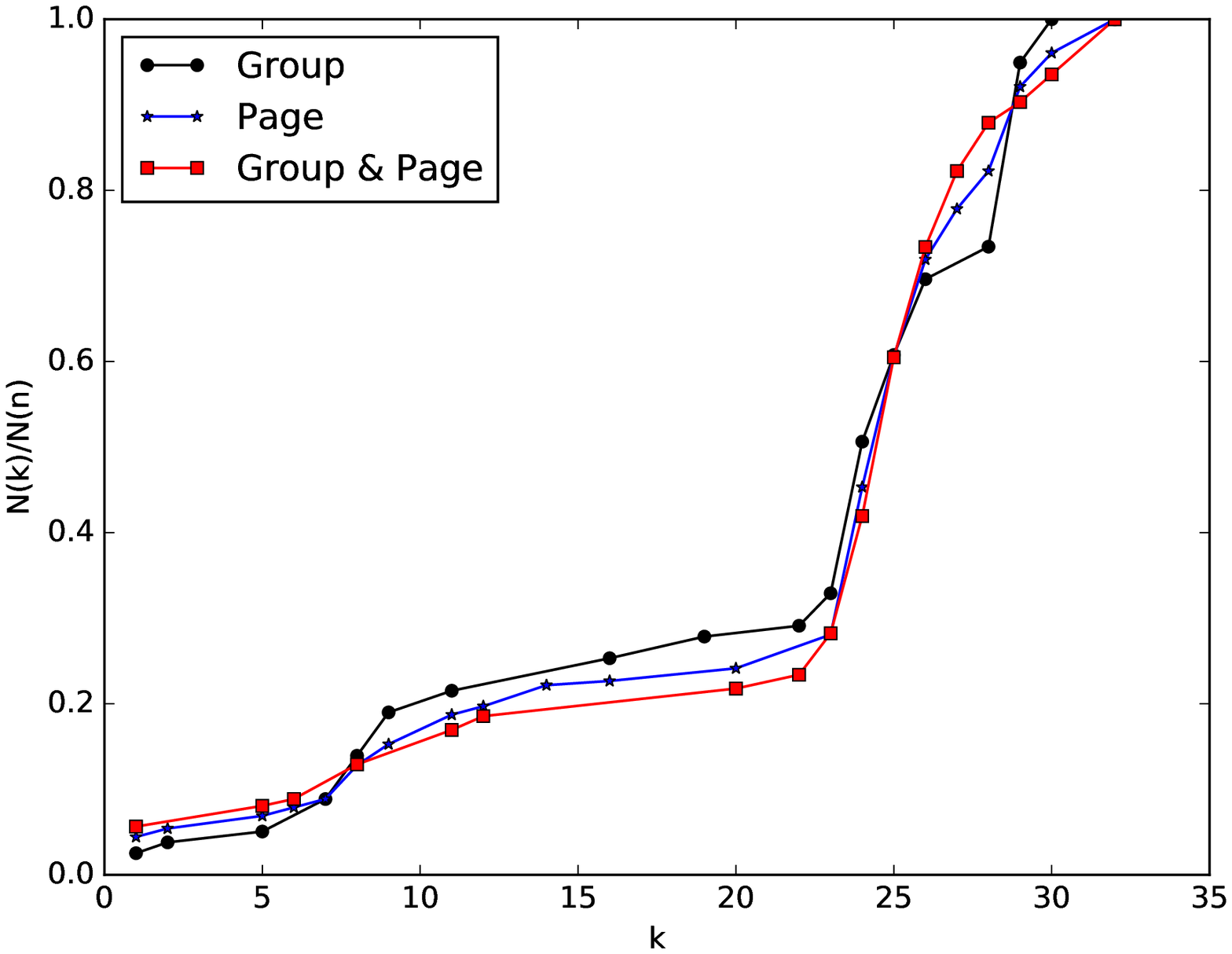} & \includegraphics[width=70mm]{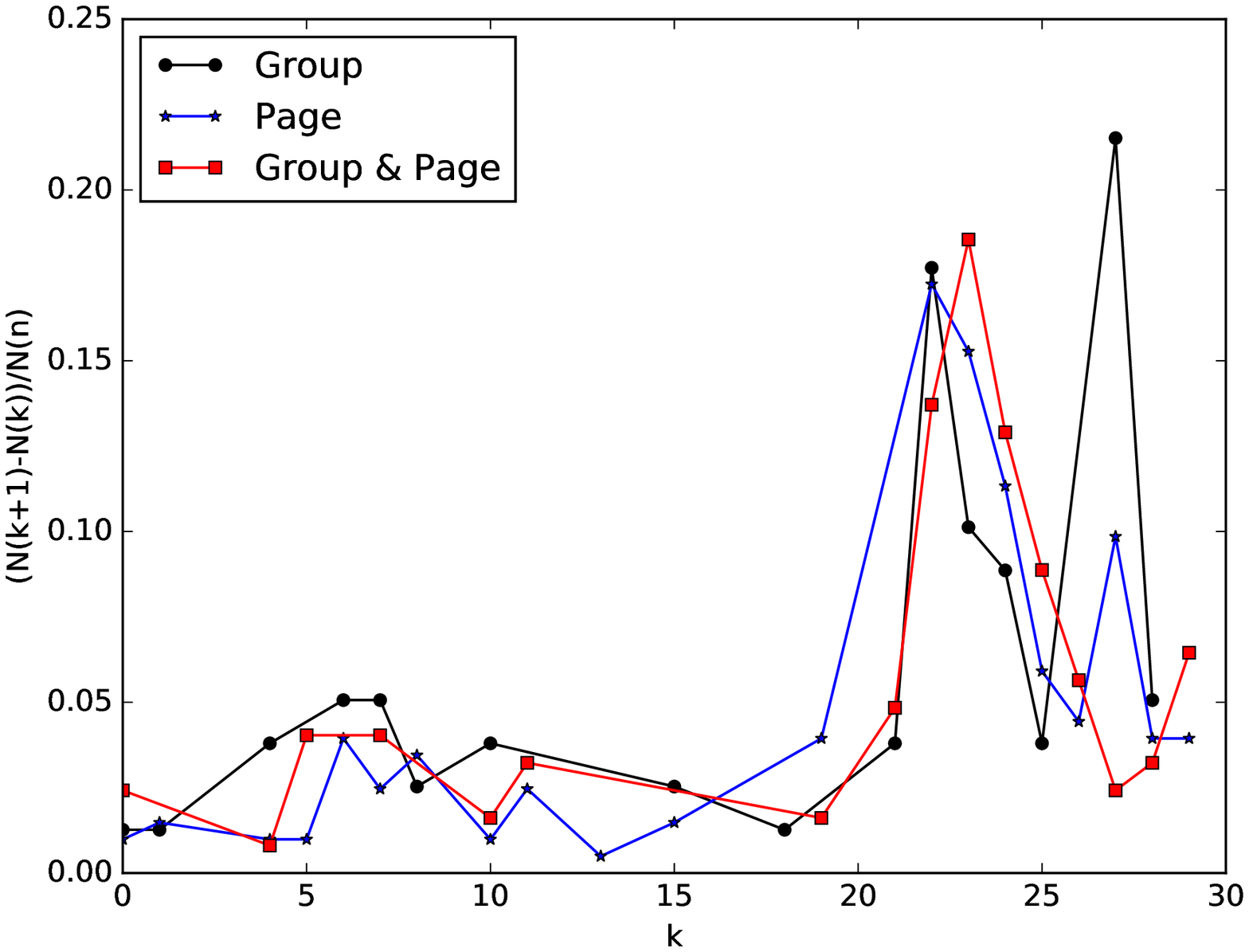}   \\
(a) & (b) \\
\end{tabular}
\caption{The growth of (a) $N(k)$/$N(n)$ over time; (b) ($N$($k+1$) - $N$($k$))/$N$($n$) against $k$.}
\label{fig1}
\end{figure}

Let us now plot the data, to understand the accelerating rate of growth of Facebook pages and Facebook groups in number during the mobilization period. Fig.~\ref{fig1}(a) depicts the growth of $N(k)$/$N(n)$, which represent the normalized number of activated pages/groups, over time. Next we plot the normalized rate of increase in actual number of pages/groups created during each time slot $k$. That is, ($N$($k+1$) - $N$($k$))/$N$($n$) is to be plotted against $k$.  Fig.~\ref{fig1}(b) shows such graphs. Now, the above results depict the following dynamics of the page / group creation. 

\begin{itemize}
\item[1.] There are sharp rises in the graphs of Fig.~\ref{fig1} after $k$ = $20$, which represents the rise of page/group creation after December $12$, $2019$. The Citizenship Amendment Bill was passed in Parliament of India on $12$ December, 2019. Further, the new group creation counter reaches its peak during the first half of January, $2020$ (see Fig.~\ref{fig1}(b)). This is an indication that the people are increasingly joining the protest movement after December $12$, $2019$.

\item[2.] It is observed that before December $12$, $2019$, Facebook groups are mostly West Bengal and Assam based. That is, people of these states were more concerned about the implication of NRC \& CAA. In Assam, the procedure of NRC dragged on for six years and finally a list of citizens was published in August, $2019$ \cite{news5} excluding more that $1.9$ million people residing in the state of Assam. Whereas, West Bengal is one of the most affected state of India, having been a victim of the partition of British India \cite{chatterji2007spoils,chatterji2019partition,sengupta2015partition}. Assam also had borne the brunt of partition.
The bitter experience and resultant mood of these two state are reflected in the above observation. Note that, in the field responses, the {\em impact of Assam NRC} ($14.94$\%) and {\em migration after $1971$ and related problems}($13.91$\%) are pointed out by people with significantly high intensity.

\item[3.] Here, these $123$ pages are associated with total $2,73,179$ likes during the entire period of time. Moreover, the $79$ groups contain total $4,66,294$ members with highest membership of $1,39,547$ in a single group. The total number of posts per day in the $79$ Facebook groups is $44,917$, which reflects high volume of activity. A possible interpretation of the difference in volume of activity is that more people have felt attachment with the Facebook groups than with the Facebook pages. This also points to the fact that the role of Facebook groups seem to be more important than Facebook pages in the protest movement against NRC \& CAA. In fact, in a Facebook page, only the `admins' (administrators) can post, whereas, any group member of a Facebook group is enabled to post in the group.  So the group can sustain the spontaneity of the members. In contradiction, several researchers \cite{Jost78,Tucker2016BigDS,Onuch14} have pointed out the importance of only Facebook pages during earlier mass mobilization period of protest/movements. According to \cite{Tucker2016BigDS}, the most popular Facebook page on Turkish protest movement in June, $2013$ was liked $6,43,951$ ($7000$/day) times over a three month period. Similarly, the most popular Facebook page on Ukrainan protests of $2014$ was liked more than $1,25,000$ times during two weeks period of time \cite{Onuch14}. 
\end{itemize}

\section{Membership proliferation in the groups}
\label{section2}

This section studies the growth in membership of the newly created Facebook groups during mobilization period of CAA \& NRC protest. To do so, we have chosen $11$ groups out of $79$ groups of Section~\ref{section1}. Choice of such $11$ group is diverse: based on number of members, creation date (before/after December $12$, $2019$) of the group, participation of parliamentary political party in the group and location (state in India) of the group. Note that, although the data of original $79$ group had been collected for the period from mid-September, $2019$ to mid-January, $2020$, now, we have slightly modified the period due to unavailability of membership data for the month of September.  The modified period is October $11$, $2019$ to January $21$, $2020$. 

These $11$ groups contain total $1,98,463$ members with highest (resp. lowest) number of member in the group being $1,39,547$ (resp. $1,276$). In the context of activity, the number of posts per day in the groups is associated with maximum (resp. minimum) count of $10,000$ (resp. $100$). Here, five groups out of $11$ are West Bengal based and are created in the month of September, October and November. However, the rests (outside of West Bengal) are created after $12$ December, $2019$ which was the passing date of CAA in Parliament of India.

This Section shares similar notations with Section~\ref{section1}. Here, $M$($k$) indicates number of members in the corresponding Facebook group at instant $k$. Total time of the study is divided into slots of $2$-days interval. Here, $M$($n$) depicts the total number of member in the corresponding Facebook group at the end of our time window, i.e. January $21$, 2020. 

\subsection{Results and Observations}

\begin{figure}[!htbp]
\begin{tabular}{cc}
\includegraphics[width=70mm]{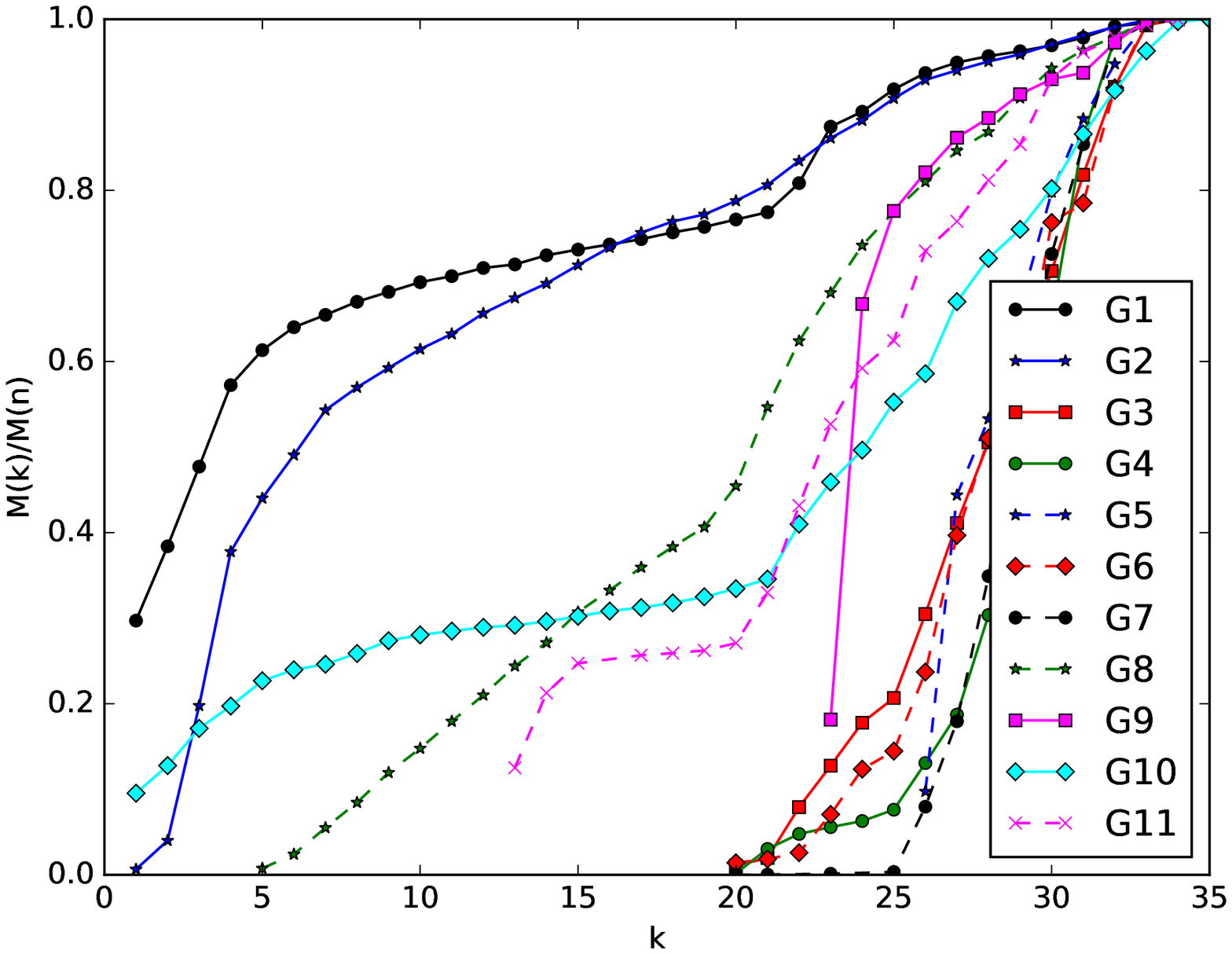} & \includegraphics[width=70mm]{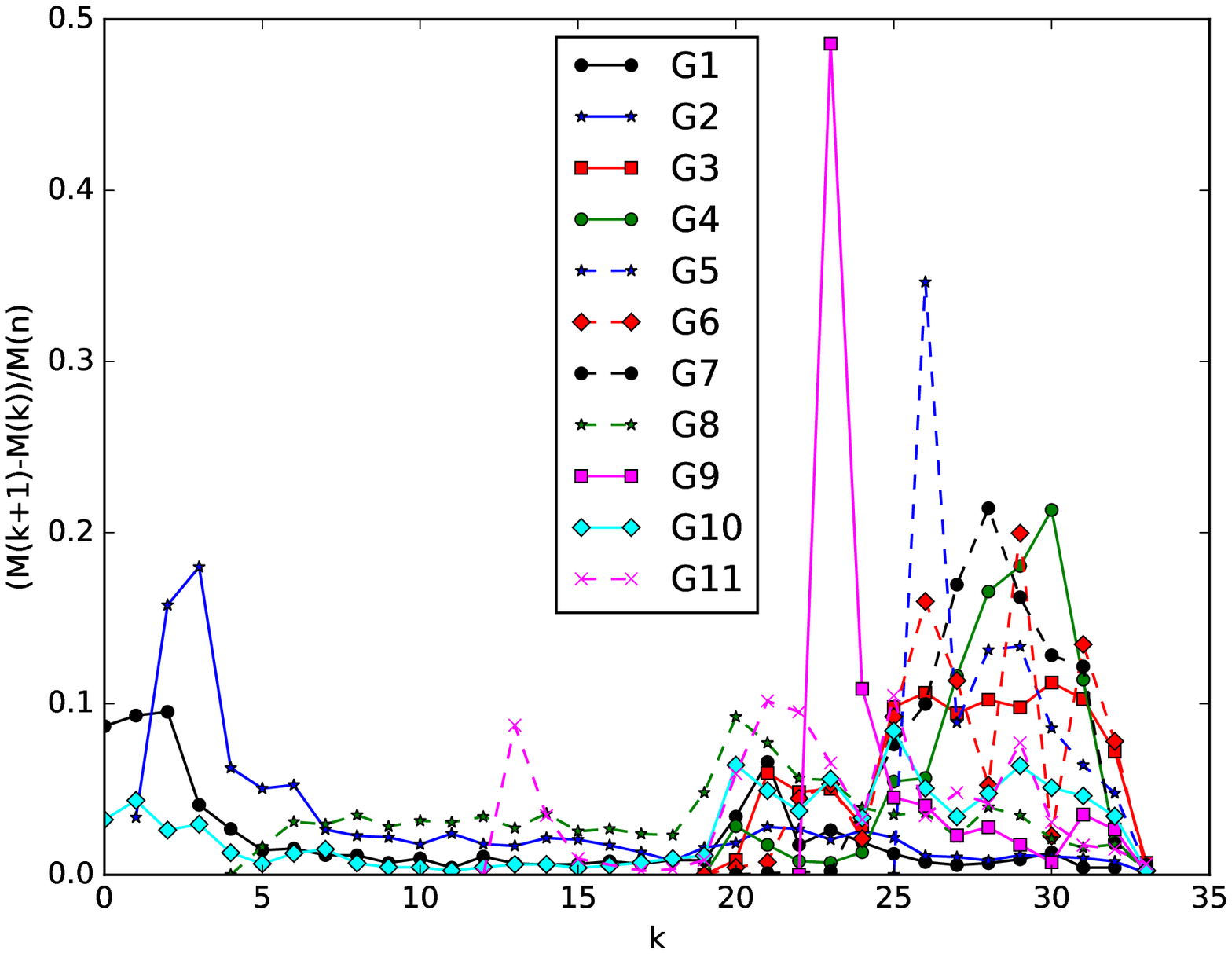}   \\
(a) & (b) \\
\end{tabular}
\caption{The growth of (a) $M$($k$)/$M$($n$) over time; and (b) ($M$($k+1$)-$M$($k$))/$M$($n$) against $k$.}
\label{fig2}
\end{figure}

To understand the dynamics of member joining rate in groups, we have plotted number of member in groups against time during the period of October $11$, $2019$ to January $21$, $2020$. Fig.~\ref{fig2}(a) depicts the growth of $M$($k$)/$M$($n$), which represents the normalized number of joined members over time. In Fig.~\ref{fig2}(b), we plot ($M$($k+1$)-$M$($k$))/$M$($n$) against $k$. Now, the above results depict the following dynamics of the number of members joining in the groups.

\begin{itemize}
\item The plots show following two sharp rises in the member joining in the groups:
\begin{itemize}
\item Firstly, for the groups located at West Bengal, a sharp rise is observed within $k = 0$ to $k = 5$, which represents starting of October, $2019$. Note that, this corresponds to the time when Home Minister of India Government had announced during a visit to kolkata that the procedure of NRC would take place in West Bengal too~\cite{news6}. This reason is definitely responsible for the rise in membership and this observation is also reflected from the field data, for example, see part of field interviews
\footnote{``After the speech of Amit Shah on 1 st October, 2019 in Kolkata the scenario has changed drastically. There was no space to walk around the court area; mothers with their 4-6 months child, elderly people were seen in the court for the process. There has been long queue for about 15 days in the court." : GF-RM-EM-AU-DNM-L4-O2-A2.}
\footnote{``There used to be a queue of about thousands and thousands (4000-5000) of people outside the Block office. As there has been a deadline there was a huge crowd. A person has also lost his life due to stampede in the crowd. Everything started after the speech of Amit Shah in Kolkata on 1st October, 2019 and then Dilip Ghosh said that the Muslims would be thrown out ruthlessly.  And now some other political parties in their conducted meetings and rallies said that they would not allow NRC to get implemented so that the fear among the people has decreased." : GM-RM-EM-AR-DNM-L3-O2-A3.}.
In this context, we noticed that $11.34$\% of those interviewed responded citing the problem of {\em confusing statements by political leadership}.

\item  A second sharp rise in new membership for all the groups has been observed after $k = 20$, which represents December $12$, $2019$, the date of passage of CAA in the parliament.
\end{itemize}
\end{itemize}

The most popular Facebook group, based in West Bengal, has touched $1,39,547$ members within only six month after its creation ($6$ September, $2019$), which is remarkable. Moreover, Contra-NRC/CAA/ NPR (National Population Register) physical protest rallies, meetings, street meeting, cycle rallies were organized by these Facebook groups. Moreover, new movemental organization was created through these Facebook groups by common people without presence of traditional political leadership. According to one of the `admins' (administrators) of this Facebook group, interviewed on $22$ February, $2020$, in Kolkata,

\begin{quote}
``\textit{Around six months ago we created a Facebook messenger group against the NRC, CAA and NPR and then a Facebook group through it. With the creation of the group the member count started to increase at an exponential rate. We had completely no idea about  how the member count could increase in such a manner.  The members were continuously pressing for an Anti-NRC, CAA, NPR movement programme on the road so that the local people could be involved. To understand the process of how to involve local people in the movement, we arranged a convention meeting including the members. At the meeting we reached a decision that on 19th December we will be organising an Anti-NRC, CAA, and NPR rally in Kolkata, announcing it through the Facebook group.} 

\textit{On 19th December, 2019 in Kolkata an anti-NRC, CAA, NPR rally was organised on behalf of ``NO NRC Movement" Facebook group, according to the police report there was estimation of around 60-70 thousand of people.}

\textit{People have built up a social media centric organisation, which started initially through a chat of 2-3 people. And now the people themselves are organising their committee, arranging programmes against NRC."}

\end{quote}

\section{Structure of the groups}
\label{section3}

In this section, we concentrate on the West Bengal (specifically Kolkata) based Facebook groups to understand the administrative hierarchy within the groups that may be conducive to understand the dynamics of a fledgling mass movement. The reason behind  this choice for the study is that West Bengal was one of the most affected states during the Partition of India in 1947 \cite{chatterji2007spoils,chatterji2019partition,sengupta2015partition}. Therefore, the ingredients  of mass movement against NRC \& CAA is already present in West Bengal. Altogether  $37$ Facebook groups have been identified, out of the $79$ groups discussed earlier in Section~\ref{section1}, those are demographically located within Kolkata and its suburbs. Dataset of unique Facebook member identification has been collected  for these $37$ Facebook groups till January $21$, $2020$. 
Total member count of these $37$ groups is $46,153$.

The overlap of members in these groups i.e. the number of common members in these $37$ Facebook groups can give some insight regarding the mass-movement dynamics.
It may be possible  to understand the presence of centralized control in the movement against NRC \& CAA from the number of common members. The amount of overlap may also be useful to understand sectoral identity in the movement against NRC \& CAA.

\subsection{Methodology}

In this section, number of common members for these $37$ Facebook groups are computed considering every possible combination. Here, $m = 37$ is the total number of Facebook groups. Now, the total number of possible combinations after considering $n$ arbitrary groups, out of $m$, is $^mC_n$ where the combination index can be represented by  $i$ = \{$1, 2, \cdots, ^mC_n$\}. Let us represent, the member sets of $n$ groups for $i^{th}$ index combination by ${q_i}_1, {q_i}_2, \cdots, {q_i}_n$. Therefore, the total number of common members of $n$ groups together in the $i^{th}$ index combination is 
\begin{equation}
\nonumber
S_{n,i} = \{{q_i}_1 \cap {q_i}_2 \cap \cdots \cap {q_i}_n\}
\end{equation}
Now, 
\begin{equation}
\nonumber
 \mathcal{S}_n  = \sum_{i=1}^{^{m}C_n}  S_{n,i} 
\end{equation}
depicts the total number of common members for all $^{m}C_n$ combinations of $n$ groups together. Therefore, the average number of common member for $n$ groups is 
\begin{equation}
\nonumber
A_n(m) = \frac{1}{^{m}C_n}  \mathcal{S}_n 
\end{equation}
where $n$ = \{$2,3,\cdots,m$\}. As an example, $A_2$($37$) indicates the average number of common members after considering all possible combinations of $2$ groups taken together. A graph representation of the members, with edges formed based on the groups they belong to, can depict a qualitative association among the members.
\subsection{Results and Observations}

\begin{figure}[!htbp] 
\centering 
\includegraphics[width=2.5in]{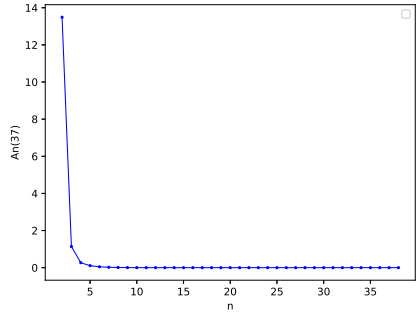} 
\caption{Average number of common members $A_n$($37$) as a function of $n$ where $n$ = \{$2,3,\cdots,37$\}.}
\label{fig3}
\end{figure}

Fig.~\ref{fig3} depicts the average number of common members $A_n$($37$) as a function of $n$. According to Fig.~\ref{fig3}, the average number of common members considering all combinations of two groups together, i.e. $A_2$($37$), is $13.49$ which implies that only $0.00029$\% of total number of members in all these $37$ Facebook groups are present in two different groups. Moreover, $A_4$($37$) $\approx$ $0$, i.e. the average number of common members in four groups considering all possible combinations almost vanishes. 

\begin{figure}[!htbp] 
\centering 
\includegraphics[width=6.1in]{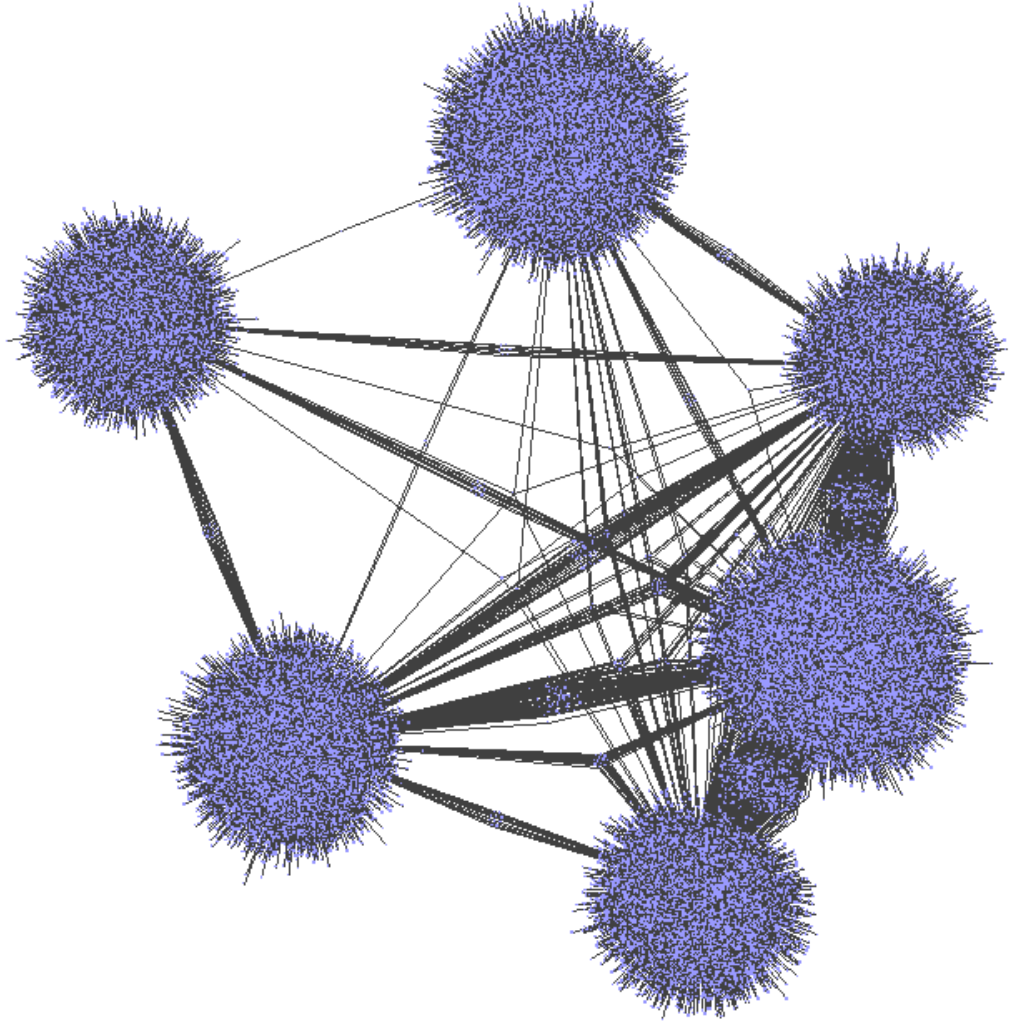} 
\caption{Dynamics of $6$ large Facebook groups containing total $30,699$ members where $A_2$($6$) = $64.93$; $A_3$($6$) = $7.5$; $A_4$($6$) = $2.26$; $A_5$($6$) = $0.66$; and $A_6$($6$) = $0$. Here, the nodes (in blue) represents the unique member Facebook identities/unique Facebook group identities. Moreover, the nodes (representing unique member Facebook identities) connects with corresponding nodes (representing unique Facebook group identities) via edges.}
\label{fig4}
\end{figure}

Fig.~\ref{fig4} depicts the membership association of $6$ Facebook groups together containing total $30,699$ members. In Fig.~\ref{fig4}, the common members are duplicated in respective groups and hence associated with more than one edge. Therefore, plotting all members together across the groups as nodes reflects on the number of common members in these groups where $A_2$($6$) = $64.93$; $A_3$($6$) = $7.5$; $A_4$($6$) = $2.26$; $A_5$($6$) = $0.66$; and $A_6$($6$) = $0$. 

At first glance, less common members in the groups indicate towards localized mass movement situation with less centralized control. Whereas, large count of common members across the groups depict activist dependent centralized control in the movement. In this study, $A_2$($37$) = $13.49$, $A_3$($37$) = $1.14$, $A_4$($37$) $\approx$ $0$ which indicates less centralized control.

However, less common members in the groups may also lead towards divergent sectoral identity dynamics in the movemental situation. That is, different people with different sectoral identity may share different sectoral movement over the Facebook group platform and may not be comfortable with cross-sectoral platform which is not a very ideal situation for any movemental dynamics. As for example, students of some elite university may not share with Facebook group discussion over people's movement or people from refugee Hindu religious community may not be ready to share same platform Facebook group with people having Muslim religious identity.

Now, to understand the sectoral identity, we study the existence of posters in these groups related to few big protest events at Kolkata. Here, we consider following protest events in Kolkata - (1) The protest event that took place on $11$ January, $2020$ when Indian Prime Minister had planned to visit Kolkata; (2) The protest rally on $21$ December, $2019$ which was organized by the general students of the city; (3) The protest rally on $19$ December, $2019$ which was organized by common people of a Facebook group named `no-nrc movement'. The commonness between these protest events is that none of them was organized by traditional political organizations and leadership. However, these protest events were the most successful protest events in Kolkata in terms of mass mobilization. Members of our research team hand-coded the picture content of these $37$ groups to identify the existence of posters in these groups related with above protest events.

As a parameter, posters and discussions about the chosen protest events in most of the groups would indicate towards convergence dynamics with lack of sectarian practice, whereas, the lack of such sharing would indicate otherwise. Now it has been observed that  $78.37$\% of the chosen Facebook groups are associated with the posters of chosen protest events. This has a tilt towards higher sharing of protest information among the active groups in spite of differences in member orientation. This  in our opinion indicates a tendency towards mass-movement dynamics leaving aside the baggage of sectarian identity.  

\section{Contagion chain dynamics}
\label{section4}

One important aspect of the study is to consider how the contra NRC \& CAA Facebook posts flow to the members and see whether any dynamics of religious participation exists therein. This section studies the contagion chain (`like/share') dynamics for somewhat randomly chosen $14$ viral Facebook posts related with protest against NRC \& CAA. These posts have an information catchment containing total $11,642$ Facebook profile users. All these chosen  posts belong to one single contra NRC \& CAA Facebook group. Here, seven posts which involve $5,138$ Facebook profile users were posted before December $12$, $2019$ which was the passing date of CAA in Parliament of India. An equal number ($7$) of posts involving $6,504$ Facebook profile users are chosen from those posted after December $12$, $2019$. Here, this study focuses on the following aspects. 

\begin{itemize}
\item[1.] This study compares the participation of Muslim and Non-Muslim Facebook users in the `like/share' dynamics of the chosen  Facebook posts. Such comparison reflects the community participation around Facebook posts before and after December $12$, 2019. This religious angle is important to understand the movement dynamics as the CAA is biased with religious  overtones in determining citizenship of India. 

\item[2.] Moreover, the study can bring out religious dependency in the `like/share' dynamics, i.e. whether Facebook users tend to like/share posts from Facebook users of their own religion can reflect upon the secular attitude prevalent in the society. 

\item[3.] In addition to the above objectives, following the methodology of Section~\ref{section3}, the overlap of Facebook profile users in this `like/share' dataset can give an idea about the role of activist or any sort of centralized dependency in shaping the movement dynamics.

\end{itemize}

Here, the raw dataset contains the source Facebook profile user (i.e. from whom the post originated), the destination Facebook profile user (i.e. who liked/shared the post). The members of our research team hand-coded the religion of profile user from the user's profile name and surname as Muslim or Non-Muslim. Note that, in case of quantitative and qualitative analysis, to identify the number of common user profiles in the like/share dynamics for these $14$ Facebook posts, this study follows the same methodology of earlier Section~\ref{section3}.

\subsection{Results and Observations}

Table~\ref{table1} depicts the like/share statistics for the $14$ Facebook posts. This result indicates that the participation of Muslim community in the movement increases ($16$\%) as a community after the passage of CAA in the parliament. It may be observed that, the fear within Muslim community is also reflected from the field data, for evidence, see selected part of field responses 
\footnote{``The leaders of RSS are saying that new bill is going to come and that it will help the refugee Hindus in the process of getting citizenship. It will not even create problems for Muslims but their right to vote will be taken away and hence will deprive them of several opportunities.": GM-RH-EHR-AU-DM-L4-O2-A3}
\footnote{``There was a great fear among the Muslims. A Muslim man gives me his LIC premium of rupees 1.5 Lakh in a month, he owns a rice mill. After all this issues of NRC he has stopped to do so, saying what will happen by giving premium, they want to chase us from the country. Let's save money, when they will chase us we can sell our houses and go.": GM-RH-EHR-AU-DM-L3-O4-A2}.

\begin{table}[!htbp] 
\centering
\scriptsize
\begin{tabular}{l|ll|ll|ll} \hline
             & Before CAA &          & After CAA  &         & Total   &  \\   
 Source-Destination pairs & Number     & \%       & Number     & \%      & Number  & \% \\ \hline \hline    
  Muslim origin    & 3078       & 60\%  & 4951       & 76\% & 8029    & 69\% \\ 
  Non-Muslim origin   & 2060       & 40\%  & 1553       & 24\% & 3613    & 31\% \\ \hline \hline  
  M-M     & 1852       & 36.04\%  & 3158       & 48.55\% & 5010    & 43.03\% \\ 
 NM-NM     & 1566       & 30.47\%  & 984        & 15.12\% & 2550    & 21.90\% \\ 
 Co-religion total & 3418 & 66.52\% & 4142 & 63.68\%  & 7560    & 64.93\% \\ \hline \hline
 NM-M     & 1227       & 23.88\%  & 1794       & 27.58\% & 3021    & 25.94\% \\
 M-NM     & 493        & 9.59\%   & 568        & 8.73\%  & 1061    & 9.11\% \\ 
 Cross-religion total & 1720 & 33.47\% & 2362 & 36.31\%  & 4082    & 35.06\% \\
\hline \hline
\end{tabular}
\caption{Religion based like share dynamics in the chosen $14$ Facebook posts. Here, M-M (resp. NM-NM; NM-M; M-NM) represents Muslim (resp. Non-Muslim; Non-Muslim; Muslim) profile users post liked/shared by Muslim (resp. Non-Muslim; Muslim; Non-Muslim) profile user.}
\label{table1}
\end{table}

\begin{figure}[!htbp]
\begin{tabular}{c}
\includegraphics[width=130mm]{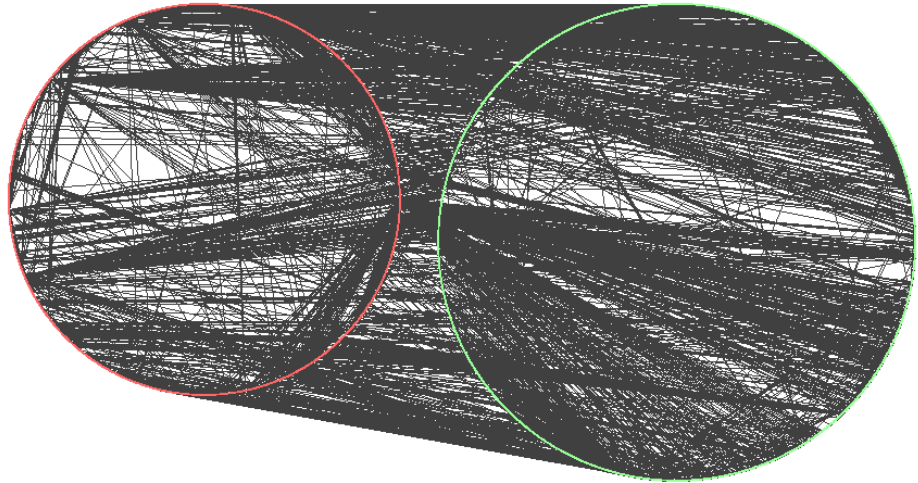}\\
(A) Before December 12, 2019 \\
\includegraphics[width=130mm]{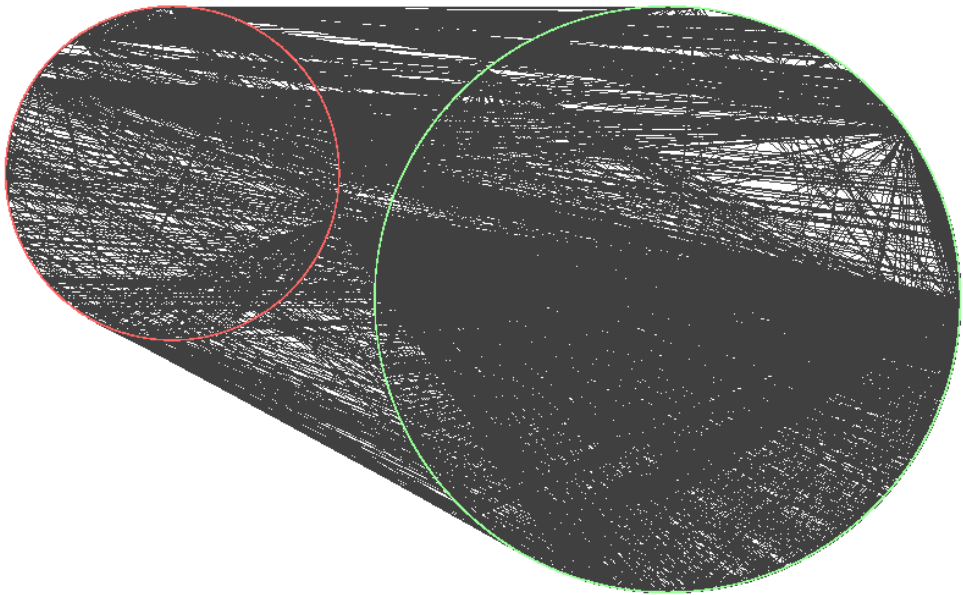}\\
(B) After December 12, 2019 \\
\end{tabular}
\caption{Like-share dynamics of $7$ Facebook posts containing $5138$ (resp. $6504$)  Facebook profile users before (resp. after) $CAA$ was passed in Indian Parliament. This Figure shows the group attribute layout based on religion. The Non-Muslims are marked by red and Muslims are marked by green, i.e. the Red circle is for Non-Muslim users and Green circle is for Muslim users. The edges among them shows the relationship.}
\label{fig5}
\end{figure}

According to Table~\ref{table1}, for $64.93$\% cases Facebook user profile liked/shared Facebook post from Facebook user profile of same religion, i.e Non-Muslim profile user liked/shared Non-Muslim profile user's post or Muslim profile user liked/shared Muslim profile user's post. This statistics remain almost same if we separately calculate for Facebook post before and after $12$ December, $2019$.  It is very interesting to note that, Muslim profile user liked/shared Non-Muslim profile user's Facebook post for $25.94$\% cases, however, Non-Muslim profile user liked/shared Muslim profile user's Facebook post for $9.11$\% cases. Fig~\ref{fig5} depicts the group attribute layout based on religion for the chosen viral Facebook posts before and after December $12$, $2019$ separately. In Fig.~\ref{fig5}, two Facebook users are connected via edges if one liked or shared another's post. Such social-media result indicates co-existence of both the secular and communal identity in the society. 

\begin{figure}[!htbp] 
\centering 
\includegraphics[width=3.1in]{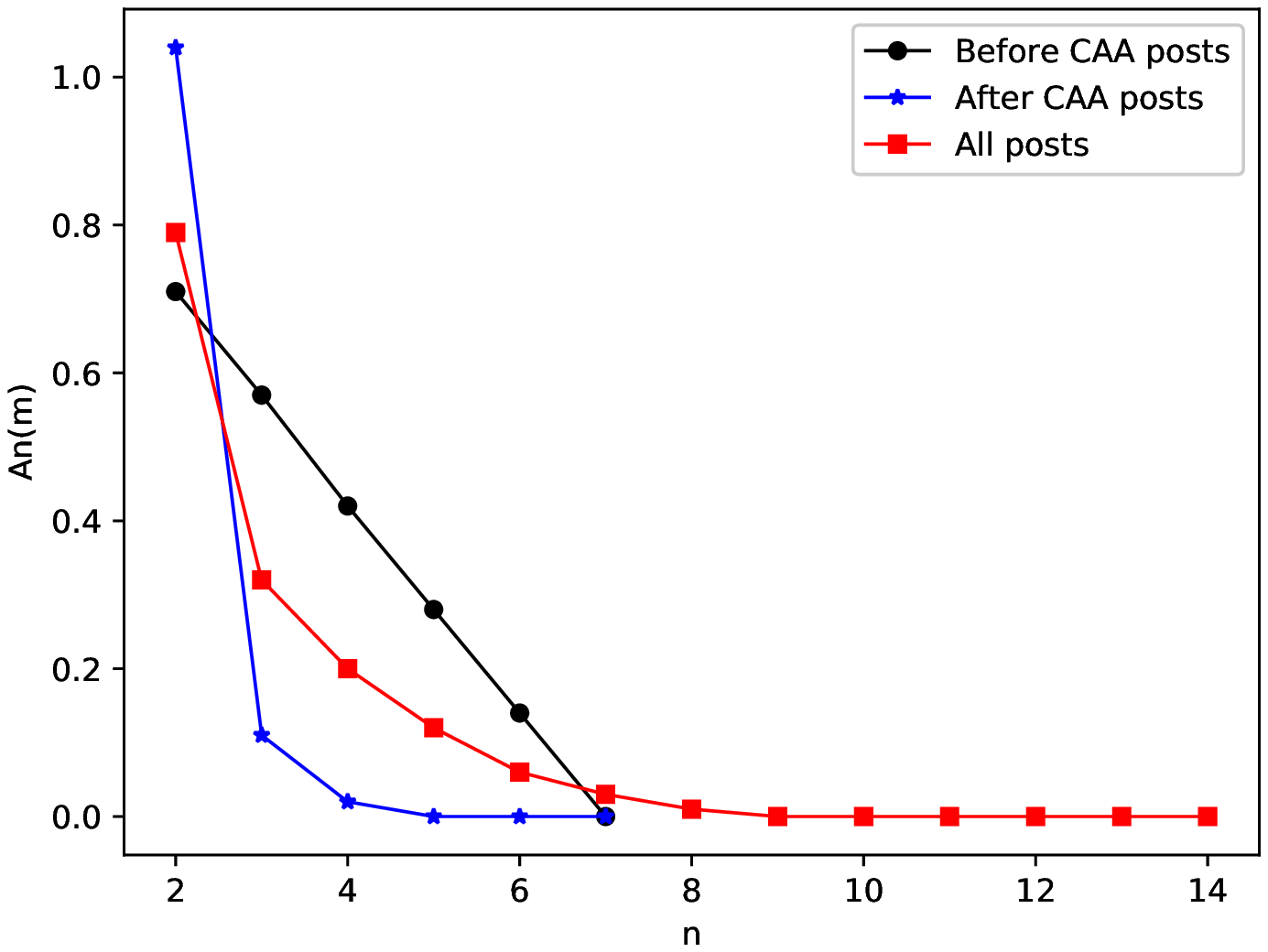} 
\caption{Average number of common profiles $A_n$(all) (resp. $A_n$(before)/$A_n$(after)) as a function of $n$ where $n$ = \{$2,3,\cdots,14$\} for all post (resp. $n$ = \{$2,3,\cdots,7$\} for before/after CAA Facebook posts separately).}
\label{fig6}
\end{figure}

\begin{figure}[!htbp] 
\centering 
\includegraphics[width=7.0in]{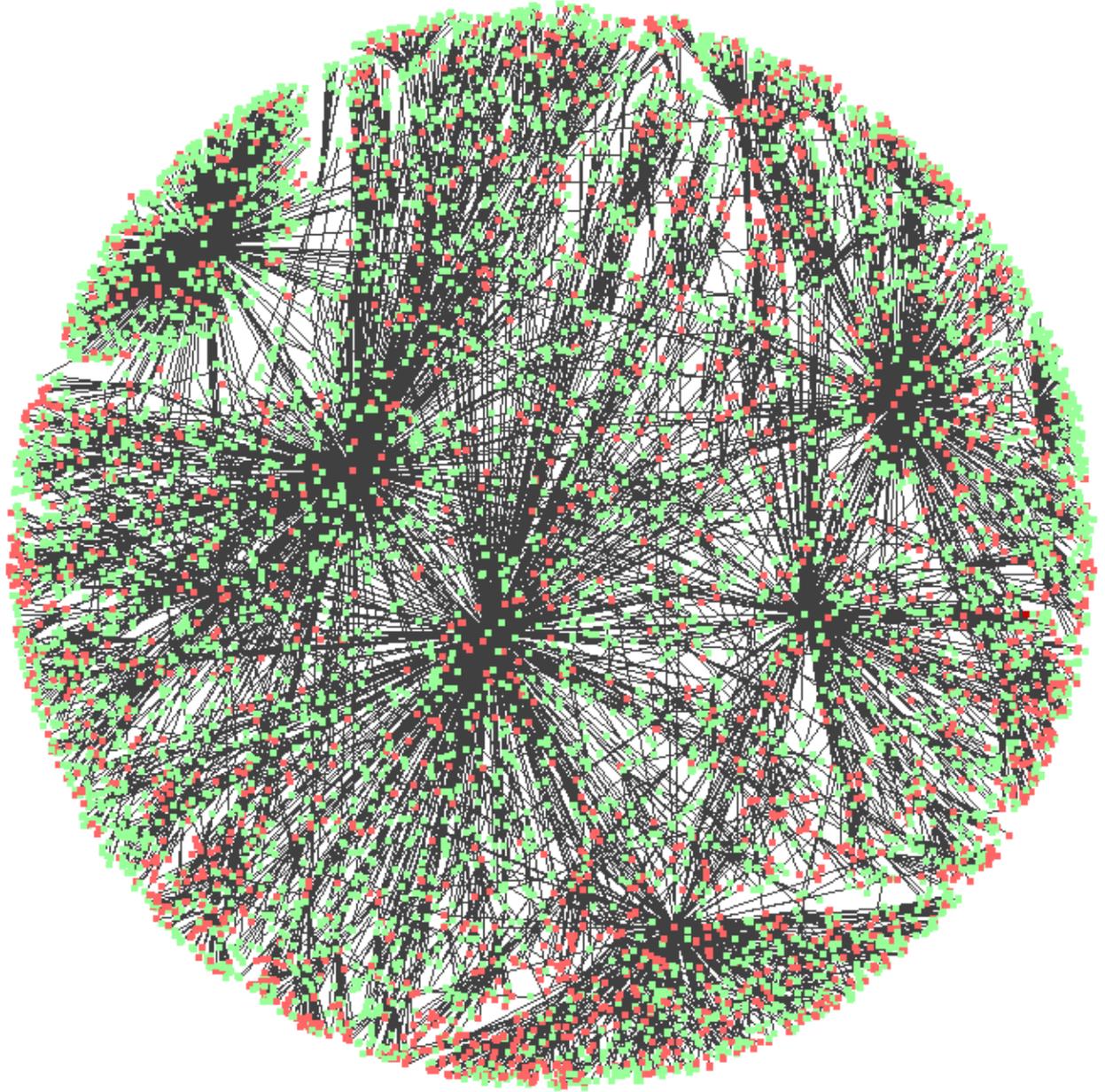} 
\caption{Like-share dynamics of $6$ Facebook posts ($3$ from before and $3$ from after $CAA$) containing $7980$ Facebook profile users where the non-Muslims are marked by red and Muslims are marked by green.}
\label{fig7}
\end{figure}

Now, from a completely different perspective, Fig.~\ref{fig6} depicts the average number of common profiles for all possible permutations of the $14$ Facebook posts, for before CAA $7$ posts, and for after CAA $7$ posts. In a more tranquil socio-political situation, a Facebook post by an activist traditionally gets like or share only from activist community. However, for a mass movement dynamics, it is not true. Therefore, less common profiles in these posts indicate towards localized mass movement 
gravitating with less centralized control. Whereas, large count of common profiles across the posts depict activist dependent centralized control in the movement.
To identify this dynamics, we have collected all the $14$ contra NRC \& CAA posts from same Facebook group. According to Fig.~\ref{fig6}, the average number of common members profile considering all combinations of two Facebook posts together approaches $1$ which indicates again towards the phenomenon of mass movement with less activist dependency. Fig.~\ref{fig7} depicts like-share dynamics of $6$ viral Facebook posts containing total $7980$ Facebook profile users where the nodes depict the unique Facebook profile user identification. Therefore, two nodes representing Facebook profile user are connected via edges if one liked or shared another's post. This  qualitative result shows the visual evidence of the quantitative result. Moreover, the qualitative result of Fig.~\ref{fig7} depicts that vast majority of contagion chain die soon which shows similar signature behaviour with the findings of \cite{Sandra11,Bakshy11,1Haewoon,Sun2009GesundheitMC}.

\section{Analysis of group post content}
\label{section5}
Informational and motivational content of social media posts during protest movements have been briefly investigated in the literature of social media \&  social protests \cite{Tucker2016BigDS,Onuch14,Langer18}. Social scientists have widely discussed about social psychological factors like – moral outrage (i.e. anger or indignation at perceived injustice) \cite{Barbalet98,John12,Kawakami95,Stefan,Nicole11,Wakslak07,Zomeren}, social identification (i.e. a strong sense of group belonging and shared interests) \cite{Smith15,McGarty14,Klandomans97,Kelly96,Drury09,John12} and group efficacy (i.e. beliefs about group efficacy or empowerment) \cite{Mazzoni15,Martijn12,Nicole11}. Here, this section reflects upon the content analysis of Facebook posts during the mobilization period of NRC \& CAA Protest. To do so, we have collected a quadrilingual dataset of $3200$ random Facebook posts from eight contra NRC \& CAA Facebook groups which are in English, Bangla, Hindi and Asomiya language. We have conducted a quantitative analysis of these $3200$ Facebook posts. 

\subsection{Methodology}

To investigate information and motivational contents of these Facebook posts during the mobilization period of NRC \& CAA protest, we have categorized the Facebook posts into following classes. 

\begin{itemize}
\item[] \textit{Class 1:} Information about contra CAA and NRC meeting and protest, posters, slogans. Moreover, we have further divided this class into following sub-classes. 
\begin{itemize}
\item Participation of Female along with Children and trans-gender community in the protest; and
\item Participation of Student community in the protest.
\end{itemize}
\item[] \textit{Class 2:} Emotional and motivational content like poems, songs in the anti CAA and NRC protest event;
\item[] \textit{Class 3:} Knowledge information about CAA and NRC which is about legal and document related complication;
\item[] \textit{Class 4:} General anti - government post about economy, employment and communal politics of government;
\item[] \textit{Class 5:} Miscellaneous content. 
\end{itemize}

Members of our research team hand - coded the classification for the selected posts. Here, every post was coded by three different research assistants and the category of the Facebook post was decided from the decision of the majority.

\subsection{Results and Observations}

Here, $48.40$ \% of the Facebook posts contained exchange of protest information, such as details about date, time, location of protest rally, pictures of mass mobilization in the protest rallies. Moreover, $23.92$ \% of the Facebook posts within this class contained information related to participation of Females in the contra NRC \& CAA protest. In reality, the participation of women in the protest, especially in Shaheen Bagh movement, have been widely discussed in the literature \cite{Nigam,Seema}. Women in the Shaheen Bagh area, Delhi began their peaceful protest on December $16$, $2019$ which has spread to more that $115$ places all over India. According to \cite{Nigam}, from around $20,000$ to $1.5$ lakhs people gathered at the Shaheen Bagh protest site without any patriarch leader. In reality, majority of women in the patriarchal society do not own immovable property under their name and are dependent on their fathers or husbands. Even their maiden names are frequently changed after marriage. Frequently marriages are not registered and getting a birth certificate is a tedious process. With no documents to prove citizenship, women know the problems they are bound to face once the law will be implemented. This reason behind women resistance against NRC \& CAA is also reflected from the field interviews
\footnote{Problems are more in cases of Muslim women, whose life has three parts, before marriage they have surnames like ‘Khatun/Begum’ after marriage they can use ‘Bibi’. In maximum cases Muslim women according to Muslim personal law ‘Bibi’ surnames are there. Now from modern ideologies many do not use ‘Bibi’ but traditionally people used it. Automatically people used to put ‘Bibi’ according to Muslim Marriage Act. After these Muslim women became widow their surname would change into ‘Bewara’. And this was similar in case of my grandmother at first it was ‘Khatun’ then to ‘Bibi’ and later when my grandfather expired voter card was changed and written ‘Bewara’. This is a very big problem and so everyone had to stand in the line for correction of documents. : GM-RM-EM-AR-DNM-L3-O3-A2} 
\footnote{The problem that came to me was that a woman was married and had a child of around 2 years, after which she had problems with her husband and he did not give her any documents. Then the lady returned back to the village and to get new documents she applied for panchayat certificates. These types of problems were seen among women and even for that child. They are still unable to get the corrected documents.: GM-RM-EM-AR-DNM-L3-O2-A2}. 
Moreover, $21.8$ \% of the Facebook posts of this class, i.e. class $1$, contained participation of students in the protest. However, the participation of students in protest was mostly within the elite university campuses of metro cities which was not reflected in the field interviews. For evidence, according to one of the organizer of sit in protest demonstration at a semi-urban centre of West Bengal,

\begin{quote}
``\textit{In cities mainly Kolkata, there is a place for movement to develop because of the few universities of Kolkata. Those universities have got many educated people who have a better understanding about Indian political background. From that basis of political background they hold their protest. There have been protests in Jadavpur University, but how many such protests were there in Jadavpur Lok Sabha Constituency? There have been no such protests related to NRC-NPR-CAA in Jadavpur Lok Sabha Constituency. The protest that happened mainly involved students of Jadavpur University. This is main advantage in Kolkata, and so the protests go on.}"
\end{quote}

\begin{figure}[!htbp]
\begin{tabular}{cc}
\includegraphics[width=70mm]{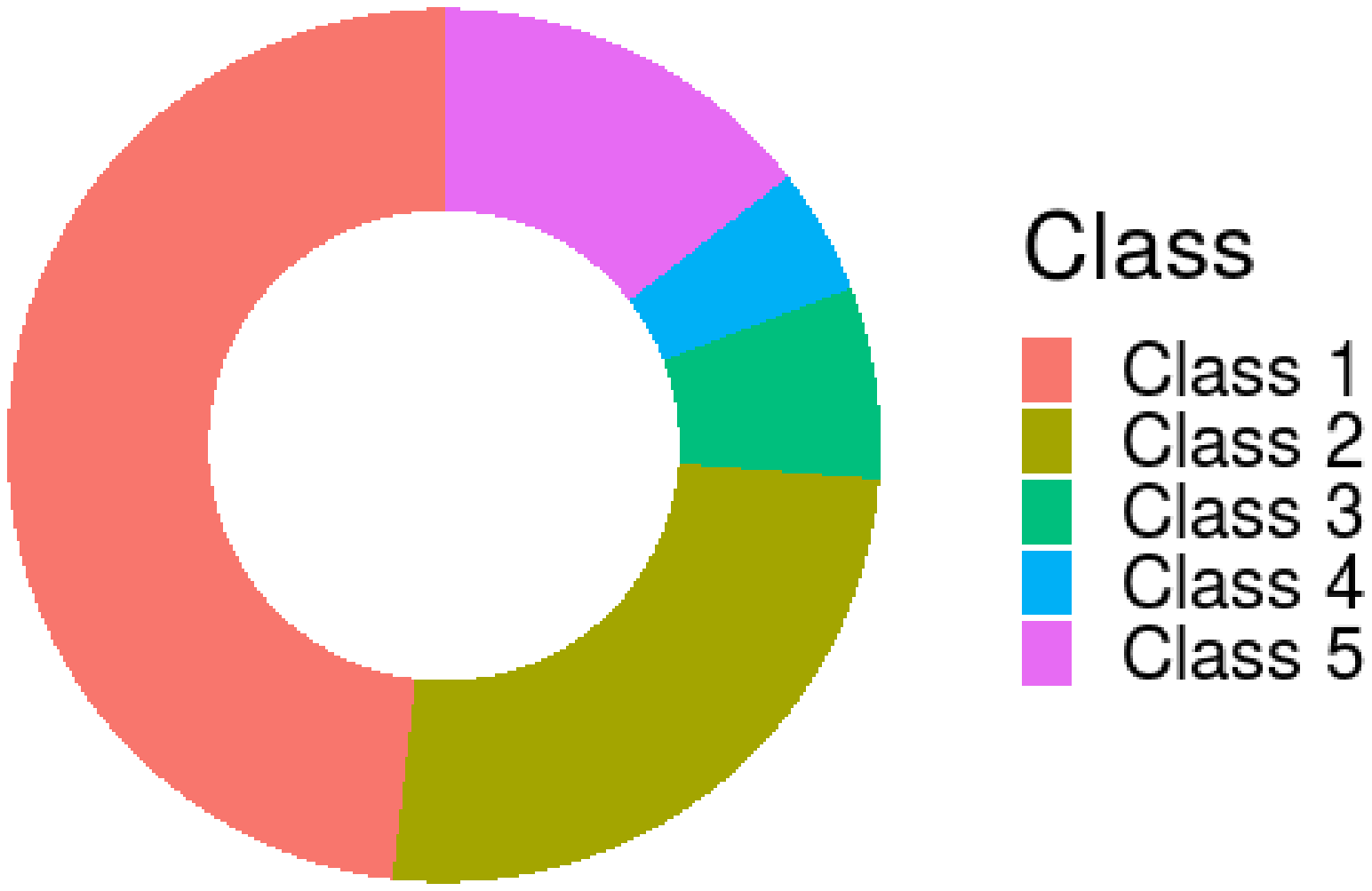} & \includegraphics[width=70mm]{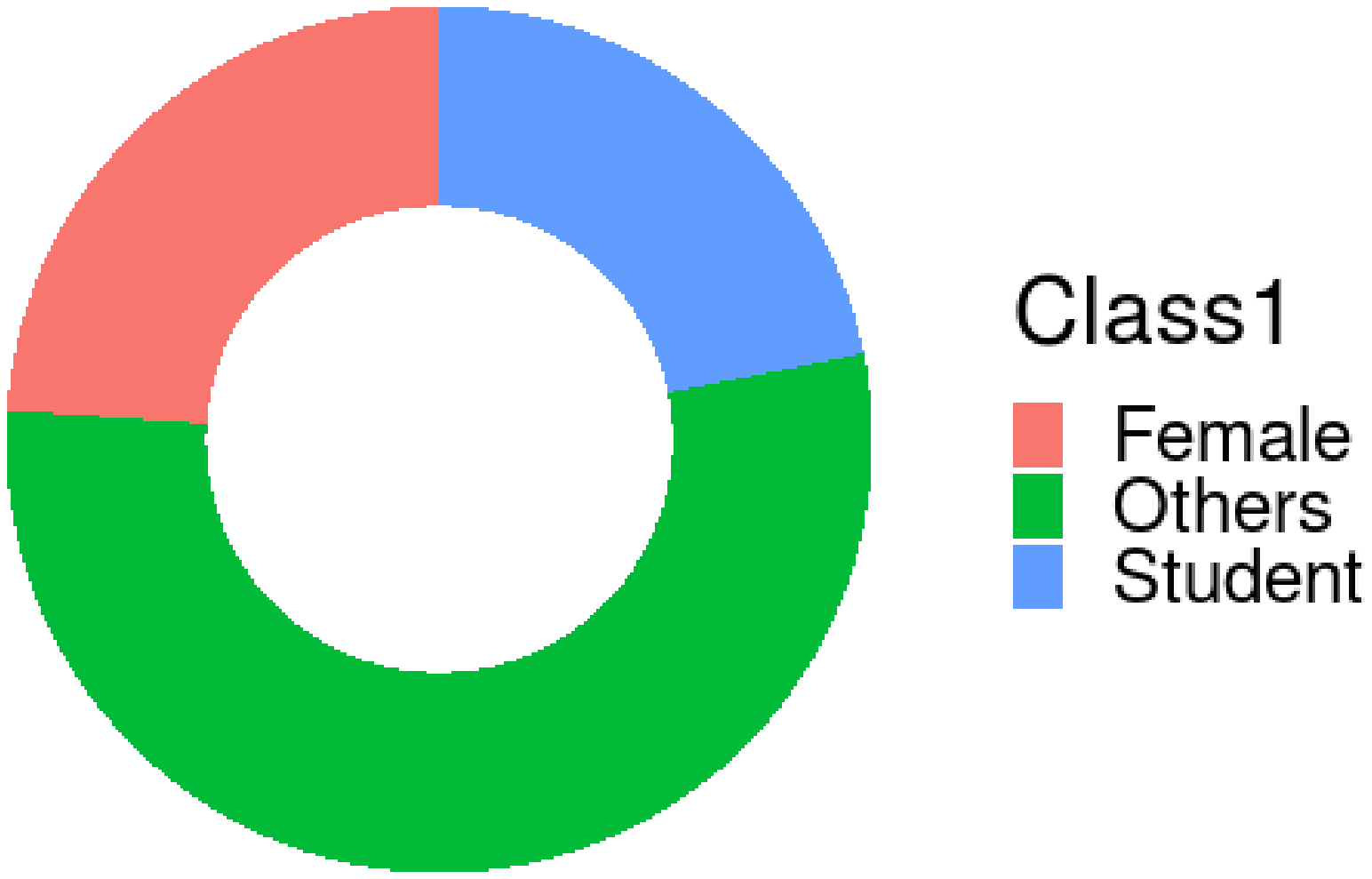}   \\
(a) & (b) \\
\end{tabular}
\caption{\% wise result of the Facebook posts (a) for the classes $1$-$5$ and (b) result within class $1$.}
\label{fig1x9}
\end{figure}

On the other hand, 25.18\% of Facebook posts contained emotional and motivational content, such as news of death of protester during protest, news of death of people in the Detention camps, poems and songs about why one should take part in the protest movement, why this process is harmful for both the original inhabitants and the migrant people, why this procedure is threat to the Muslim community etc.. In this class, a sizable number of Facebook posts contained the heartbreaking incidents of Assam after NRC implementation. In this regard, we observed that impact of Assam NRC ($14.94$ \%) are discussed by people during the field interviews. Here, $7.15$ \% of Facebook posts contained knowledge information about the Government procedure of NPR, NRC and CAA. This is corroborated by the field interviews, where the detail knowledge about complex technicalities of implementation on NRC \& CAA  ($4.63$ \%) are responded by very few people which depicts similar behaviour. Moreover, we are not able to categorize 14.53\% Facebook posts in any category which remains as miscellaneous posts. However, these Facebook posts are not always explicitly 'irrelevant' or 'spam'. Sometime, these Facebook posts are associated  with anti government feelings.

\begin{landscape}
\begin{table}[!htbp] 
\centering

\begin{tabular}{l||l} \hline 
Class  & Example of Facebook post  \\ 
\hline
Class 1 & On behalf of Joint forum against NRC, the sit-in demonstration which took place \\
&  in the Chaplin Square (New Market) which began on 21st of January has come to \\

&  an end today at 9pm through a rally to Gandhi Statue. \\
&  \\
Female &   The protest against NRC-CAA-NPR began from Rajbazaar more. There is an increase \\

Participation &  in the number of people joining the protest. There are women in the front row.\\
&  \\ 
Student &   Protest against CAB and NRC at Pune University Maharashtra \# StudentPower.\\
Participation & \\

&  \\
Class 2 & Even after being sick, Samida Khatun used to come to Park Circus for future well being\\
&  of human beings and used to protest against CAA. She passed away while protesting.\\
&  \\
Class 3 &  NPR rules do not allow Modi government to collect data on parent's birthplace \\
&  and Aadhaar.\\
&  \\

Class 4 &  \# NRC \# CAA I think that it's just a strategy to divert the attention of Indians from \\
&  the current economic status of the country\\
&  \\

Class 5 &  Cowdung is a peculiar object - when dried it becomes fuel, in contact of soil it becomes \\
&   fertiliser, wonder what happens when it enters the brain?\\ \hline
\end{tabular}
\caption{Examples of the Facebook posts for each class.}
\label{table1x}
\end{table}
\end{landscape}

Fig.~\ref{fig1x9} depicts the summarized results of 3200 Facebook posts. Moreover. Table~\ref{table1x} shows the example of Facebook post for every class. In comparison with literature, the result of Fig.~\ref{fig1x9} shows similar signature behaviour with tweet activity during Occupy Wall Street protest. Langer et. al. \cite{Langer18} have qualitatively analysied more than 7000 tweets on Occupy Wall Street protests that occurred in New Yark city where 44\% of tweets contained informational content, i.e details about protest location, safety, police presence.

Note that, here, this study conducted on Facebook posts during the mobilization period of contra NRC \& CAA protest starting from mid-December, $2019$ to mid-February, $2020$. 
However, the situation drastically changed in March after commencement of the lock-down due to the spread of COVID-$19$. For evidence, we have investigated the informational contents of Facebook posts in the contra NRC \& CAA Facebook groups during the lock-down period following the same methodology. In this study, we have collected 500 posts randomly during the last week of March (1st week of lock-down period in India). It is interesting to note that only $10.8$\% Facebook posts contained information about NRC \& CAA. Along with that, $29.4$\% Facebook posts, out of total 500 Facebook posts, contained issues related with NRC \& CAA and situation of COVID-$19$ both. However, $53.6$\% of Facebook posts contained information about issues related with only COVID-$19$, that is, general and health information related with COVID-$19$, situation of working class people and failures of government during the lock down period. 

\section{Spatiotemporal analysis of individual activity }
\label{section6}

This section studies the Twitter activity during the mobilization period of NRC \& CAA protest. To do so, we have collected tweets for $48$ hash-tags during the period from December 10,2019 to February 10, 2020. The 48 hash-tags contain total 5,05,792 tweets during the two month mobilization period of NRC \& CAA protest. The study deals with 4,90,978 tweets, out of 5,05,792 tweets and the rest are excluded from the study due to incomplete information. Here, ``\#CAAprotests" hash-tag contains highest number of 54,554 tweets. Each row of the dataset contains the unique username of corresponding Twitter profile, location (i.e. city name), country name, number of retweets, date and time of tweet, content of the tweet. In literature, most of the studies have dealt with growth and structure of Twitter network from social movement prospective \cite{Sandra11,Pablo15}.

\subsection{Dynamics of number of tweets}

To understand the dynamics of number of tweets per day, we have plotted number of tweets against time during the period of December $10$, $2019$ to February $10$, $2020$.
This section shares similar notation with Section~\ref{section1} where $N$($k$)  indicates total number of tweets present at instant $k$. Total time of the study is divided into slots of $2$-days interval. Here, $N$($n$) depicts the total number of tweets at the end of our time window, i.e. February $10$, 2020. Therefore, $N$($n$) = $4,90,978$.

\begin{figure}[!htbp]
\begin{tabular}{cc}
\includegraphics[width=70mm]{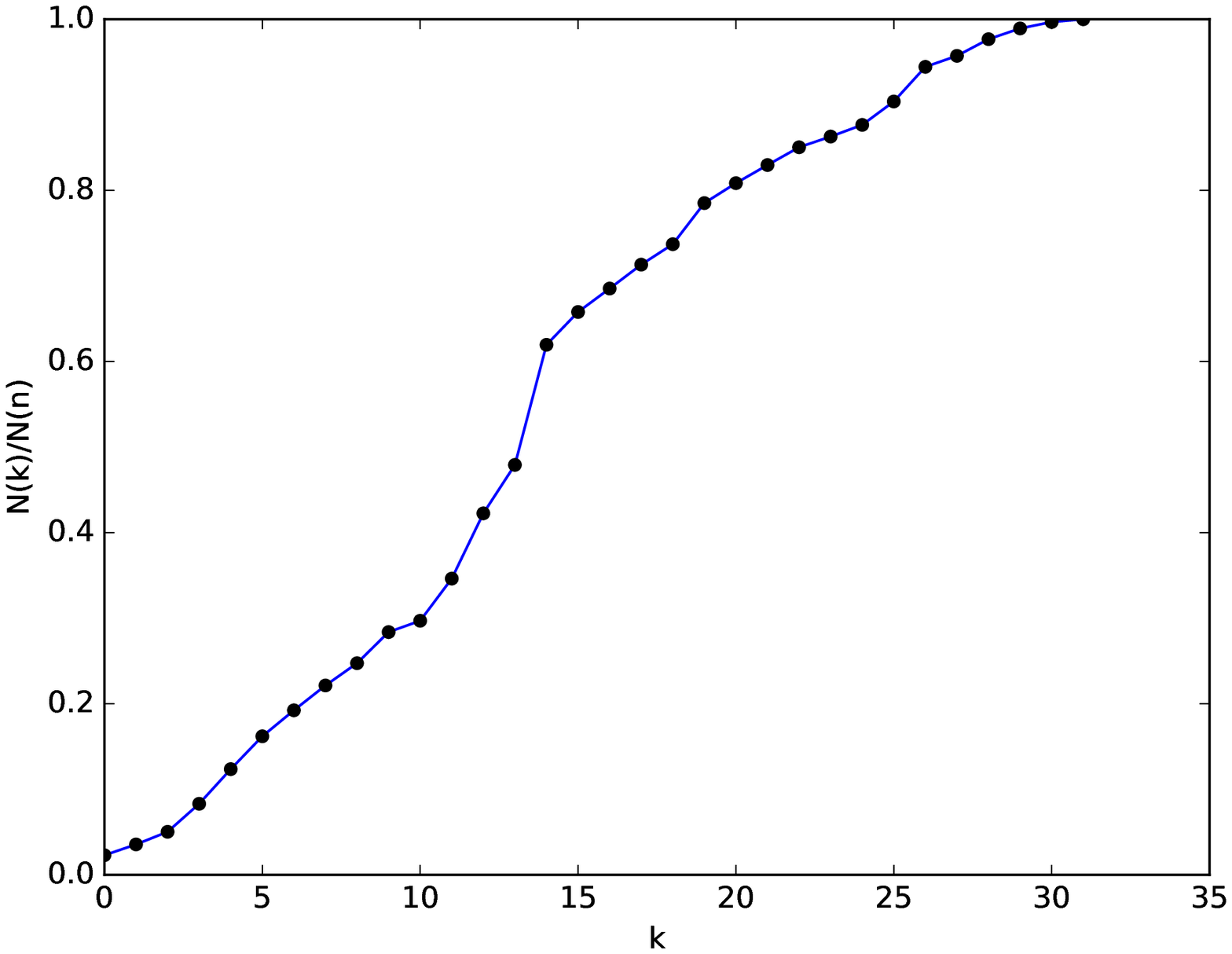} & \includegraphics[width=70mm]{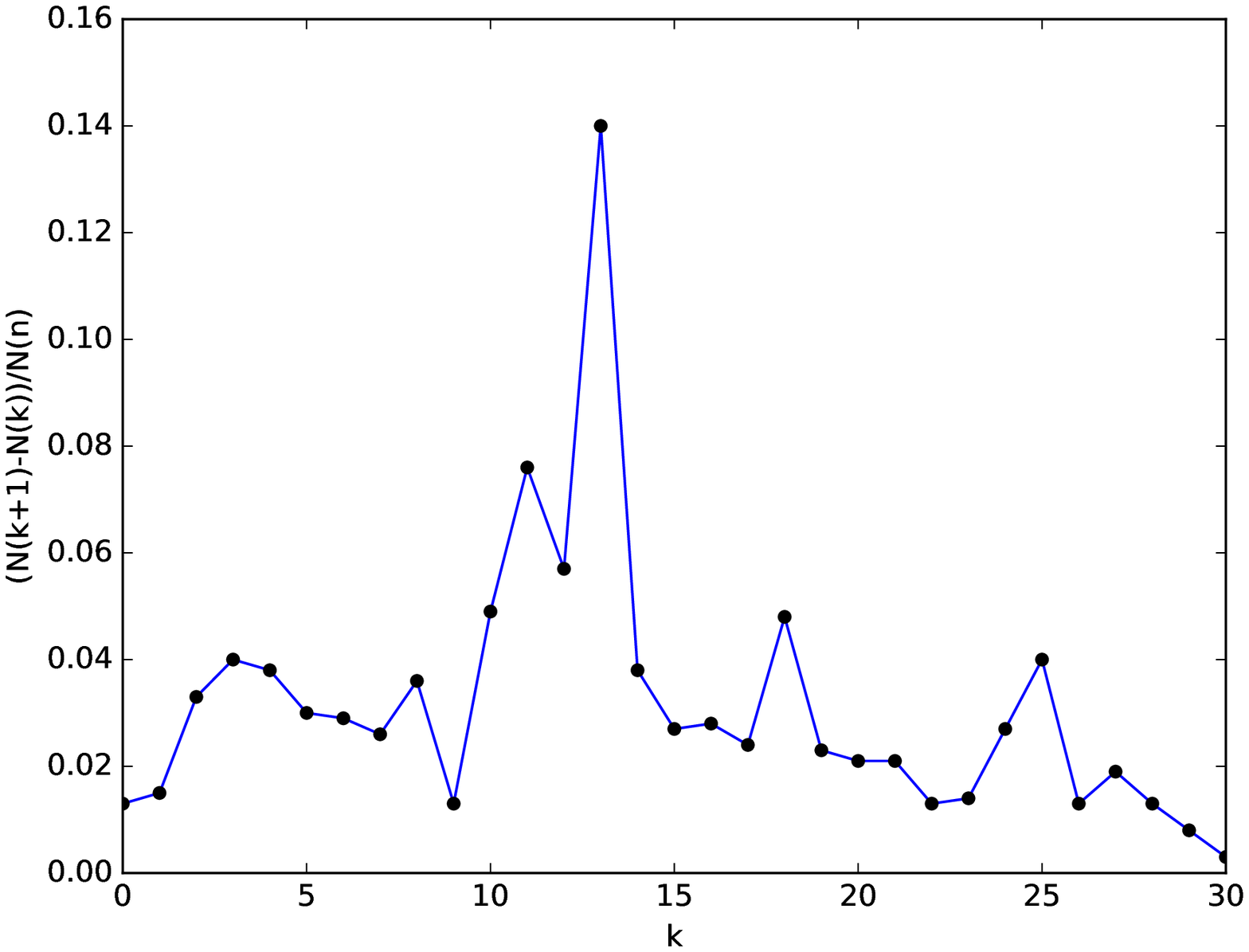}   \\
(a) & (b) \\
\end{tabular}
\caption{The growth of (a) $N(k)$/$N(n)$ over time; (b) ($N$($k+1$) - $N$($k$))/$N$($n$) against $k$.}
\label{fig1x}
\end{figure}

As a result, Fig.~\ref{fig1x}(a) depicts the growth of the normalized number of tweets.
Fig.~\ref{fig1x}(b) shows the acceleration in tweet count. These plots depict the following dynamics of tweets. 

\begin{itemize}
\item[1.] There is sudden rise (highest peak) in the plots of Fig.~\ref{fig1x} after $k$ = $13$, which represents the rise in number of tweets after January $6$, $2020$. Note that, in the first week of January $2020$, the goons attacked students in Jawahar Lal Nehru University. In fact, the higher education has becomes a battle field in the war of culture and ideologies \cite{nature}. If we follow the rise of tweets with respect to total number of tweets, we find that the peak is touched during that time (see Fig.~\ref{fig1x}). This is an indication that the people are increasingly joining the protest movement after January first week.

\item[2.] Moreover, the plots of Fig.~\ref{fig1x} show the first peak after $k$ = $3$, which represents December $16$, $2019$. In this context, the most popular Shaheen Bagh protest began on December $16$, 2019 following the violent brutal attack by Delhi police on the students of Jamia Millia Islamia University on December $15$, $2019$ \cite{Nigam,Seema}. 

The plots show the last peak after $k$ = $25$ which represents date after January $29$, $2020$. Here also, on January $30$, $2020$, an armed Hindu fundamentalist man fired at a crowd gathered in Jamia Millia Islamia University.

\item[3.] It is further observed that $10$ hash-tags, out of $48$, are associated with protest movement by students. $1,24,588$ tweets are related with these $10$ hash-tags which is $24.63$\% of total number of tweets. On the other hand, 4 hash-tags, out of $48$, are associated with protest by Muslim women which is $7.6$\% (38,472) of total number of tweets. This is an indication of participation of students and Muslim women in the protest movement. The details of Facebook posts on female and student participation in the protest and their relationship with field responses are discussed in Section~\ref{section5}. The twitter data also reflects similar signature behaviour. 
\end{itemize}

\subsection{Dynamics of location}

The dataset contains the location of twitter user profile (where available in the twitter profile). Here, the dataset depicts the list of $82$ countries across world and $47$ cities within India along with number of tweets from the corresponding country/city. The location of tweets can be able to give an insight about the relationship between ground reality and online activism. In literature, in case of Egyptian revolution of $2011$, less than $30$\% of tweets originated in Egypt \cite{Starbird12}. However, most of the tweets, out of $30$ million tweets, were sent in Turkish from inside the country during Turkish protest of $2013$-$14$ \cite{Tucker2016BigDS}. 

\begin{figure}[!htbp]
\begin{center}
\begin{tabular}{c}
\includegraphics[width=120mm]{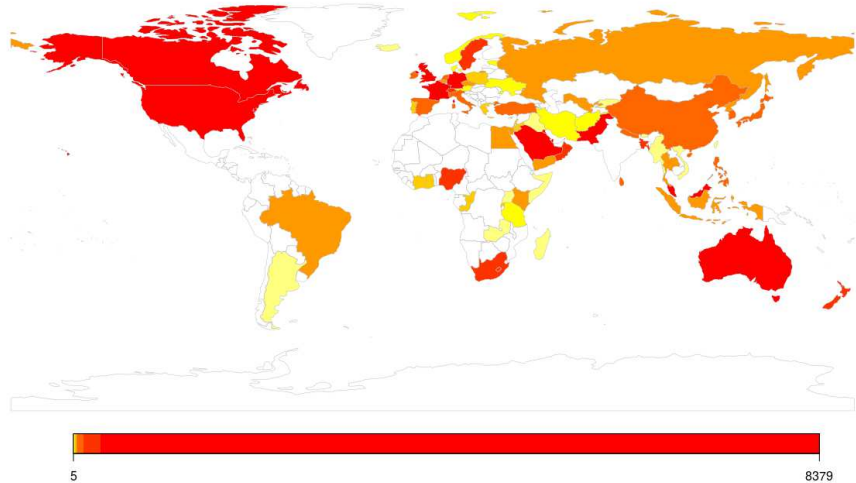}  
\end{tabular}
\caption{Countries across the world from which tweets originated during the mobilization period of contra NRC \& CAA protest (outside India).}
\label{fig2x}
\end{center}
\end{figure}

Fig.~\ref{fig2x} depicts countries (except India) across the world with number of tweets as parameter during the mobilization period December $10$, $2019$ to February $10$, $2020$ of contra NRC \& CAA protest. Note that, out of 4,90,978 tweets, 2,64,071 tweets are associated with the specific location of the user twitter profile. Out of these 2,64,071 tweets, 2,33,960 tweets (88.60\%) are from India, whereas 12,992 (4.92\%) (resp, 9,716(3.68\%); 5,045(1.91\%)) tweets are respectively associated with location in Middle east countries, USA and Europe. That is, most of the tweets ($\approx 90\%$) originated from location in India which reflects similar signature with Turkish protest of $2013$-$14$ \cite{Tucker2016BigDS}, but not with  Egyptian revolution of $2011$ \cite{Starbird12}. 

\begin{figure}[!htbp]
\begin{center}
\begin{tabular}{c}
\includegraphics[width=80mm]{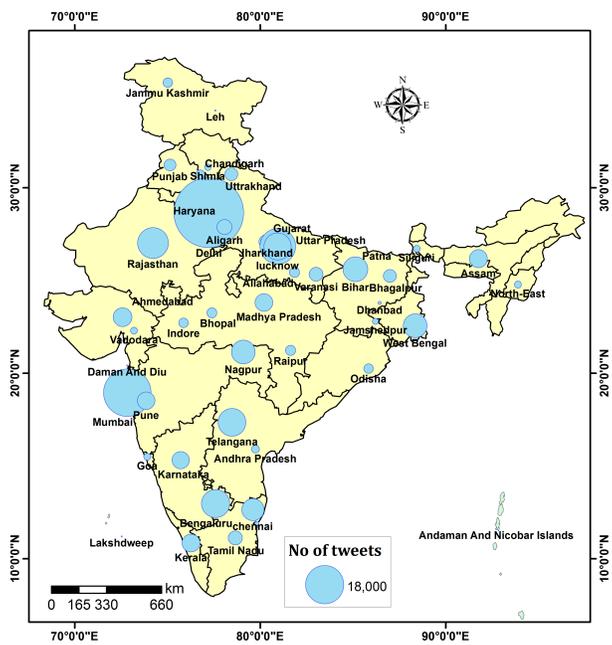}  \\
\end{tabular}
\caption{Tweets density in Indian cities during the mobilization period of contra NRC \& CAA protest.}
\label{fig3x}
\end{center}
\end{figure}

However, in India also most of the tweets are associated with location in Metro cities. Here, 61,353 (resp. 28,399; 10,016; 7,098), i.e. 26.22\% (resp. 12.14 \%; 4.28 \%; 3.03 \%), tweets are associated with metro city location Delhi (resp. Mumbai; Bangalore; Kolkata). Fig~\ref{fig3x} depicts location wise tweets density in Indian cities during the mobilization period of contra NRC \& CAA protest. In conclusion, this is an indication that the people on twitter network of contra NRC \& CAA protest are mostly from metro cities and belongs to elite class.

\subsection{Structure of twitter network}

Structure of twitter network during mass mobilization period of social protests have been widely discussed in the literature, in case of mass mobilization in Spain $2011$ \cite{Sandra11}, `united for global change' demonstration in May $2012$ \cite{Pablo15}, Istanbul's Gazi park protest in May $2013$ \cite{Jost78,Pablo15}. These studies mainly focus on dynamics of protest recruitment and information flow with respect to the role of active and moderately active twitter users. In this study, we explore the survival dynamics of online activism during the mobilization period of contra NRC \& CAA protest.

The $4,90,978$ tweets during the two month mobilization period of contra NRC \& CAA protest are tweeted by $1,23,019$ unique twitter user. That is, on average each user has twitted four times during the corresponding two month mobilization period of contra NRC \& CAA protest. To understand more detail dynamics, let us consider, $X$, $Y$ are sets with twitter users having tweeted in time step $t$ and $t+1$. The size of the sets are $x =$ $\mid X \mid$ and $y =$ $\mid Y \mid$. Therefore, we define, 
\begin{itemize}
\item[] Survive = $\mid X \cap Y \mid$ (Twitter users having tweeted);
\item[] Birth = $\mid Y  - X \mid$ (Newly active twitter users);
\item[] Death = $\mid X  - Y \mid$ (Twitter users active only in previous).
\end{itemize}

\begin{figure}[!htbp]
\begin{tabular}{ccc}
\includegraphics[width=50mm]{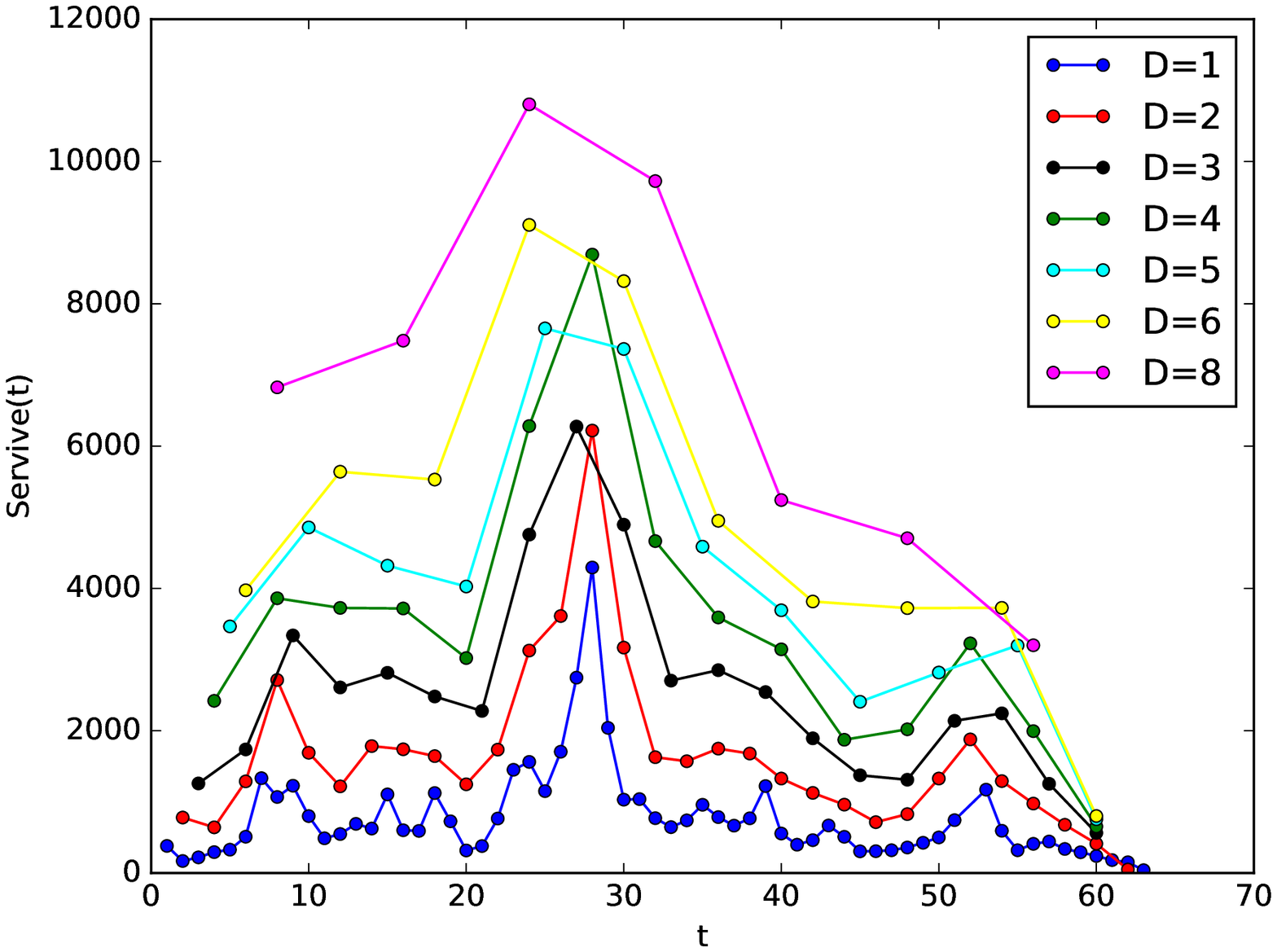} & \includegraphics[width=50mm]{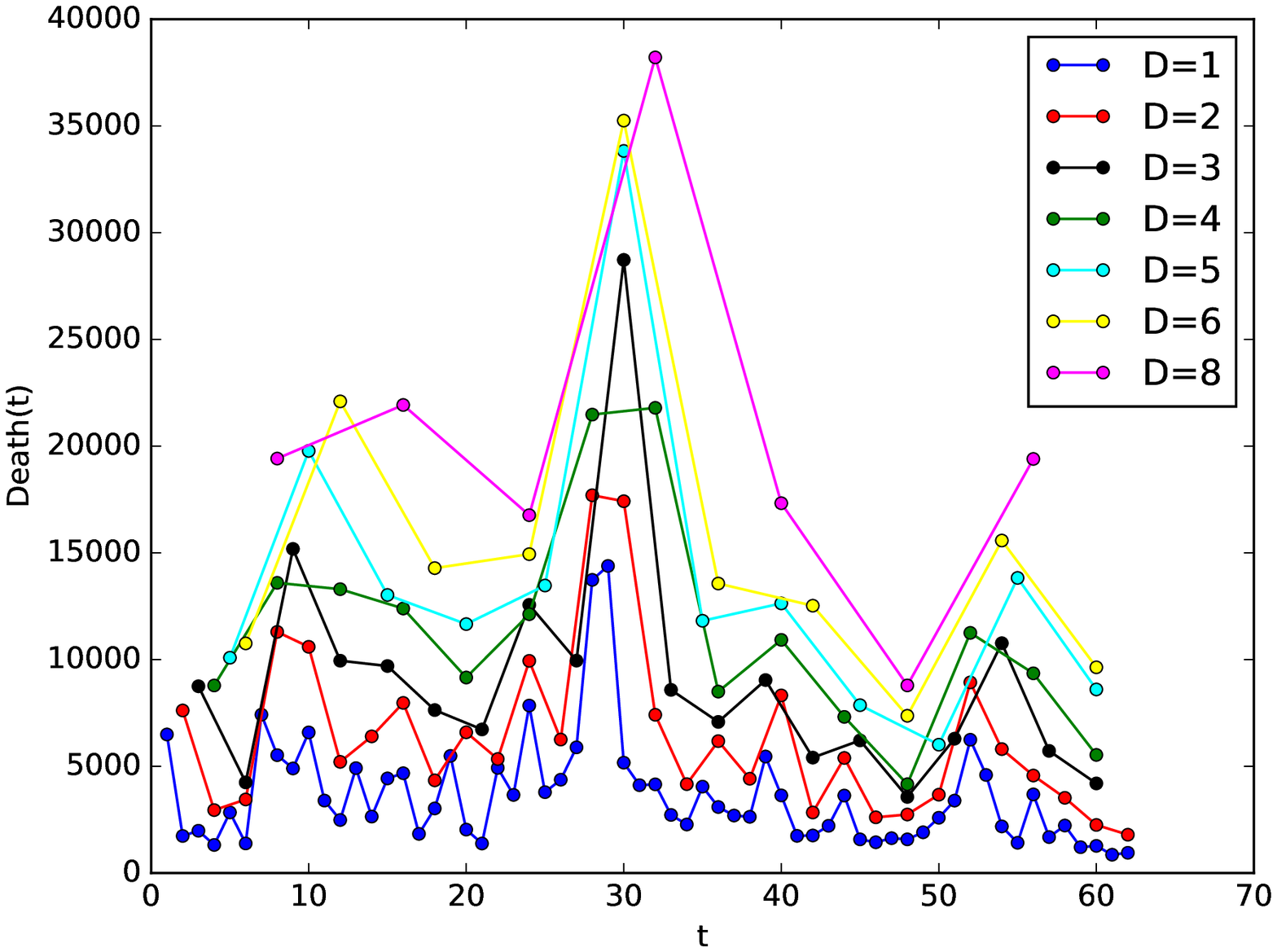} & \includegraphics[width=50mm]{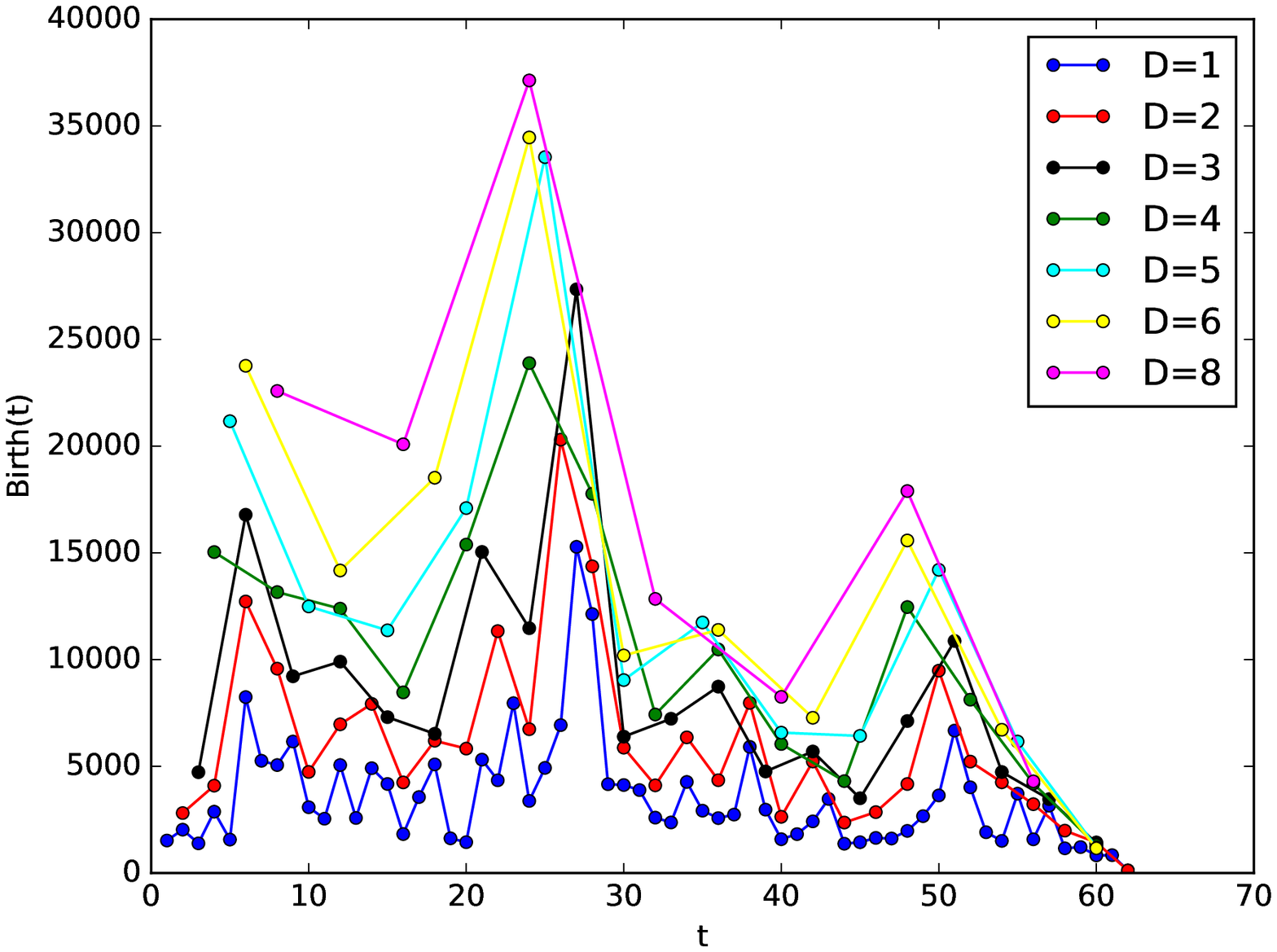}  \\
\end{tabular}
\caption{Change in Survive, Death, Birth with time for D = \{1,2,3,4,5,6,8\}.}
\label{fig4x}
\end{figure}

Here, ``$\cap$, $-$" respectively notes the notion of intersection and difference following classical set theory notation. Different size of time step is considered where $D = n$ depicts time step of size $n$ days ($n \times 24$ hours). Now, Fig.~\ref{fig4x}~(a),~(b),~(c) respectively shows change in actual number of survive, death, birth for D = \{1,2,3,4,5,6,8\}. That is, we plot Survive(t), Death(t), Birth(t) against $t$ in Fig.~\ref{fig4x}.  

\begin{figure}[!htbp]
\begin{tabular}{ccc}
\includegraphics[width=50mm]{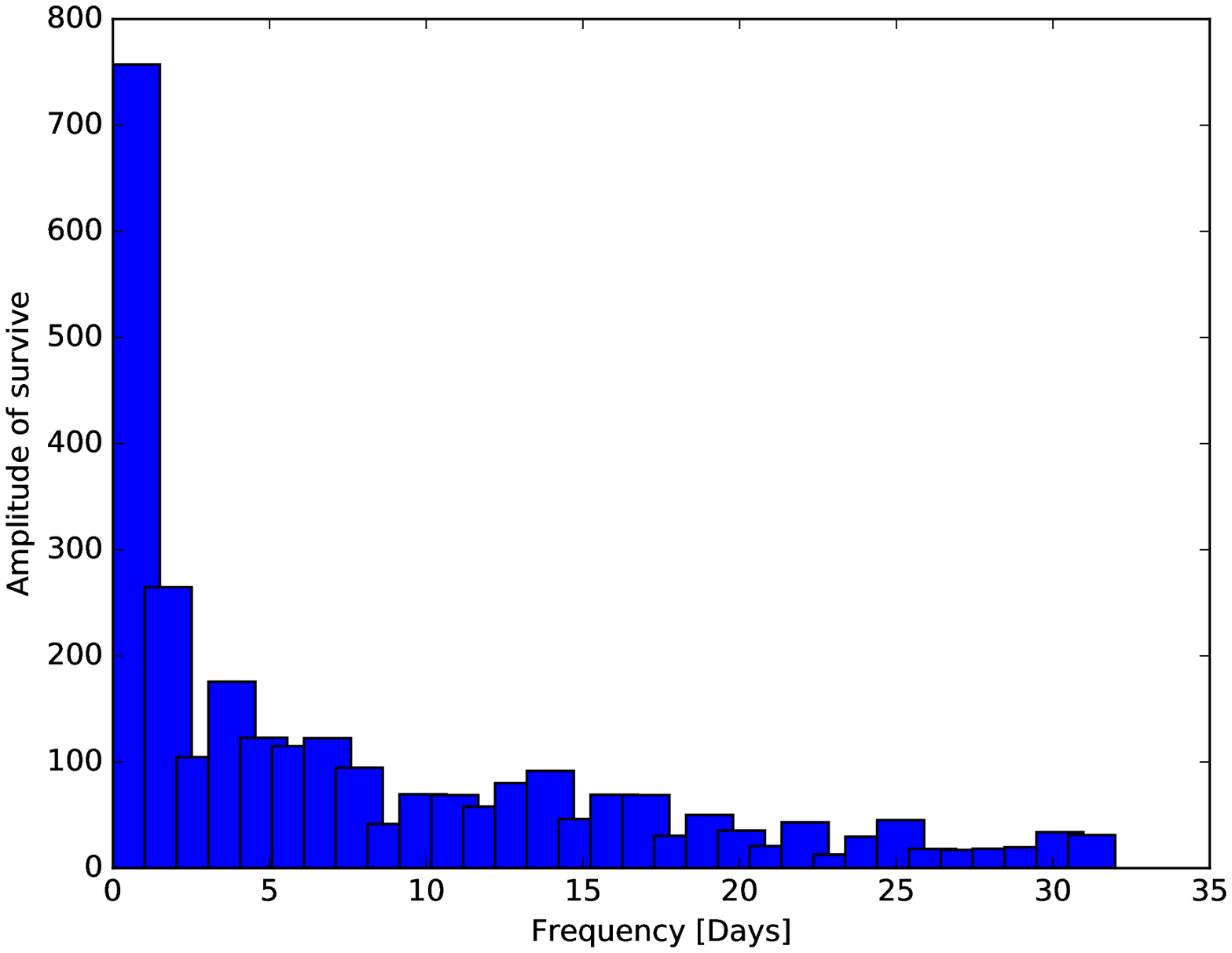} & \includegraphics[width=50mm]{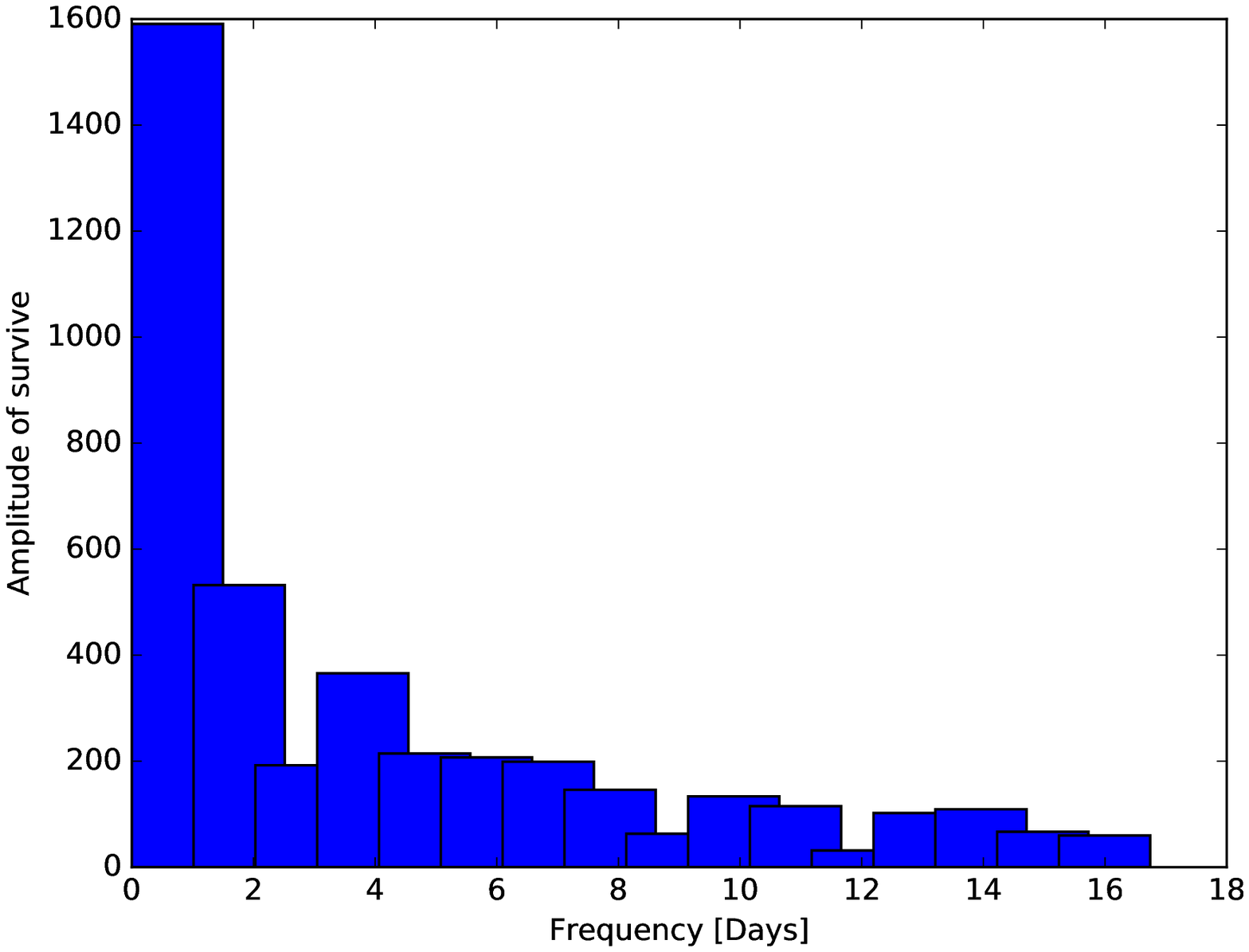} & \includegraphics[width=50mm]{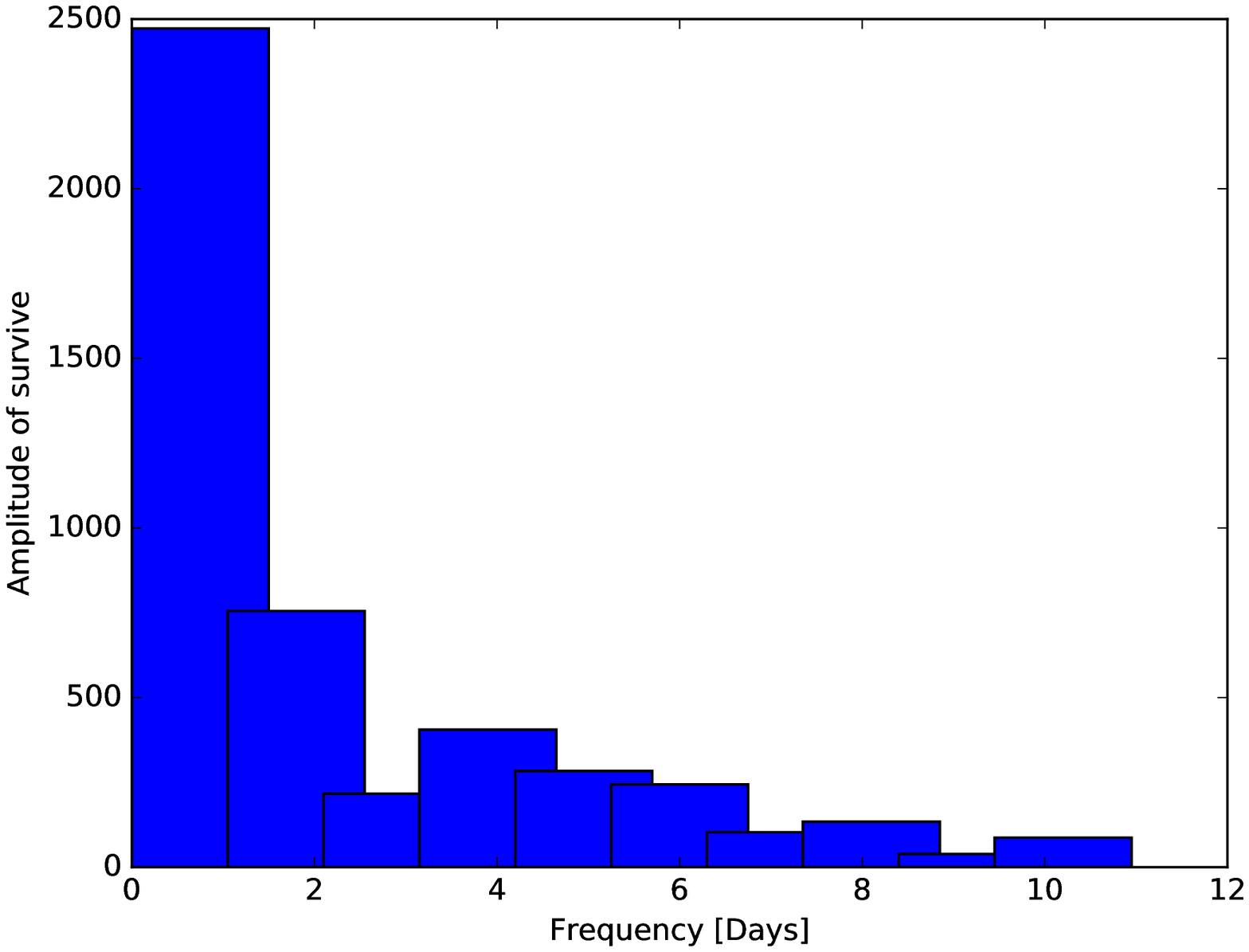}  \\
 D = 1 & D = 2 & D = 3\\
\end{tabular}
\caption{Amplitude of survive over frequency (days) for D = \{1,2,3\}.}
\label{fig5x}
\end{figure}

To understand frequencies involved, fast Fourier transformation (FFT) \cite{Charles} has been computed using standard subroutines. The FFT of a time series can provide an idea about the dominant frequencies associated with the time series data. In the parlance of signal processing, the time domain signal is expressed as sum of signals having known frequencies with varying amplitude that depends on the actual data values. Here the dataset comprises the count of unique individuals tweeting on a given day. Hence the FFT analysis computes the amplitude (modulus used for handling the complex number) for the defined frequencies. It resembles the count of unique individuals who tweet as frequently as daily, on alternate days, twice a week and so on. In conclusion, it can be seen that there is no dominant group of individual users who tweeted regularly on the issue, as is expected in case automated fake tweets are fired which contradict with the widespread usage of (fake/bot) accounts in Indian political twitter network \cite{Ganguly19}. Fig.~\ref{fig5x} depicts amplitude of survive over frequency (days) for D = \{1,2,3\}.

To summarize, the peaks of number of tweets tally with ground events during the mobilization period of contra NRC \& CAA protest. However, the twitter activity only reflects from elite class of metro cities. Moreover, though there is no evidence of presence of (fake/bot) twitter accounts in contra NRC \& CAA twitter network, the survival rate of twitter users depicts no evidence of recruitment patterns for online activism as well. Therefore, It may be true that the intention of twitter activity is to make `contra NRC \& CAA' hashtags as only twitter trends to seek the attention of global spectator. 

\section{Field interview findings}
\label{section7}

\subsection{Dataset}
\begin{figure}
\begin{center}
\begin{tabular}{ccc}
\includegraphics[width=50mm]{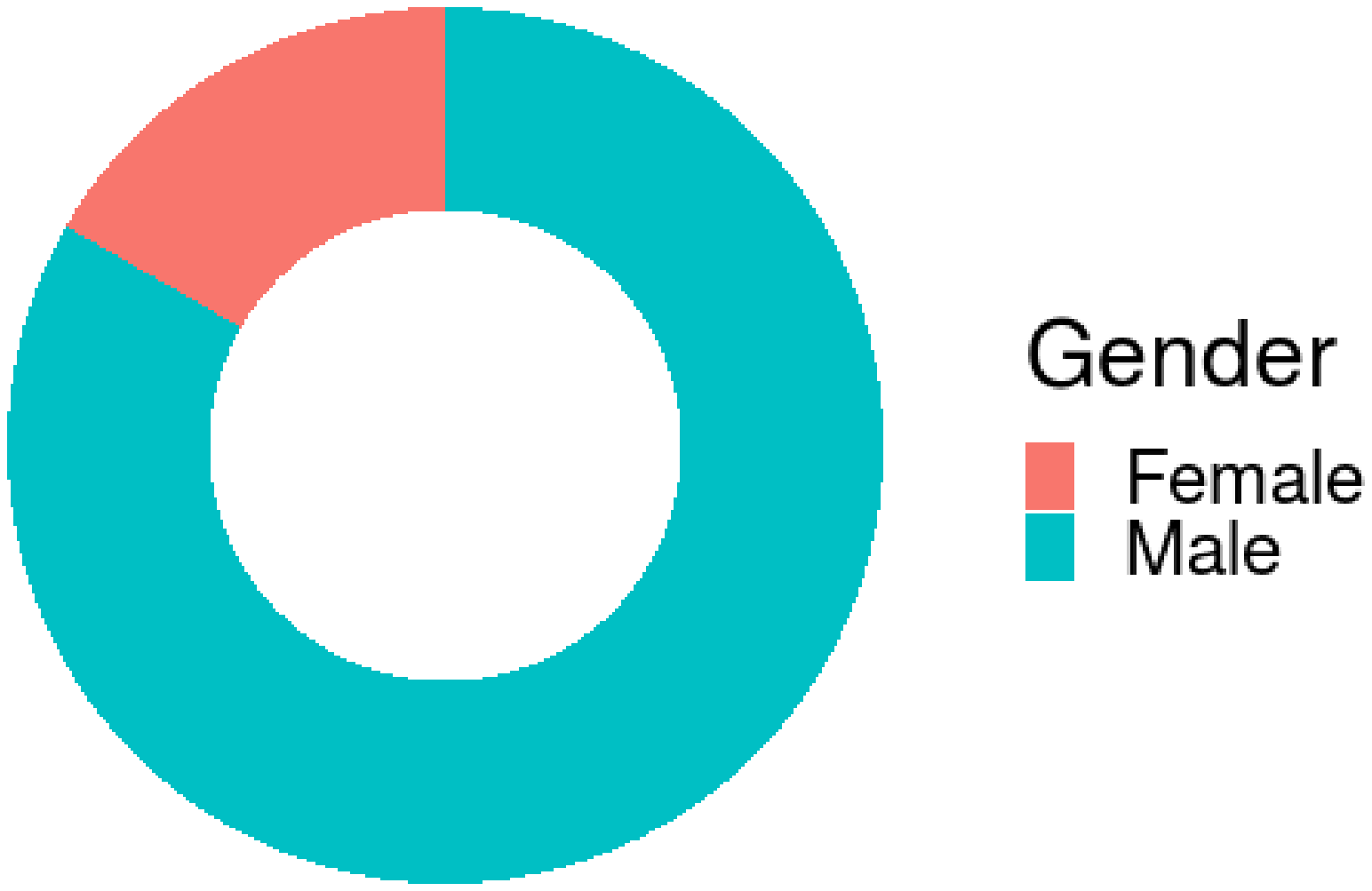} & \includegraphics[width=50mm]{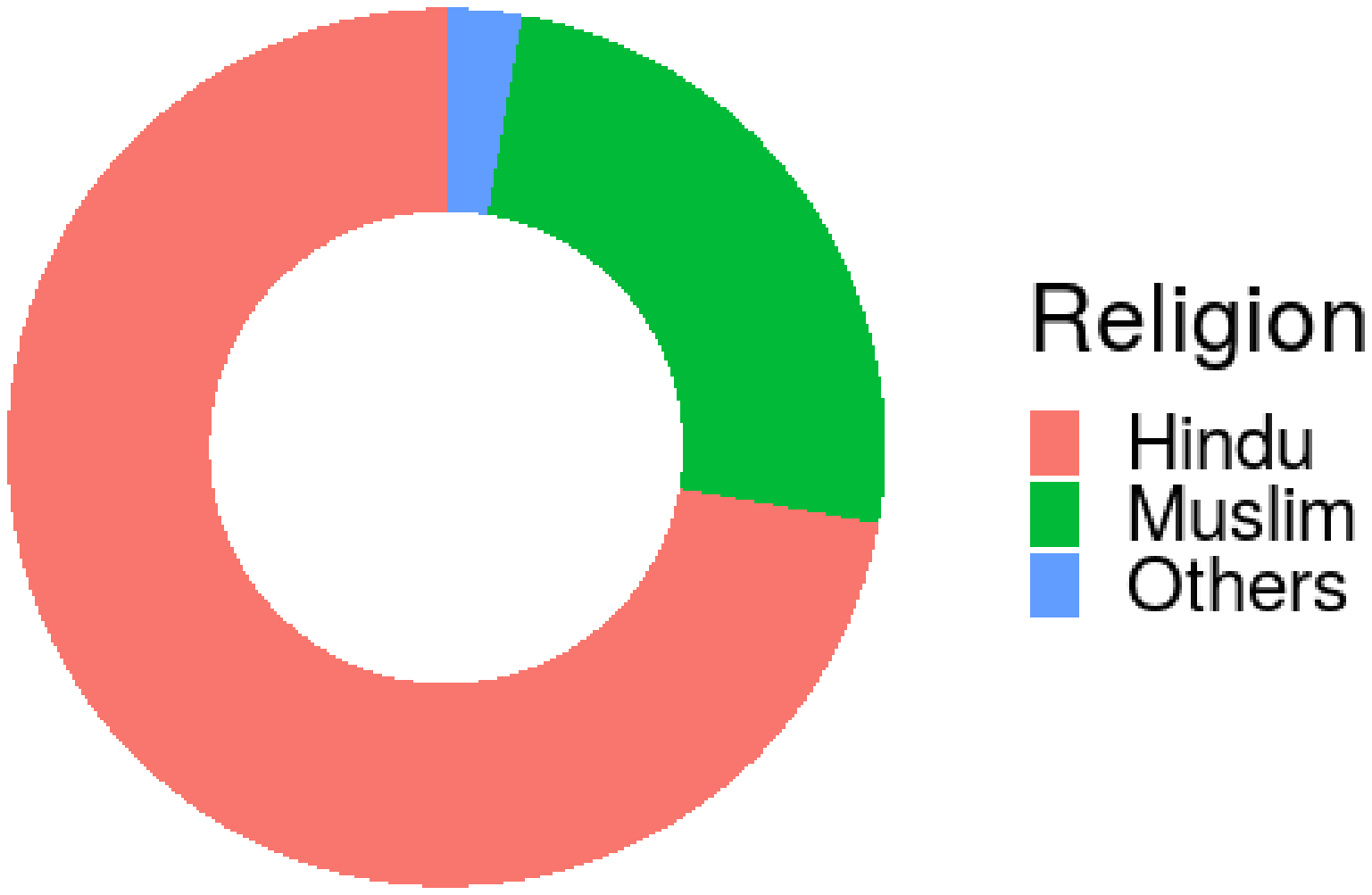} & \includegraphics[width=50mm]{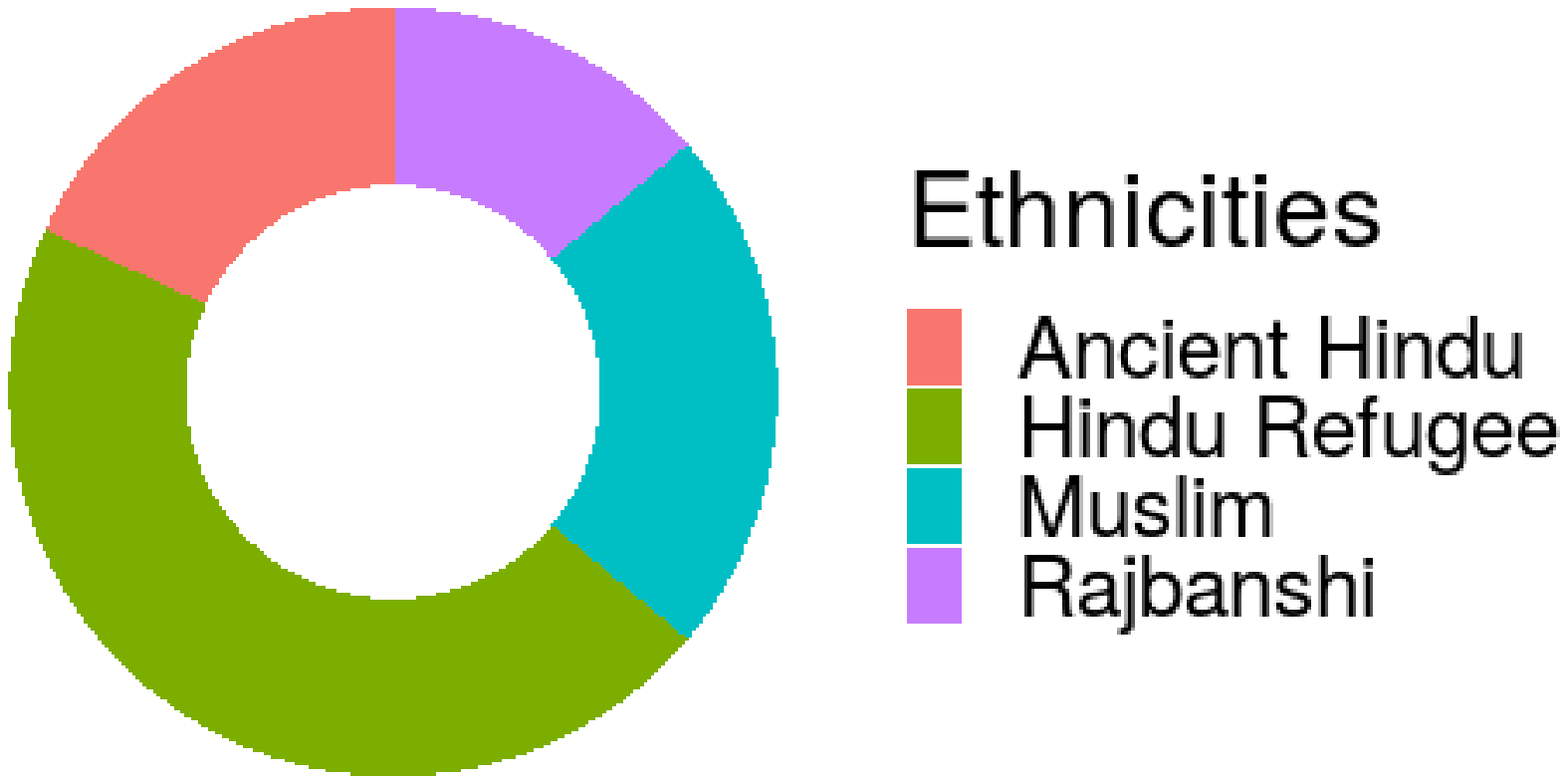}  \\
\includegraphics[width=50mm]{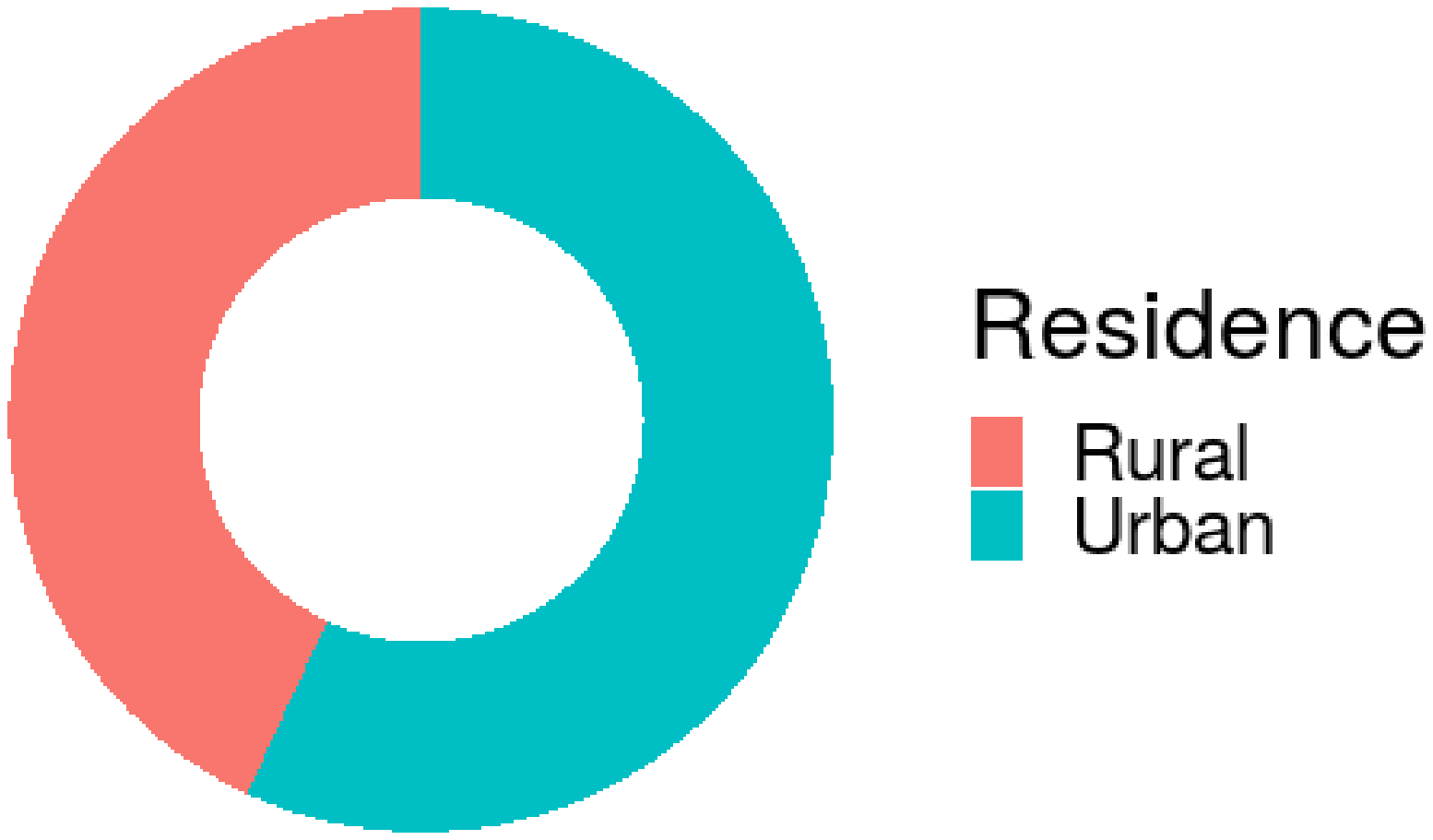} & \includegraphics[width=50mm]{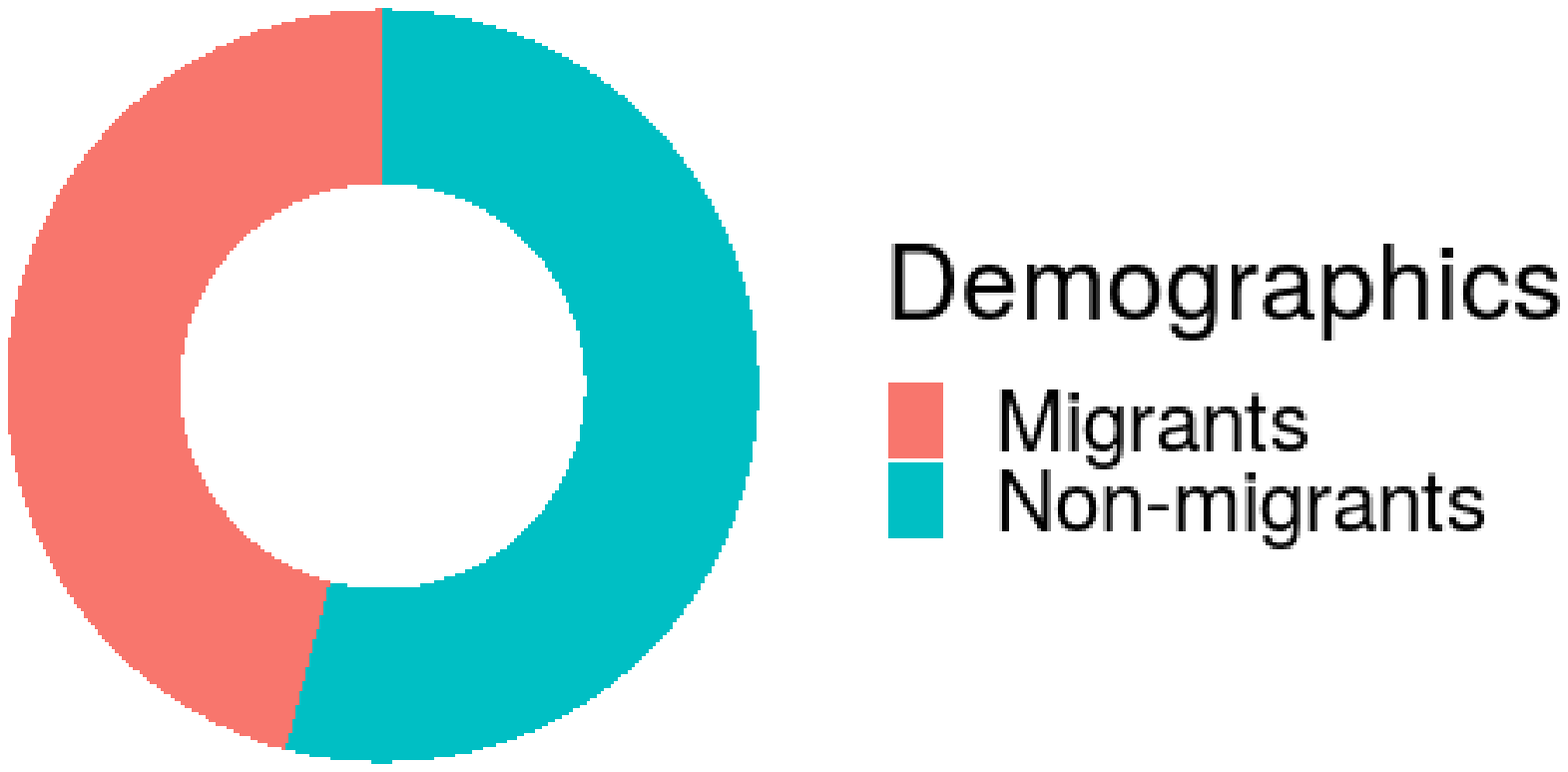} & \includegraphics[width=50mm]{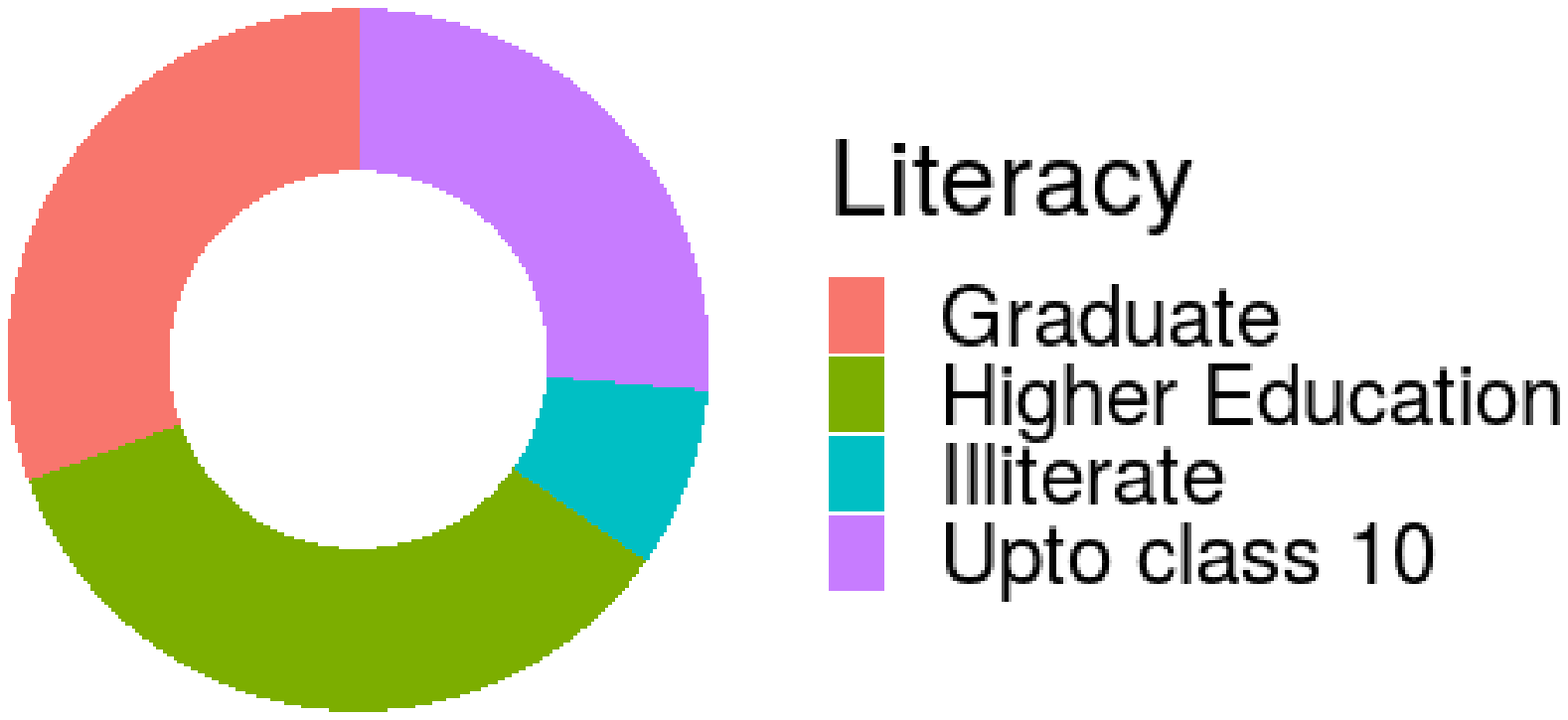}   \\
\includegraphics[width=50mm]{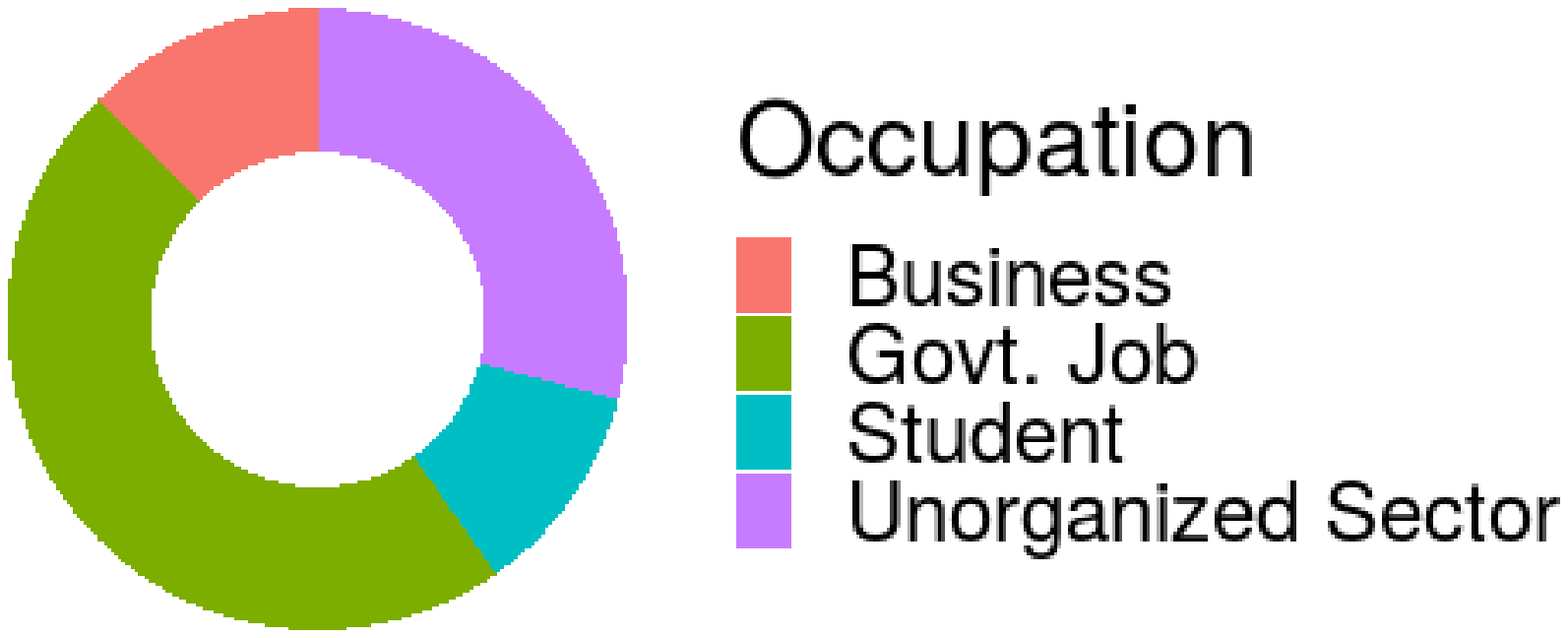} & \includegraphics[width=50mm]{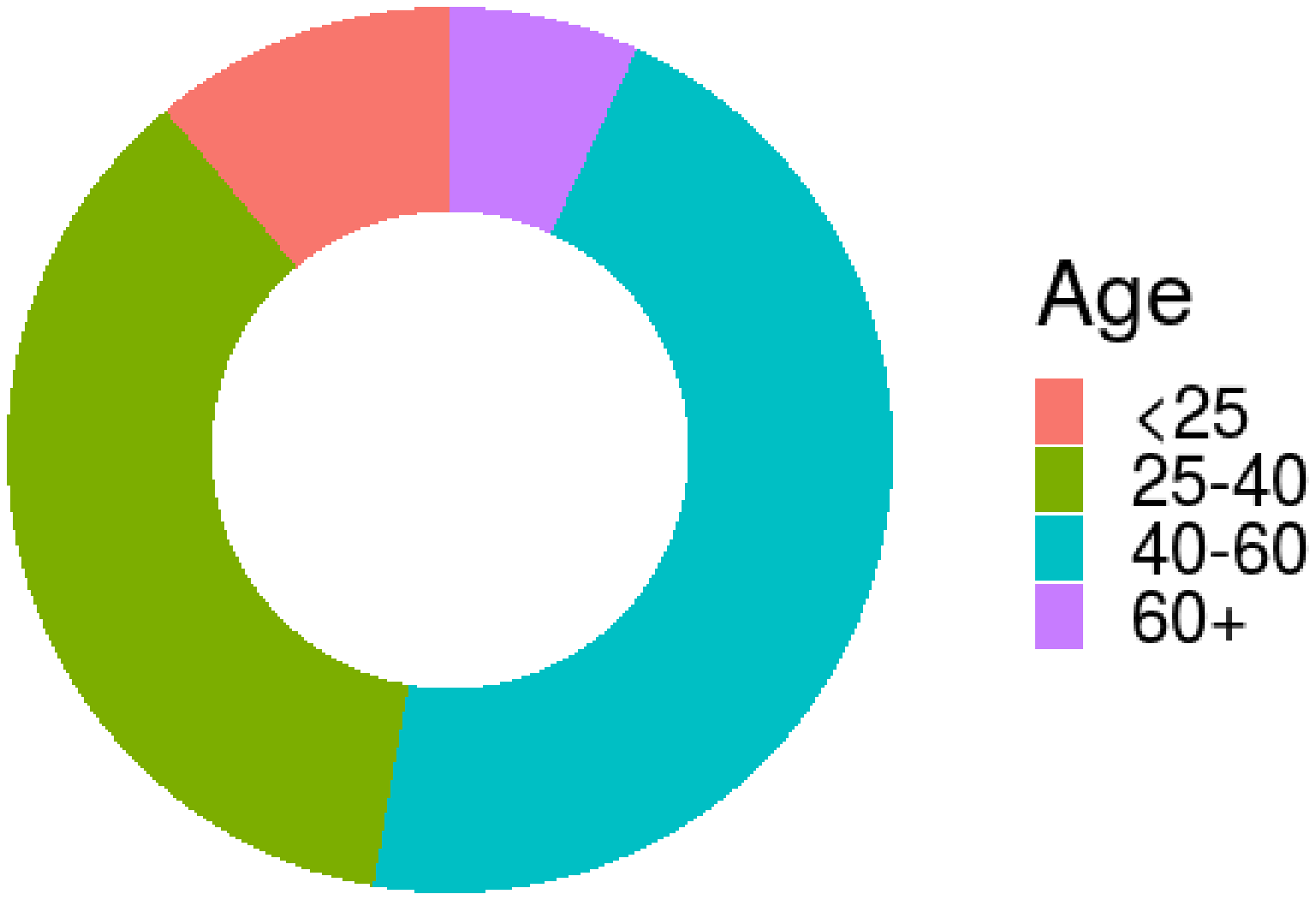} &  \includegraphics[width=50mm]{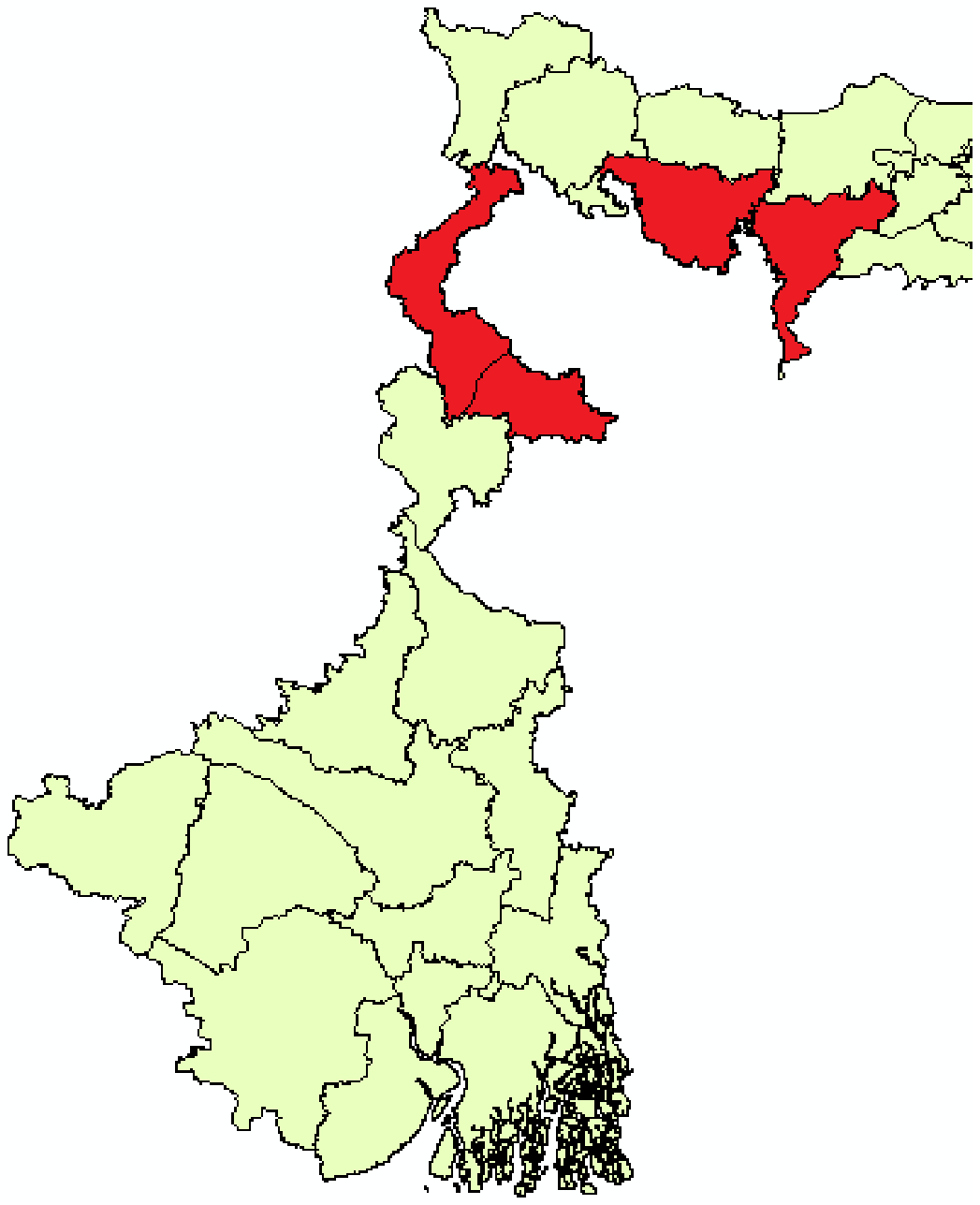} \\
\end{tabular}
\caption{Demographic details of the respondents and district of West Bengal and Assam to which the people are responding in field interviews.}
\label{S11a}
\end{center}
\end{figure}

During the field visit, we received $72$ field interviews (with a gender ratio of $5M:1F$) from $6$ districts (Dakshin Dinajpur, Uttar Dinajpur, Cooch Behar, Alipurduar of West Bengal and Dhubri, Kokrajhar of Assam). However, most of the respondents were from West Bengal. Here, $72.22$ \% of the respondents are Hindus, whereas $25$ \% are Muslims.
In terms of ethnicities, 45.83\% (resp. 18.05 \%; 22.22 \%; and 13.88 \%) responding people belongs from Hindu refugee (resp. Ancient (non-migrant) Hindu; Muslim; and Rajbanshi) community. The demography related information of the respondents are summarized in Fig~\ref{S11a}. Most of the responding people were from middle age group, belonged either to the age group $25$-$40$ ($36.11$\%) or $40$-$60$ ($45.83$\%). A majority of the respondents have no direct association with political parties.

\subsection{Results and Analysis}
To understand the field interviews, we classify the content of field interviews into following nine classes:

\begin{itemize}
\item[T1] As a community, Muslims are in Problem;
\item[T2] Impact of Assam NRC;
\item[T3] Confusing statements by political leadership/social media;
\item[T4] Migration after 1971 and related problems;
\item[T5] Problem faced by son of the soil;
\item[T6] Business/harassment in the name of Document correction;
\item[T7] Sectarian and Communal content;
\item[T8] Technicalities of implementation; and
\item[T9] Fallout of Government Failure - political/economic and protests.
\end{itemize}

\begin{figure}
\begin{tabular}{cc}
\includegraphics[width=55mm]{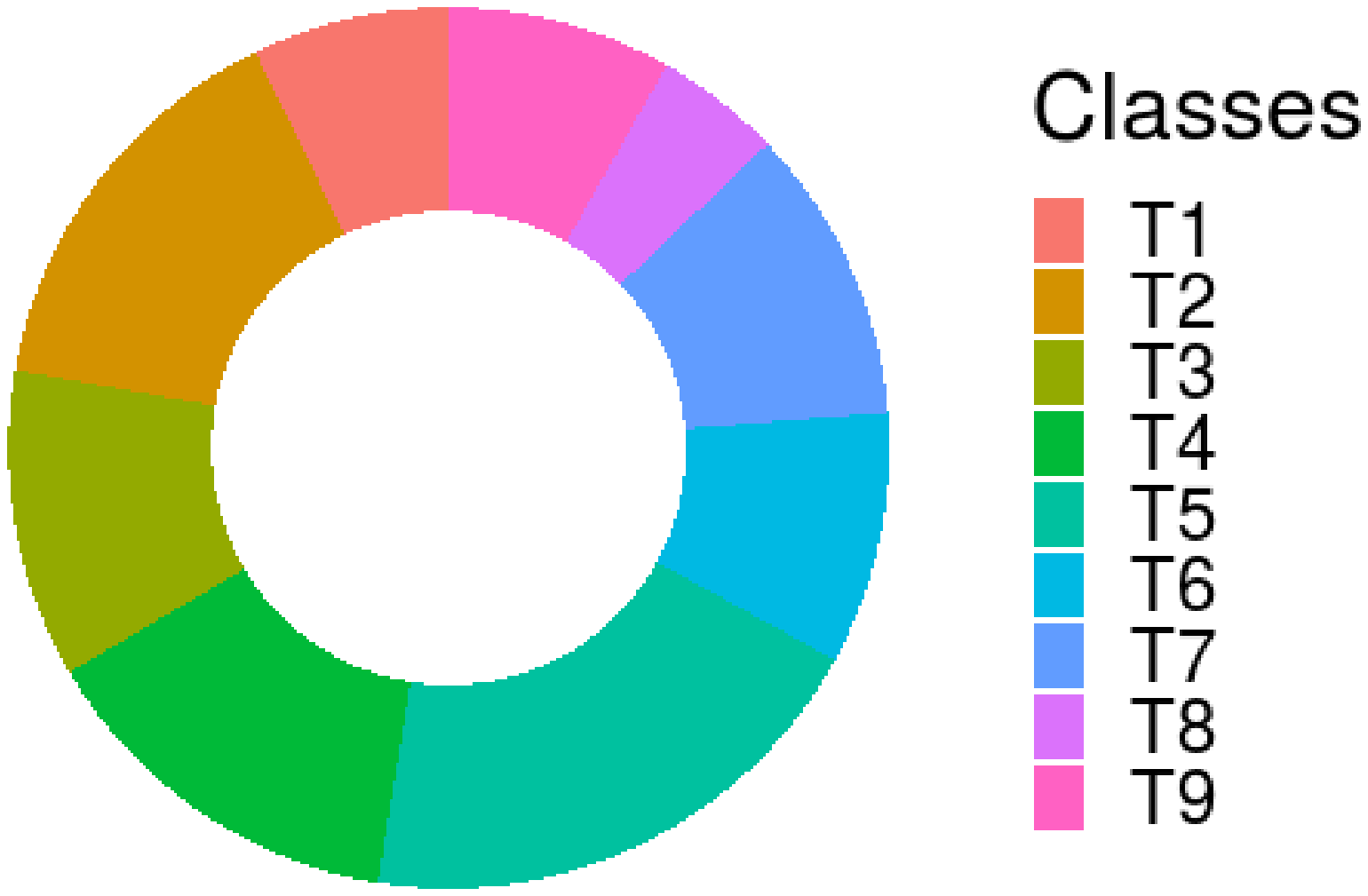} &  \includegraphics[width=55mm]{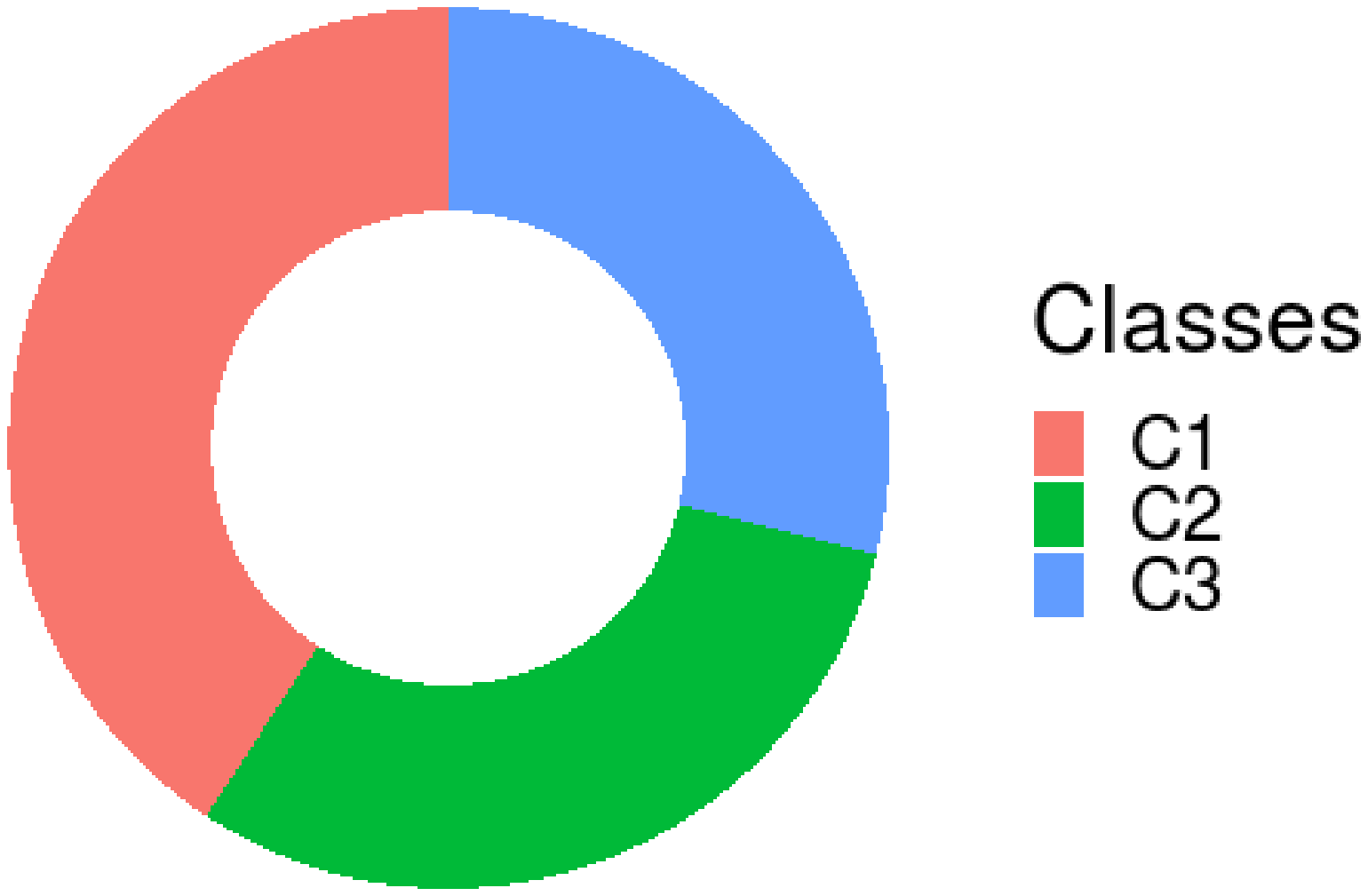} \\
\end{tabular}
\caption{ \% wise topic responses in the field interviews for the classes T1,~T2,~T3,~,T4,~T5,~T6,~T7,~T8,~T9 and major classes C1,~C2,~C3.}
\label{S11b}
\end{figure}

Now, we again put these nine classes under following three major classes: (C1) {\em Documents} [class: T1, T4, T5]; (C2) {\em Political} [class: T3, T7, T9]; and (C3) {\em Implementation} [class: T2, T6, T8]. Fig.~\ref{S11b} depicts the \% wise topic responses in the field interviews for the nine classes and three major classes. We noticed that {\em problem faced by son of the soil} (19.58\%), {\em impact of Assam NRC} (14.94\%), {\em migration after 1971 and related problems} (13.91\%) are responded by people with significantly higher importance. 

\begin{figure}
\begin{center}
\begin{tabular}{cccc}
\includegraphics[width=35mm]{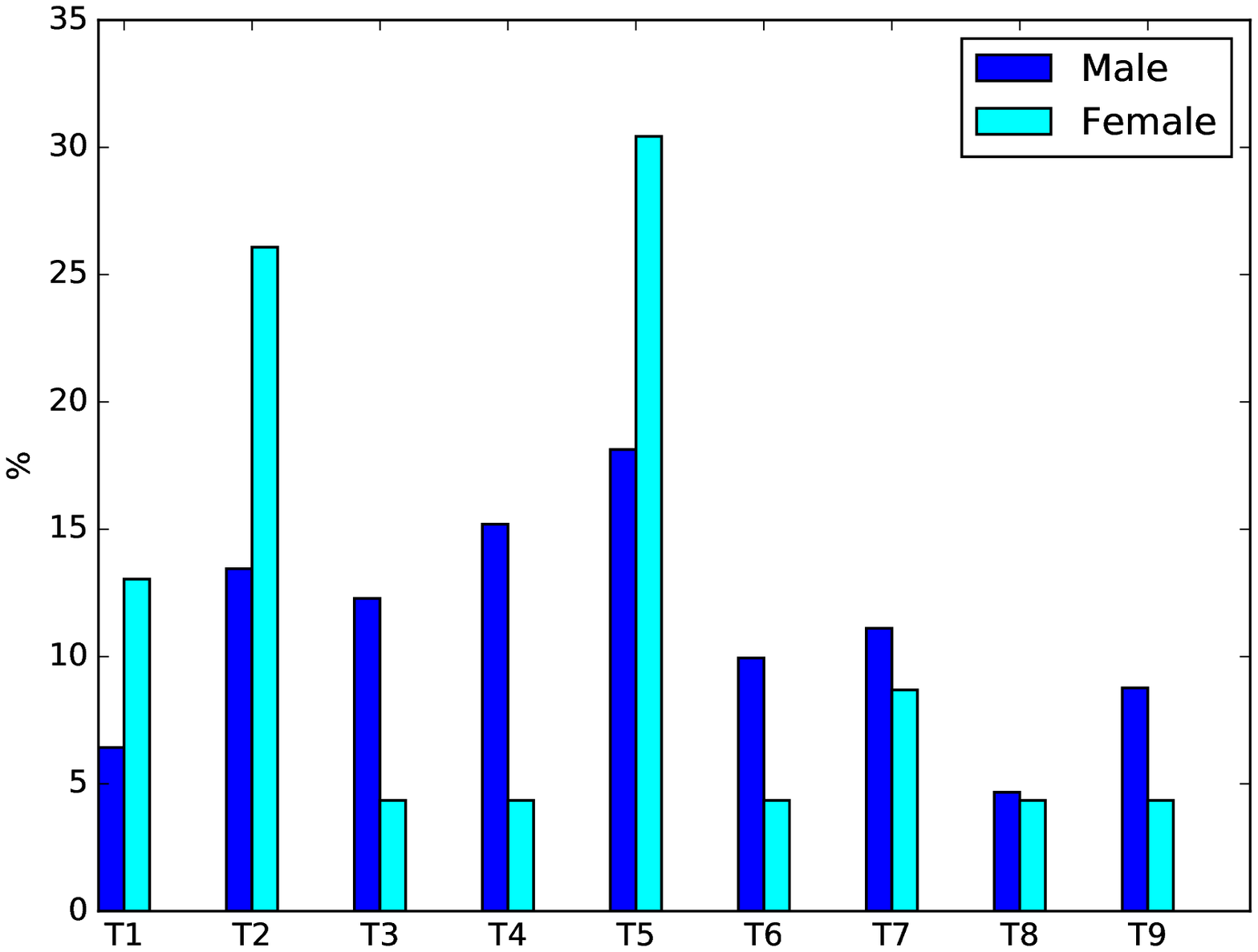} &  \includegraphics[width=35mm]{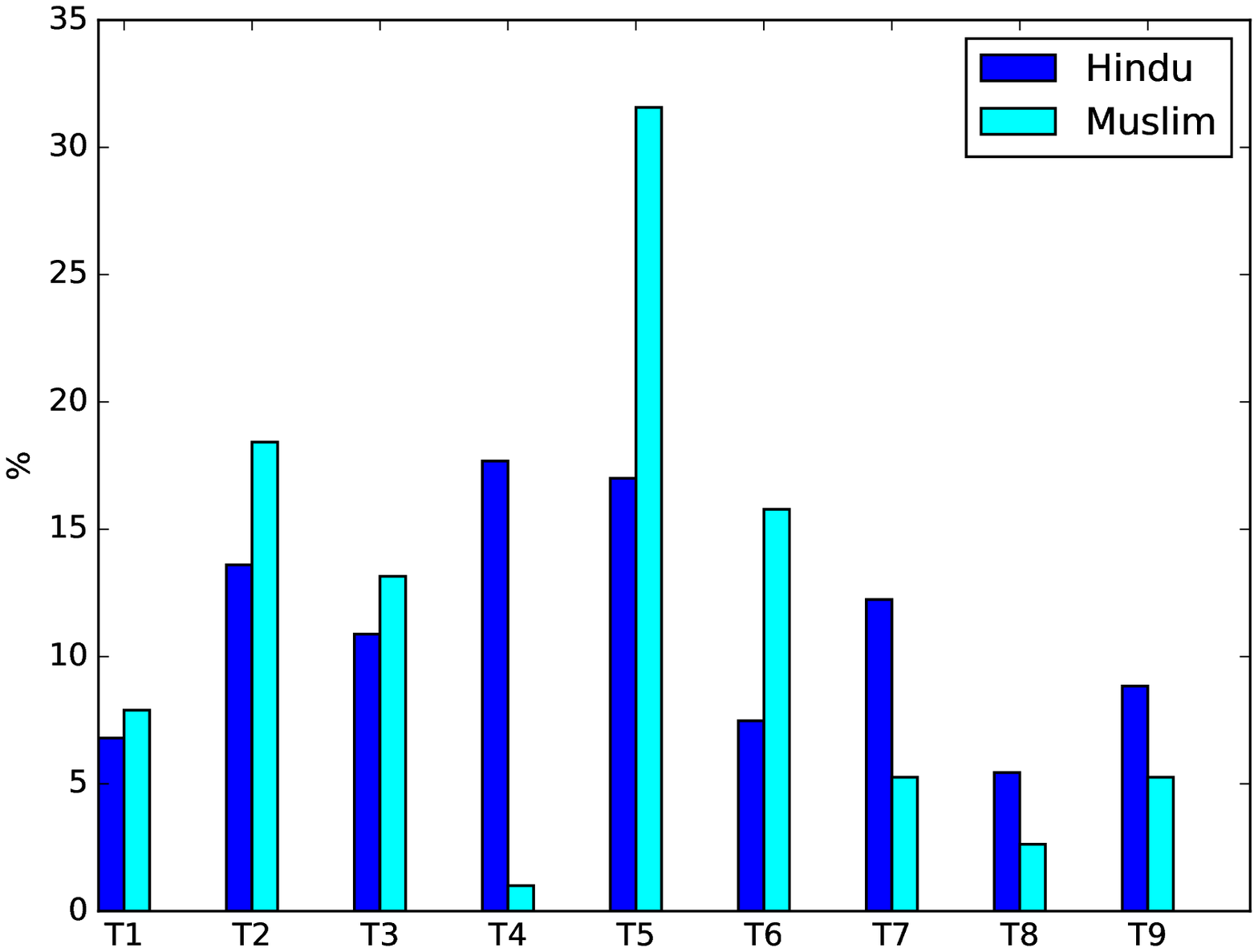} & \includegraphics[width=35mm]{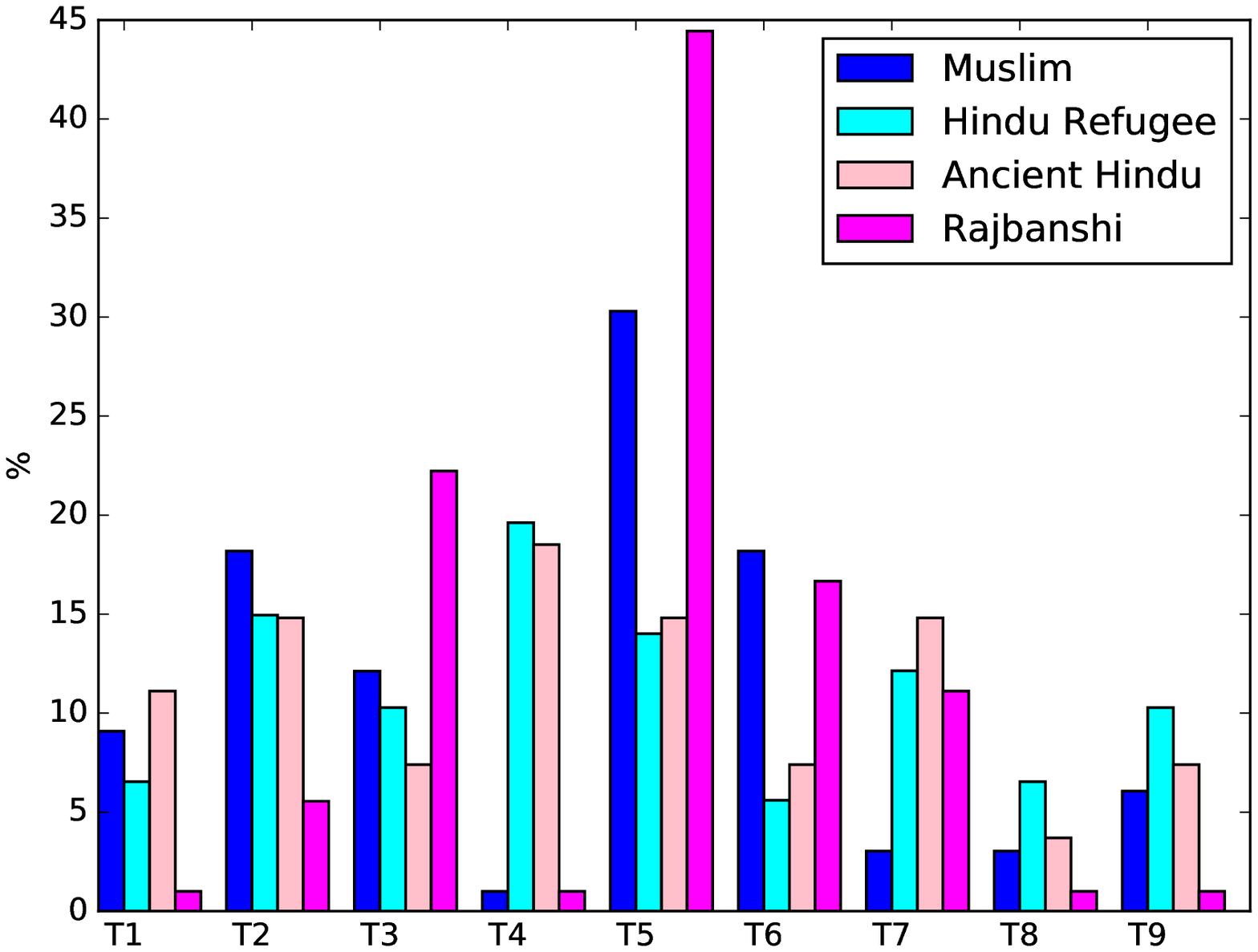} & \includegraphics[width=35mm]{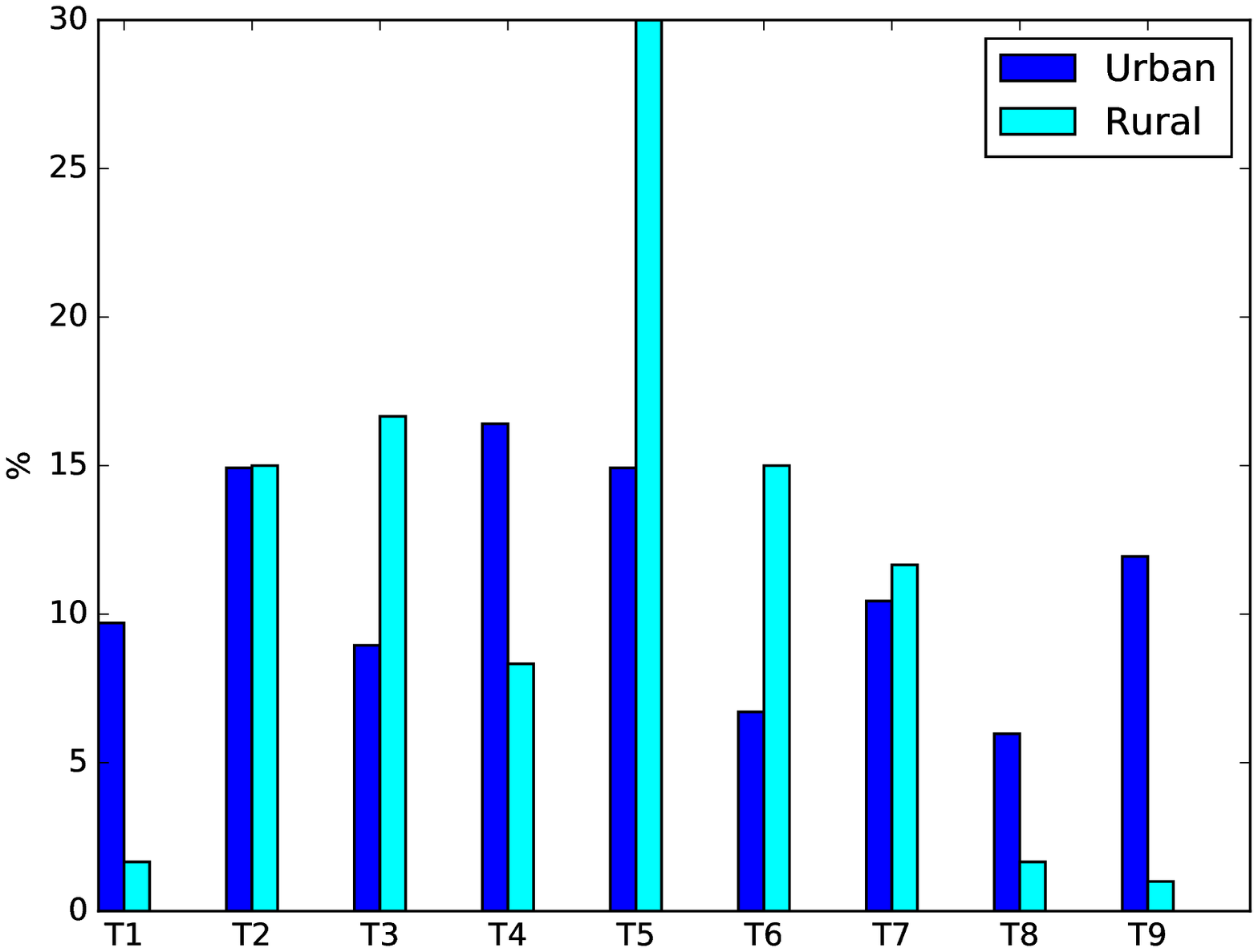}\\

\includegraphics[width=35mm]{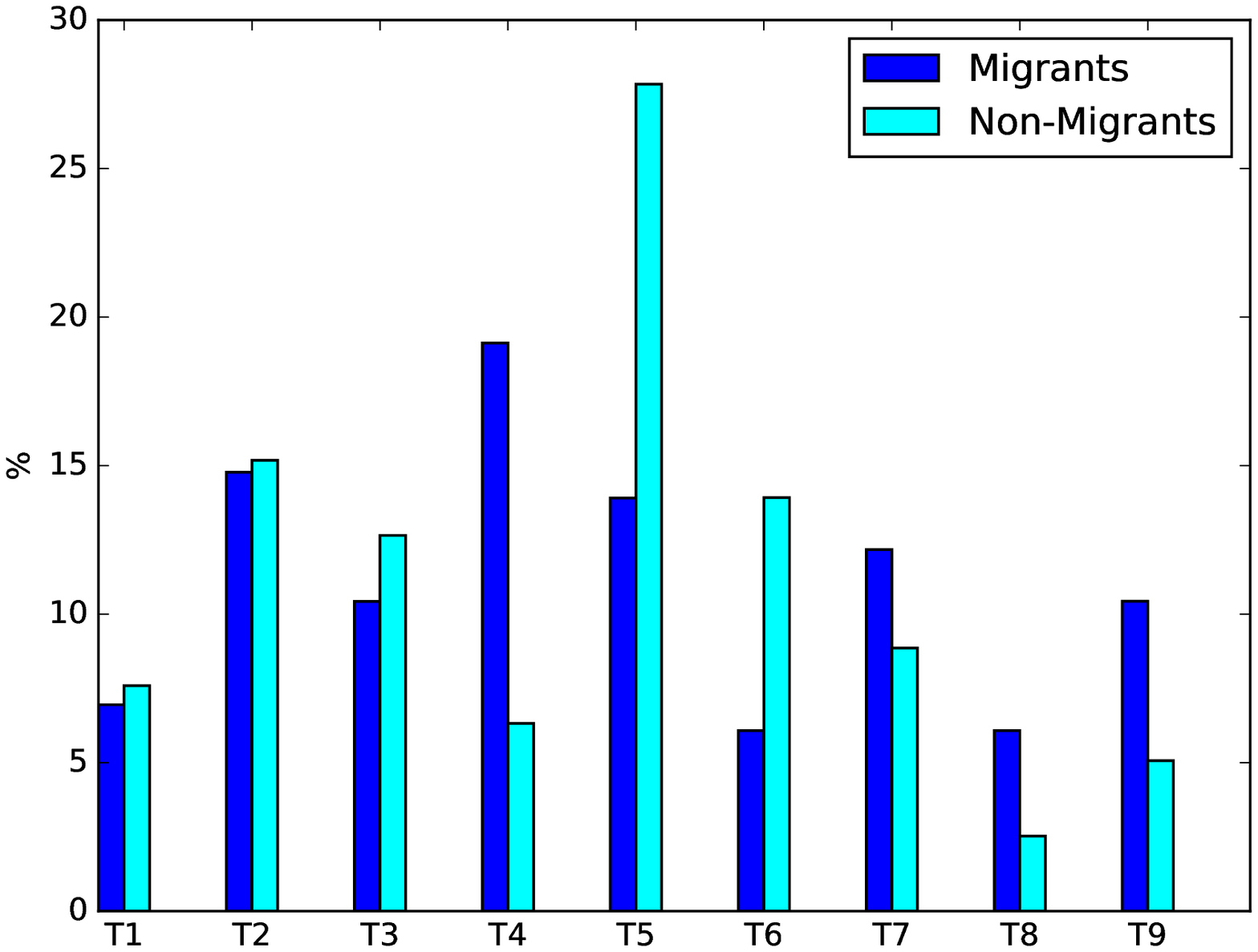} &  \includegraphics[width=35mm]{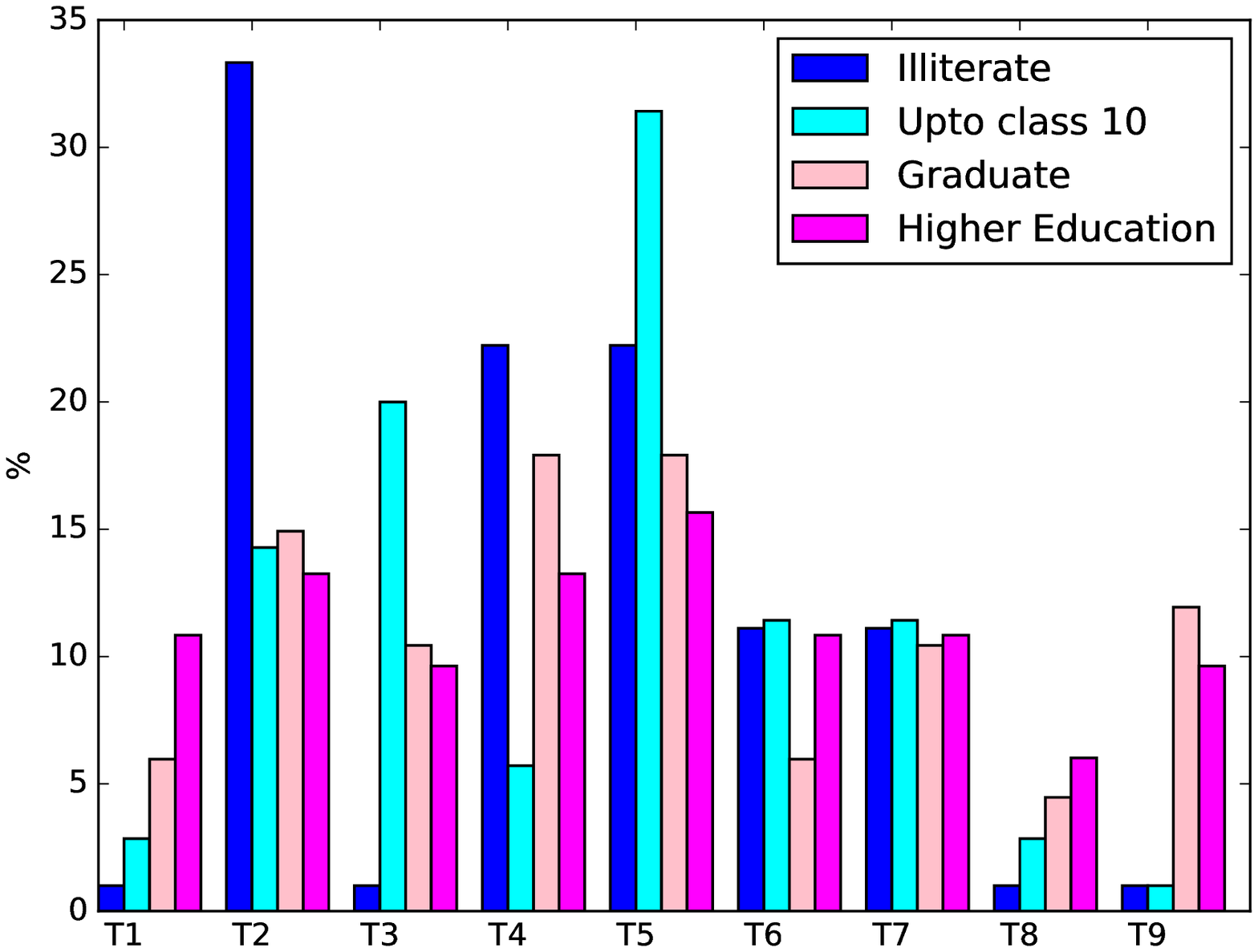} & \includegraphics[width=35mm]{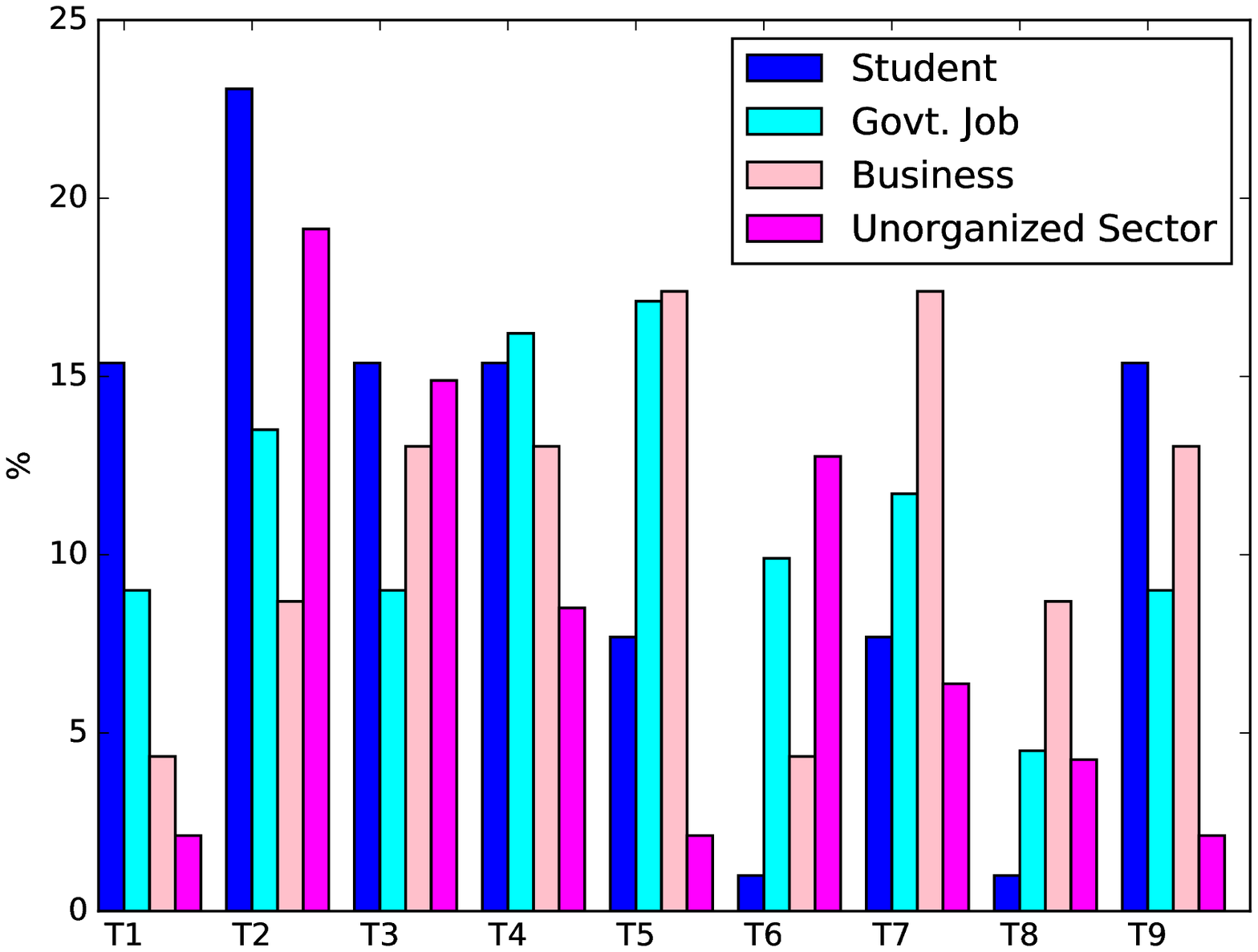} & \includegraphics[width=35mm]{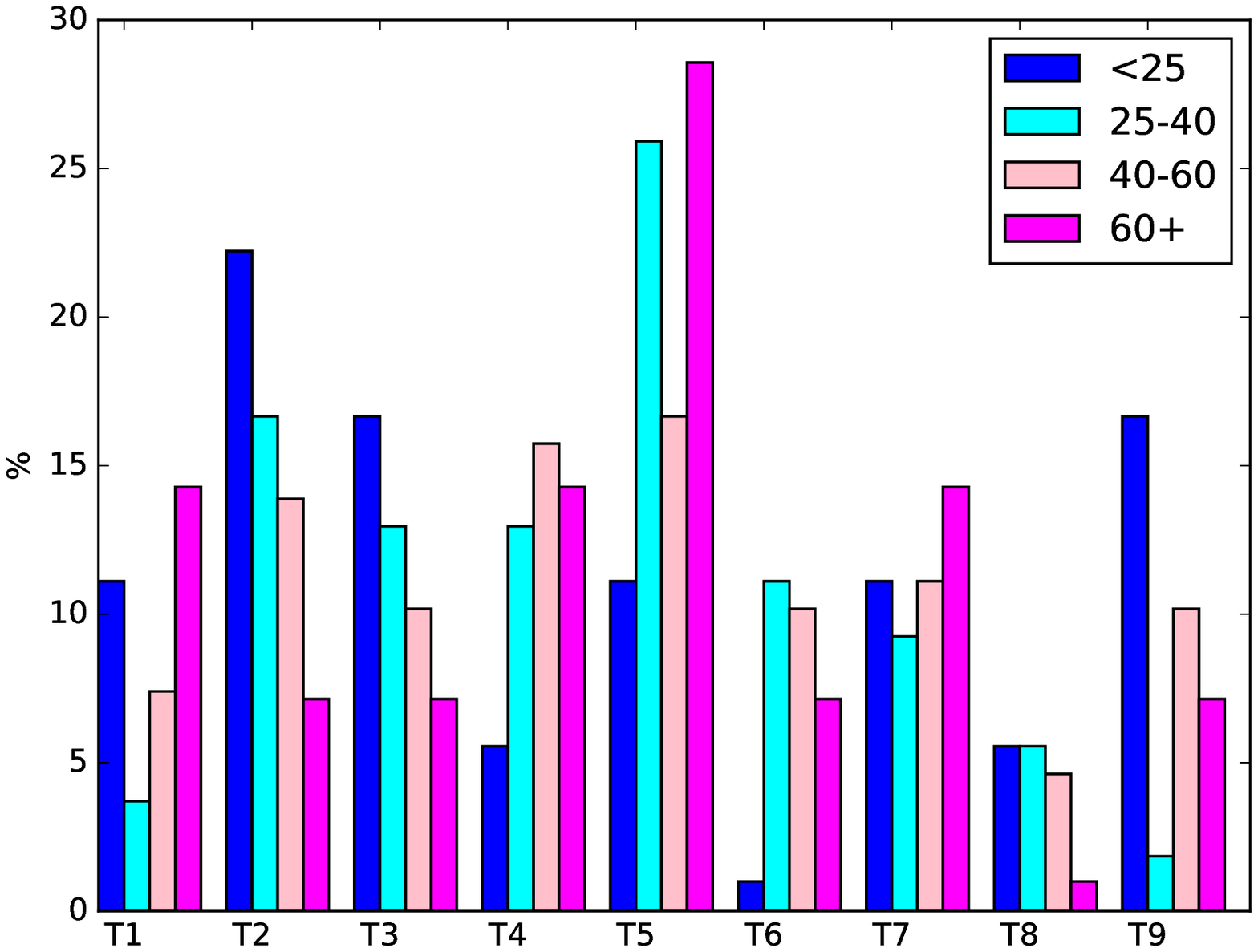}\\
\end{tabular}
\caption{Association between the different demographic factors and topic of responses.}
\label{S11c}
\end{center}
\end{figure}

{\em Bar chart} can highlight association between the different demographic factors and topic of responses in the field interview. Fig.~\ref{S11c} depicts Bar chart for different demographic factors. It is interesting to note that Muslim people, i.e. ethnic Muslim, express a significantly higher level of concern about the problems of son of the soil than the problems of Muslim community. Moreover, we found that Muslims (religion wise) have no association with the migration after $1971$ of Hindus and related problems. On the other hand, Rajbanshi community (ethnicity wise) strictly speaks about their own ``son of the soil" sentiments. We noticed that the Assam NRC has high impact on the females (gender wise), illiterate people (level of literacy wise) and working people from the informal sector (occupation wise). However, Assam NRC reflects almost same impact for Urban and Rural people (area of residence wise); and migrants and non-migrants people (demography wise). Again, it is very interesting to note that urban and educated people express significantly higher concern about economic fallout of Government in association with NRC \& CAA issue. 

Next, to understand the {\em expressiveness} of the field interviews, we have plotted a Bipartite graph G = (U,V,E), where U = \{$R_1,R_2,\cdots,R_n$\} indicates the set of field responders, V = \{$T_1,T_2,\cdots,T_9$\} depicts the set of nine topic classes and $E$ denotes the set of edges of the bipartite graph. Here, node $R_i$ and $T_j$ are connected via edge if $R_i$ responder in the field interview speaks about $T_j$ topic. Therefore, the `expressiveness' property of responder in the field interview indicates the number of cycles in the bipartite graph starting from the corresponding responder node. As an example, $R_{j1} \rightarrow T_2 \rightarrow R_{j3} \rightarrow T_3 \rightarrow R_{j5} \rightarrow T_7 \rightarrow R_{j1}$ indicates a length three cycle covering $T_2, T_3, T_7$ topics. Therefore, different length cycle is possible following this procedure. Let us consider, $c^n_{R_j}$ indicates number of $n$ length cycle for $R_j$ responder. Therefore, total number of cycle for responder $R_j$ is, 
\begin{equation}
\nonumber
C_{R_j} = c^1_{R_j} + c^2_{R_j} + \cdots + c^n_{R_j}
\end{equation}
Moreover, 
\begin{equation}
\nonumber
WC_{R_j} = 1 \times c^1_{R_j} + 2 \times c^2_{R_j} + \cdots + n \times c^n_{R_j}
\end{equation}

Here, expressiveness of a responder is depicted by parameter $C_{R_j}$ and $WC_{R_j}$ which indicates the local connection of a responder with other responder.

\begin{figure}
\begin{center}
\begin{tabular}{cccc}
\includegraphics[width=35mm]{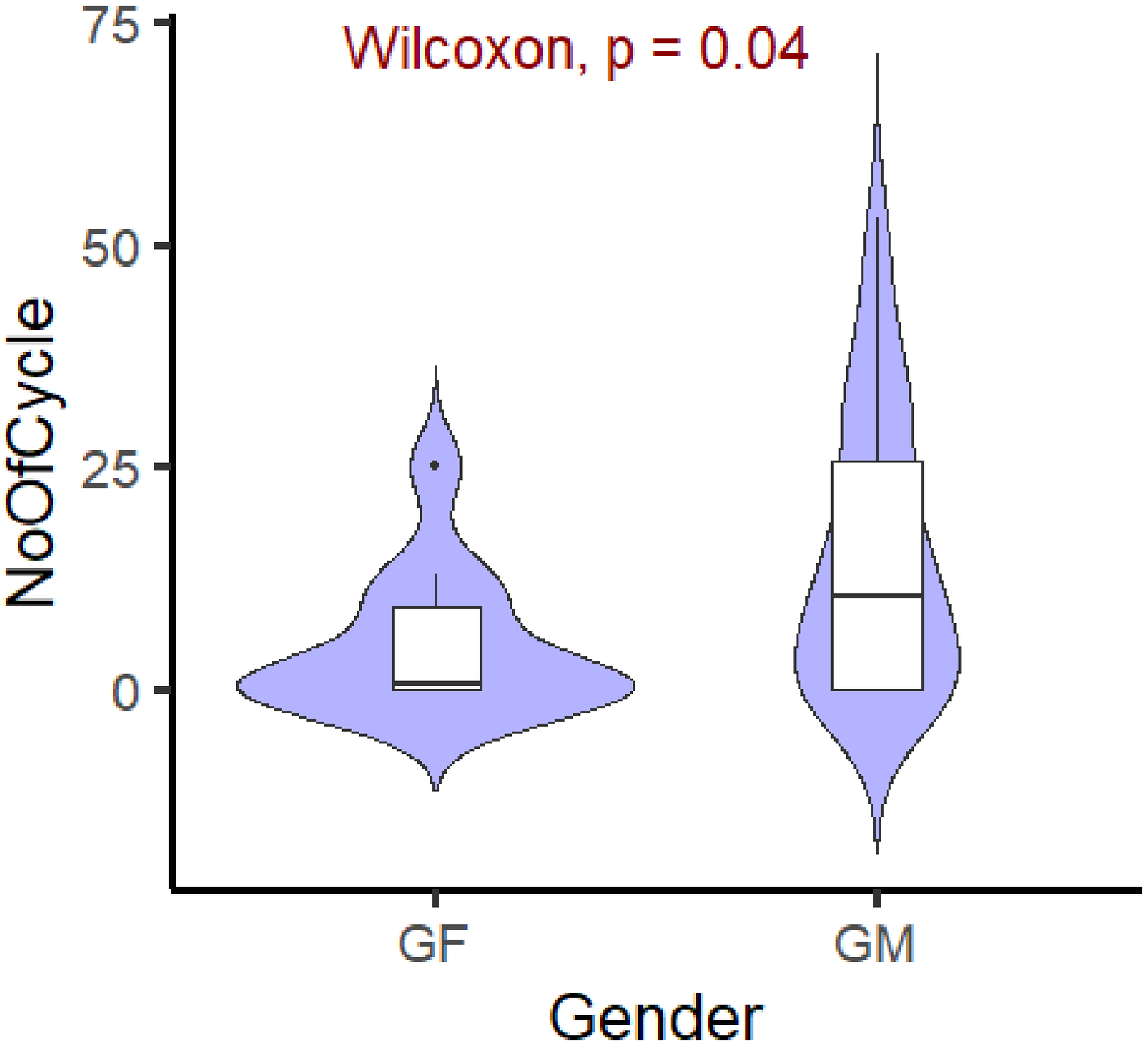} &  \includegraphics[width=35mm]{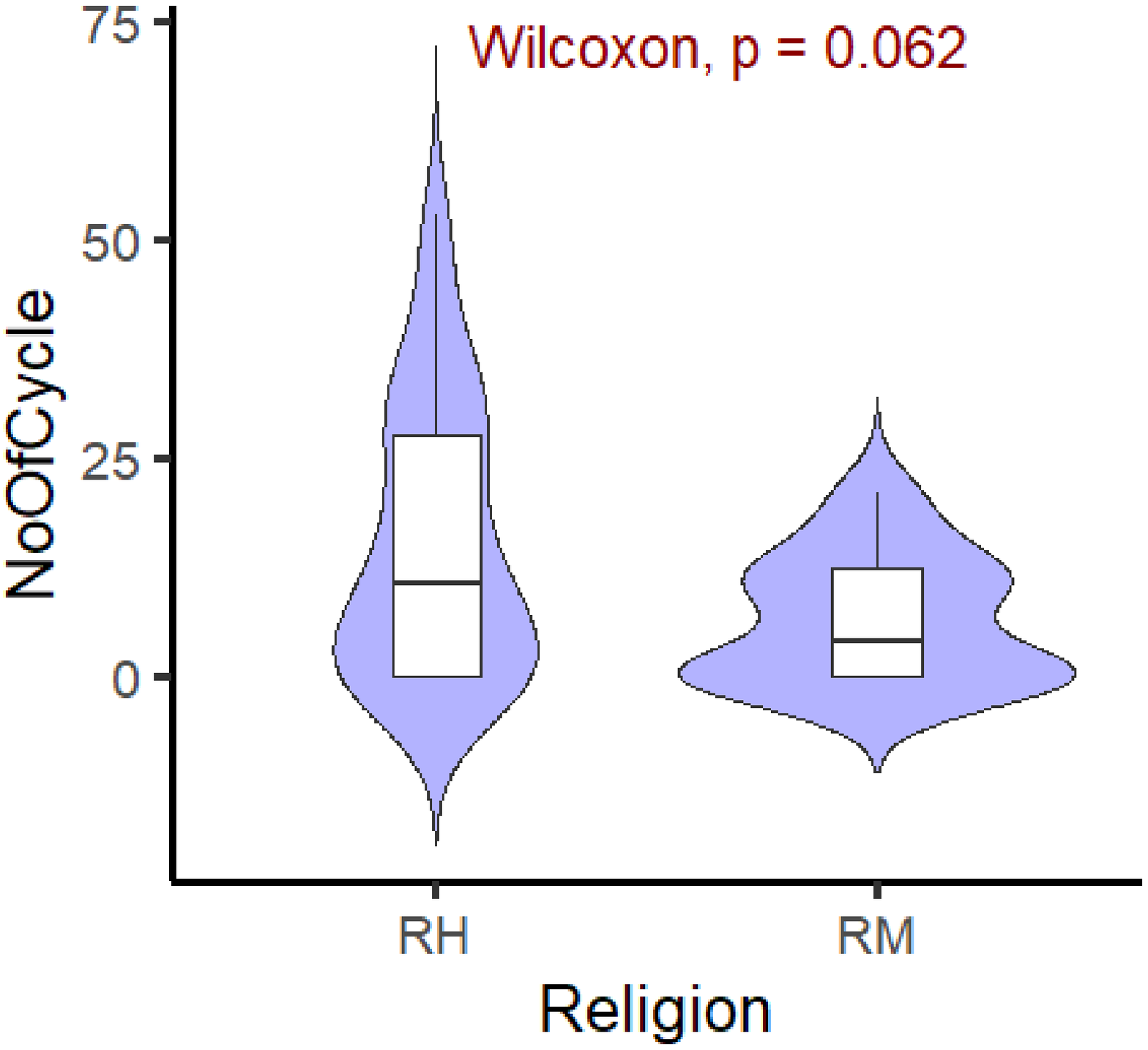} & \includegraphics[width=35mm]{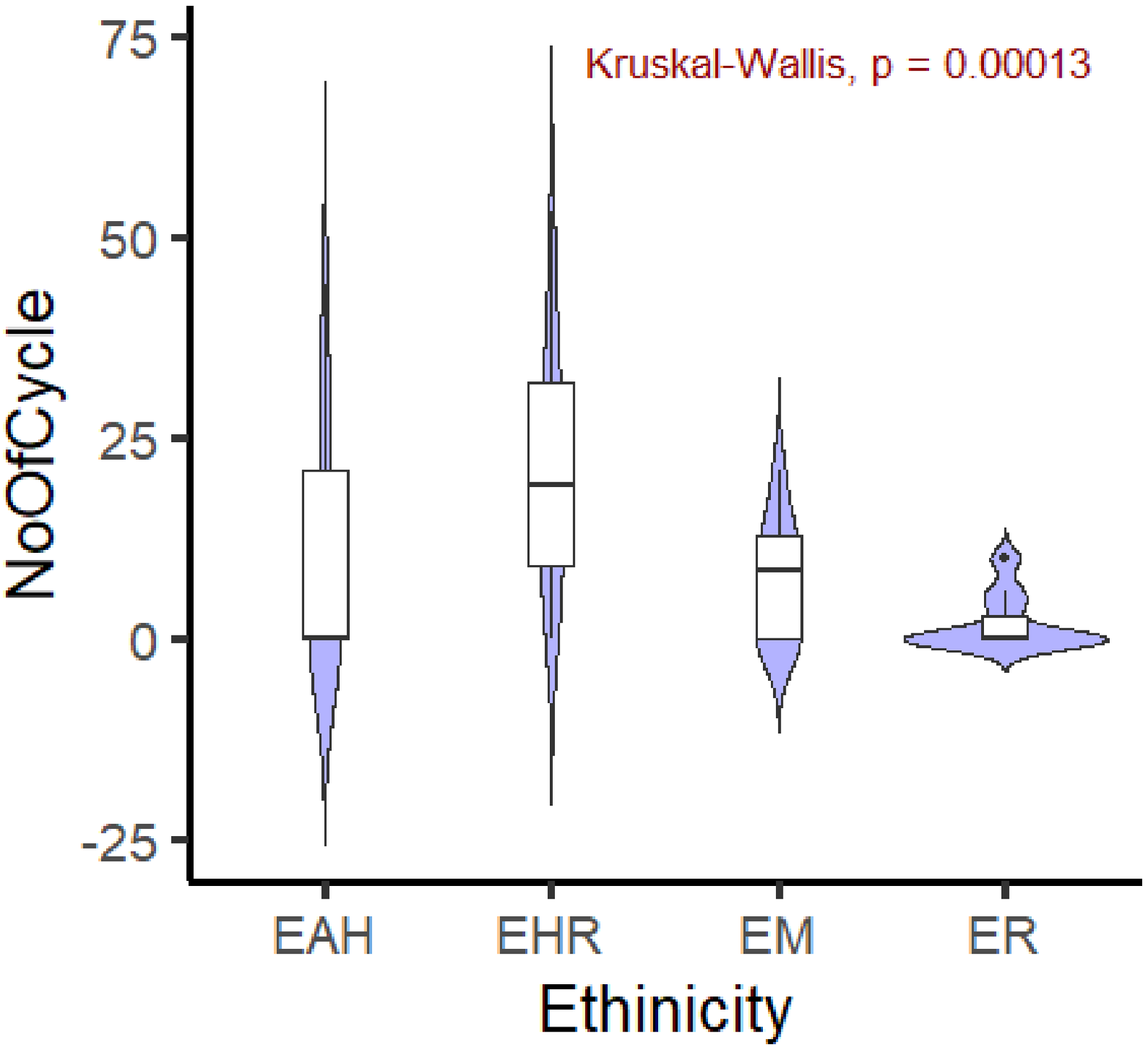} & \includegraphics[width=35mm]{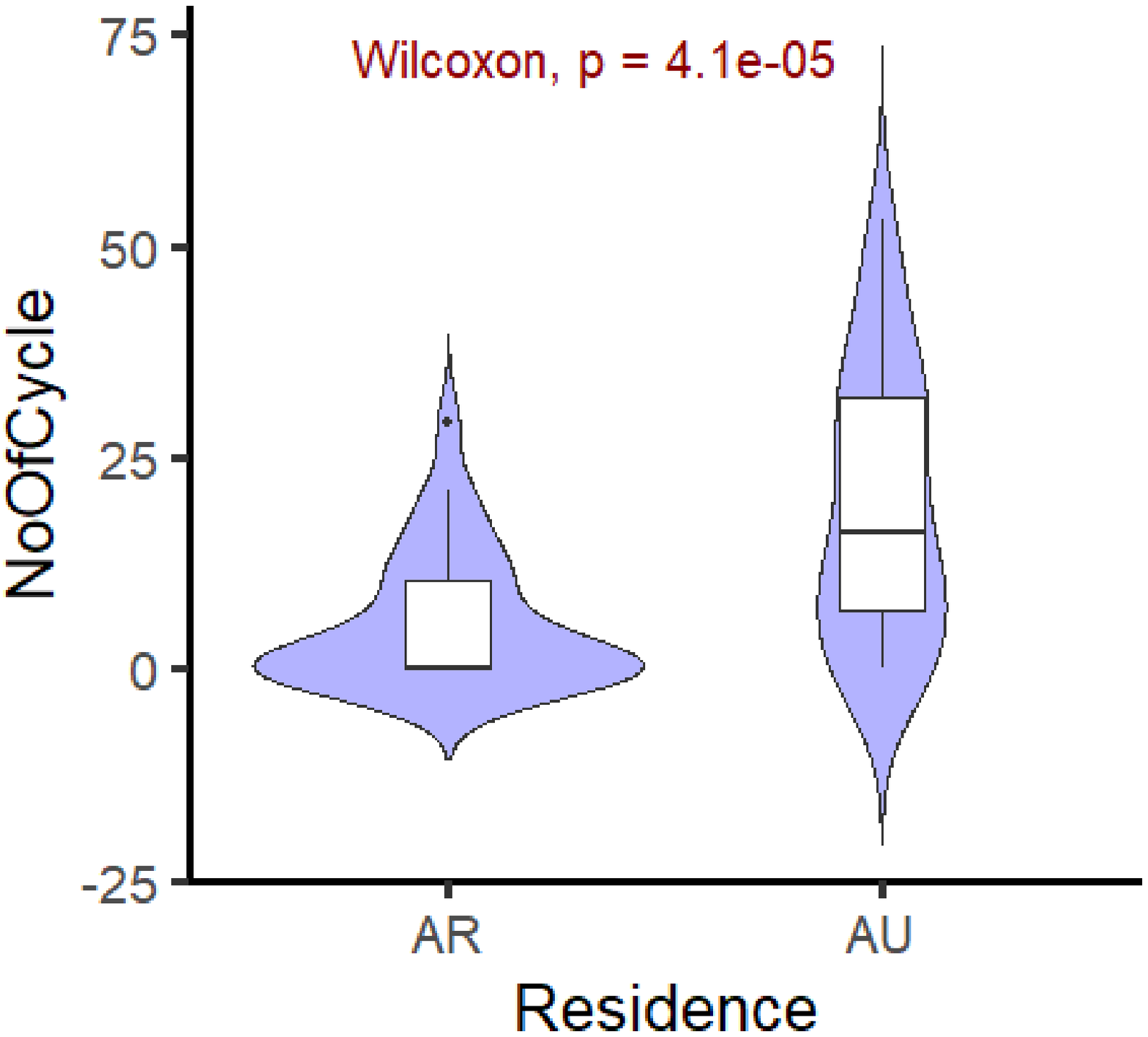}\\

\includegraphics[width=35mm]{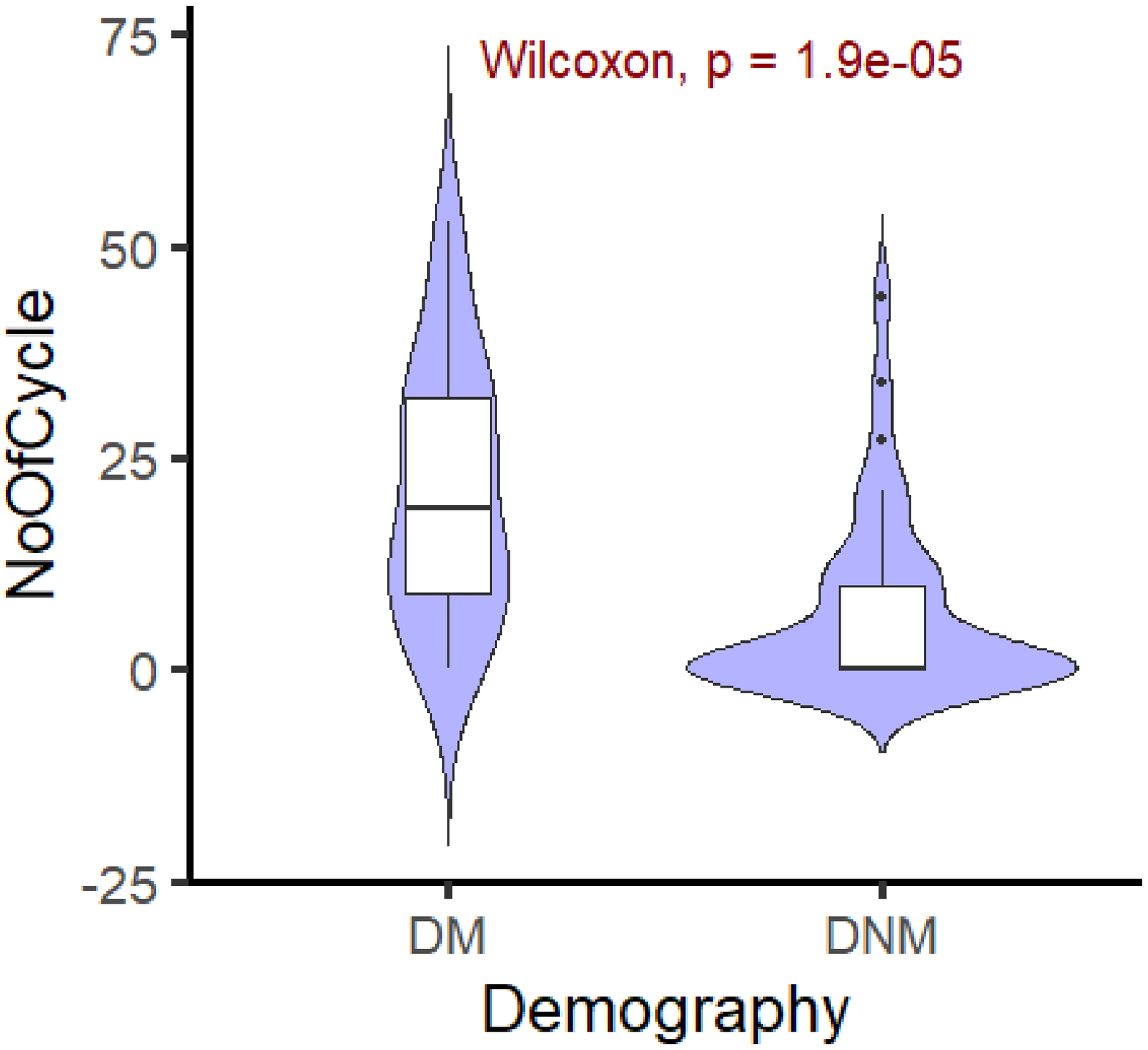} &  \includegraphics[width=35mm]{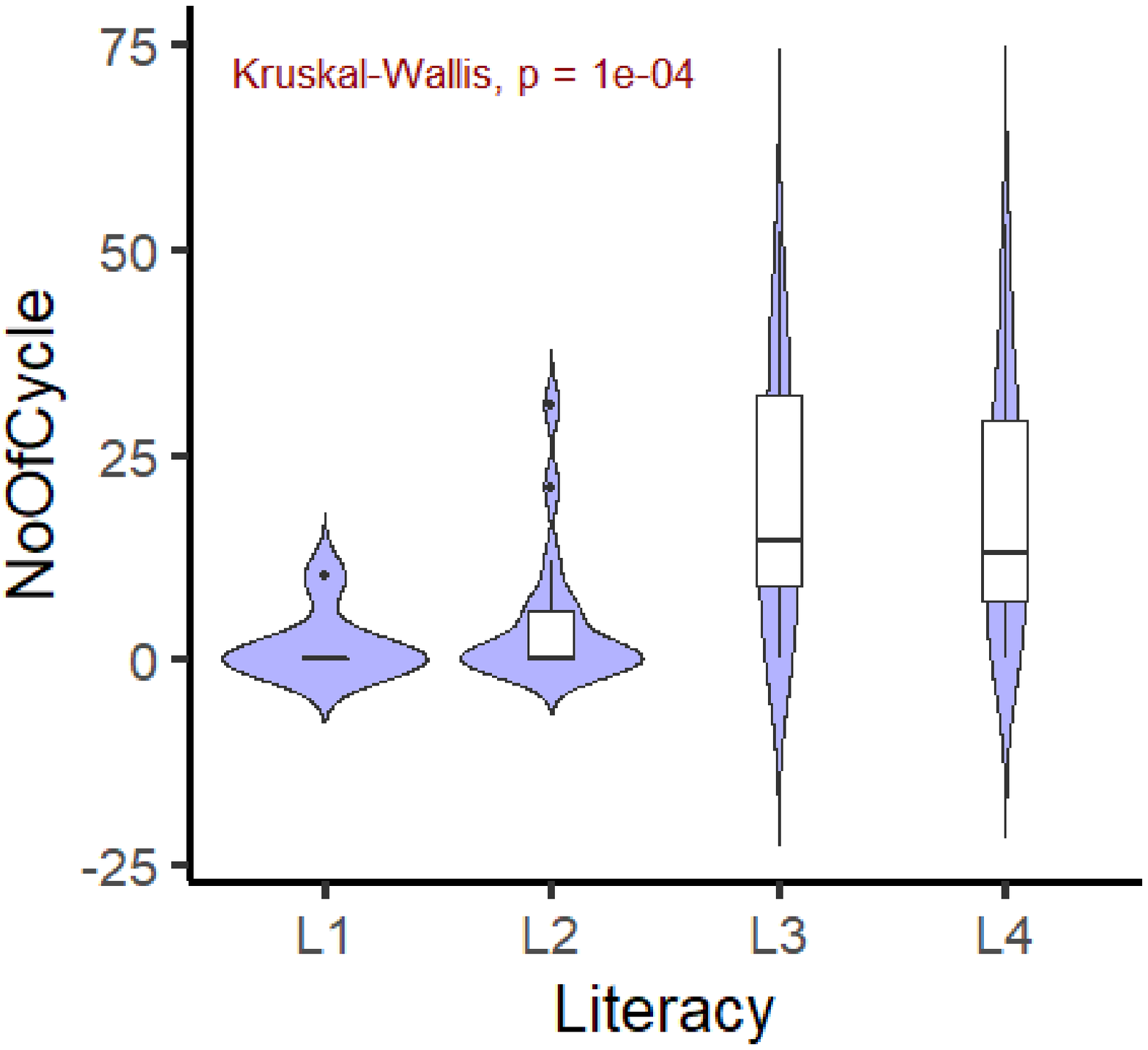} & \includegraphics[width=35mm]{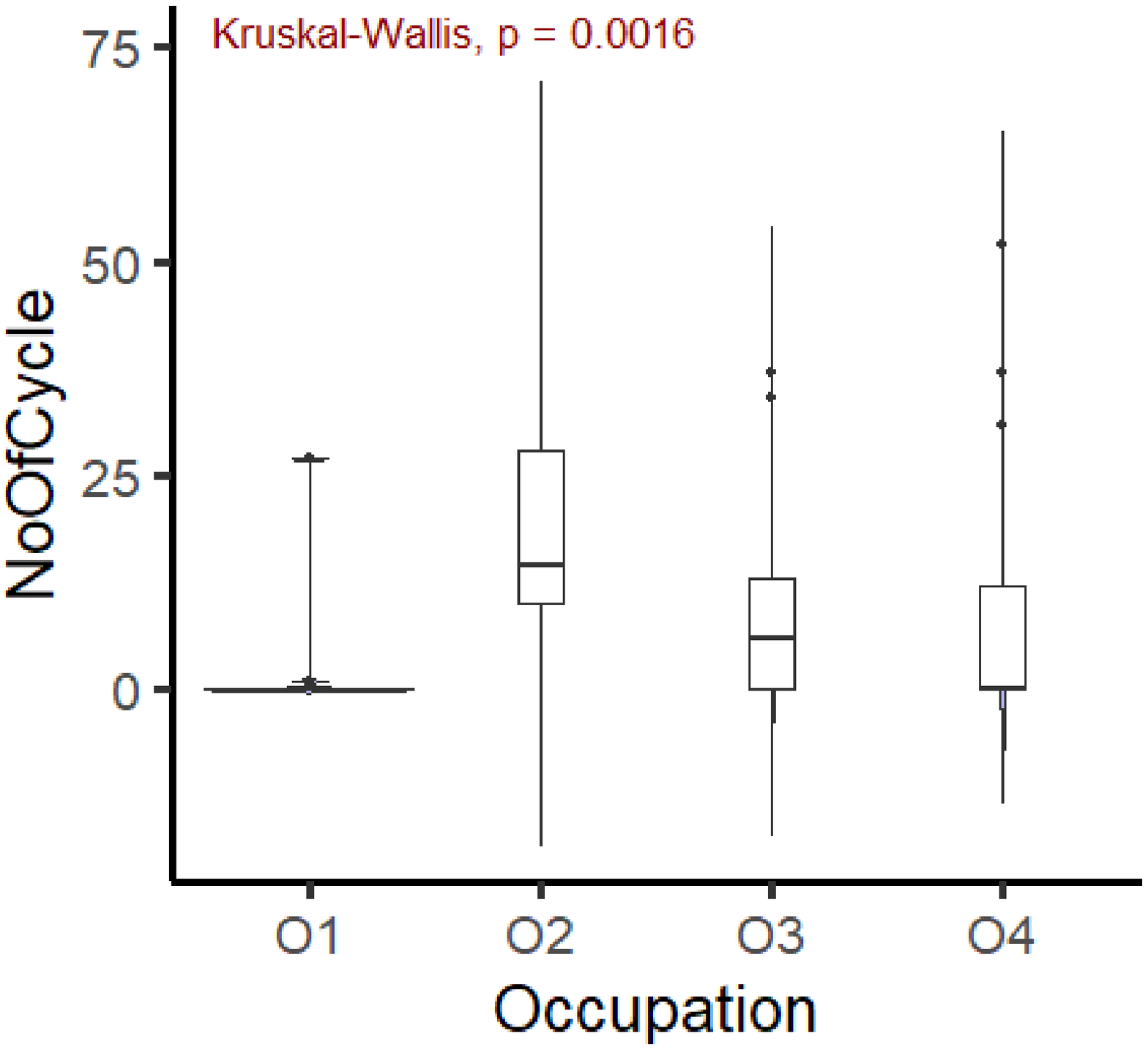} & \includegraphics[width=35mm]{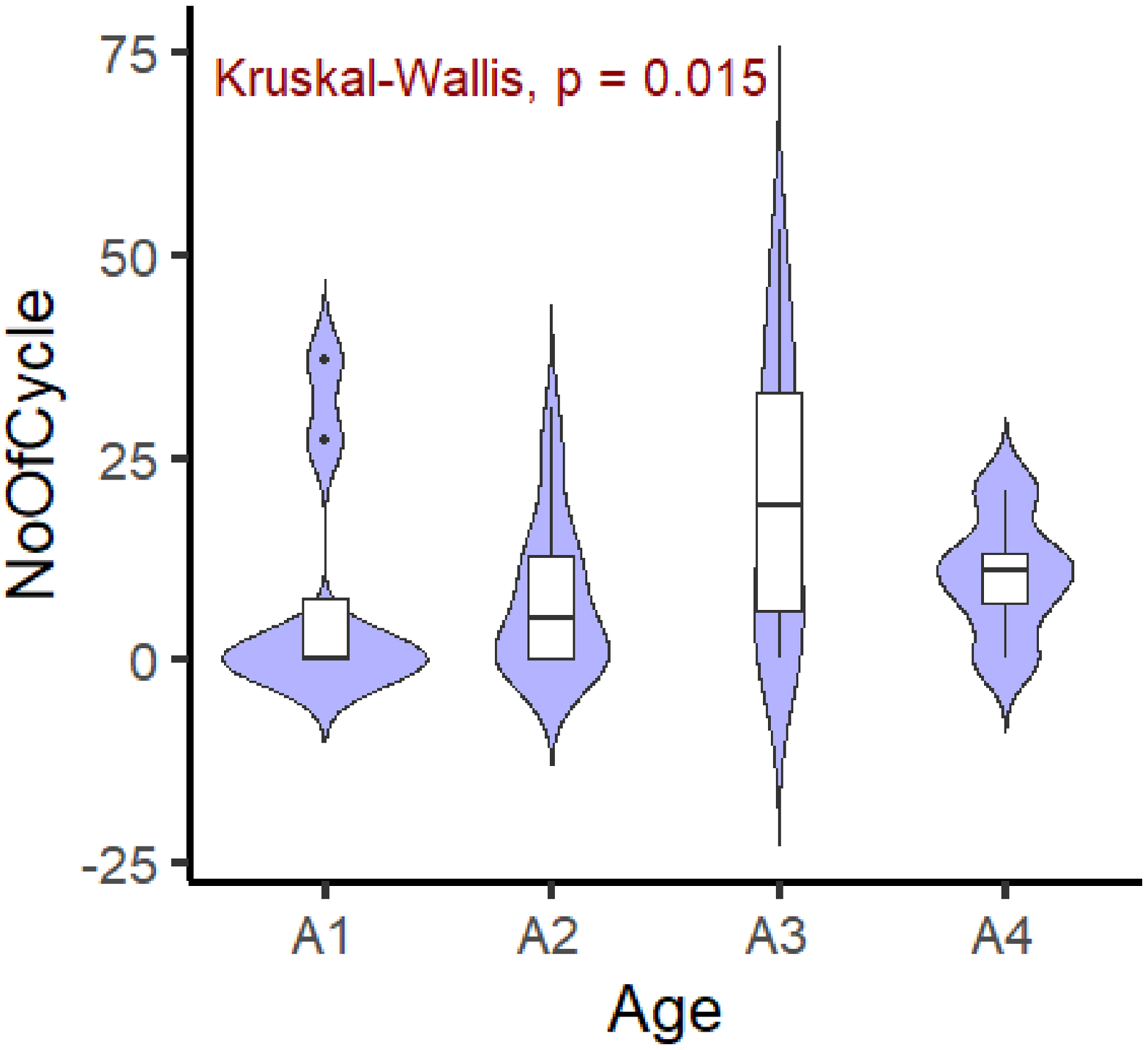}\\
\end{tabular}
\caption{Expressiveness (number of cycle in the bipartite graph) against different demographic factors. Here, for gender, Male: GM, Female: GF; for Religion, Hindu: RH, Muslim: RM; for Ethnicities, Rajbanshi: ER, Hindu Refugee: EHR, Muslim: EM, Ancient Hindu: EAH; for Area of Residence, Urban: AU, Rural: AR; for Demographics, Migrants: DM, Non-Migrants: DNM; for Level of Literacy, Illiterate: L1, Upto class 10: L2, Graduate: L3, Higher Education: L4; for Occupation, Student: O1, Govt Job: O2, Business: O3, Unorganized Sector: O4; for Age, less than 25: A1, (25-40): A2, (40-60): A3, (more than 60): A4.}
\label{S11d}
\end{center}
\end{figure}

Next, we examined expressiveness ($C_{R_j}$) of the responders for different demographic category. The violin plots \cite{Jerry559} in Fig.~\ref{S11d} highlight the expressiveness score over different subgroups of demographic factors. We noticed that the mean expressiveness level of the people belonging to the ethnic Hindu refugee group significantly higher (Kruskal-Wallis \cite{Jerry559}, p = $0.00013$) than the other ethnicity. Moreover, the mean expressiveness level of the people belonging to occupation wise govt. job is also significantly higher (Kruskal-Wallis, p = $0.0016$) than the people of other occupation. We found that people of middle age (40-60) shows a higher rate of expressiveness (Kruskal-Wallis, p = $0.015$) than others. Finally, it is interesting to note that the level of expressiveness is almost independent of the demographical factor area of residence (i.e. urban and rural).  

\begin{figure}
\begin{center}
\begin{tabular}{c}
\includegraphics[width=115mm]{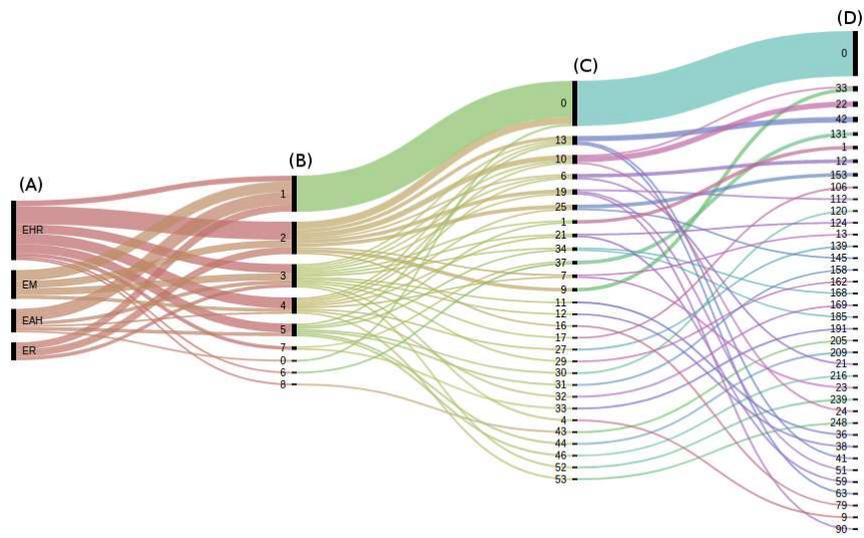}  \\
\end{tabular}
\caption{Alluvial diagram to realize the relationship between (A) ethnicity of responder where Rajbanshi: ER, Hindu Refugee: EHR, Muslim: EM, Ancient Hindu: EAH; (B)  association of a responder with number of topics (\{$T_1,\cdots,T_9$\}), (C) number of cycle in the Bipartite graph $C_{R_j}$ and (D) $WC_{R_j}$.}
\label{S11e}
\end{center}
\end{figure}

In a similar study, to realize the relationship between association of a responder with number of topics (\{$T_1,\cdots,T_9$\}, $C_{R_j}$ and $WC_{R_j}$, we used Alluvial diagram \cite{Martin10}. The four categorical dimensions from left to right in Fig.~\ref{S11e} are the ethnicity of the responder, association of responder with number of topic, number of cycle in the Bipartite graph $C_{R_j}$ and connection with others $WC_{R_j}$. Fig.~\ref{S11e} highlights that most the people from Rajbanshi community has association with only one topic and very less expressiveness.

To summarize, this section depicts the topic wise responses in the field interviews; association between different demographic factors and topic of responses; and expressiveness of field interviews against different demographic factors. Moreover, the correlation between social media dynamics and field responses is already discussed in previous sections. 

\section{Conclusion}
\label{section8}

This work has explored the contra NRC based content in social media, mainly from the forums of Facebook groups and Twitter activists. The dynamical nature of the groups with respect to their formation and membership growth has been analyzed in detail. The textual content of the deliberations in English, Bangla, Hindi and Asomiya posted by the members has been subject to scrutiny and classified into relevant categories. The like and share dynamics of some viral posts have been characterised using graph theory algorithms. The study also considers the spatial dynamics of individual Twitter activists reacting to the NRC / CAA policies of the central government. Throughout all this analysis, the goal has been to study how and whether the social media is able to build up a mass response (or movement ) against the draconian and sectarian policies of granting and retaining Indian citizenship.

Field work has been conducted to generate evidence for the social media based analysis. The following points could be established from the available evidence.
\begin{itemize}
\item There is a perceptible dichotomy in people's choice between abiding by new requirements and joining the protest movement. Looking at the affidavit / document correction crowd size and the timings as reported by the interviewed people in the court areas of both Uttar and Dakshin Dinajpur, this becomes evident. The dates when the crowd increased in number also saw huge growth in membership of social media groups against NRC. This means that people are becoming part of the protest movement and supportive of protests, but everyone is not doing so convincingly, because at the same time the crowd for taking precautions kept swelling.
\item Influence of Assam situation and NRC conducted there is quite perceptible in social media . The ground truth as evidenced by field visits to the bordering districts of Kokrajhar and Dhubri indicate similar nature. The timing of the field visit coincided with the protest against CAA in the state of Assam. The topic was being highly debated among the people. There was dissent regarding the nature of implementation of NRC in Assam and at the same time dissent was growing about the fact that the whole effort of six years was getting nullified and people's sacrifice becoming fruitless
\item Participation of Muslims before and after CAA or Ayodhya ruling by supreme court has been studied from social media datasets. Similar nature of Muslim participation is evidenced by the field visits. The dates of visit incidentally fell during two crucial rulings of Ayodhya ruling and CAA passage. The feeling of alienation and need for protest was voiced by those interviewed by us. The community seemed to be relying  on protest movement to a great extent. 
\item Nature of Leadership of the contra NRC movement has shown some contradiction.  The social media analysis shows non-hierarchical or flat leadership for movement in social media. But the findings during the field work has given evidence of hierarchical party presence. Many of the interviewees have talked about the different political parties, especially their local leadership operating under instructions from the central party leadership. 
\item The partition of undivided Bengal is another area of departure. This aspect is relatively scarce in the discussion topics of the social media groups. But during field visit it appears that the partition has left deep wound in people's mind. 
\item Another aspect that shows contradictory nature is with regard to the quality of the topics being discussed. Social media groups appear to be more progressive than some field visit findings which brought out intensive  communal nature of the participants.  
\end{itemize}

The intersection between the two sets, the members of the social media and the field interviewees, is almost null set. Hence the contradictory nature of the findings can be explained to an extent. With time, it is expected that when the government starts to push ahead with the implementation of NRC / CAA, the contradiction will either lead to withering away of the protest movement or produce one of the still unfolding mass movement with emerging non-hierarchical leadership.

\textbf{Acknowledgements} This research is partially supported by Impactful Policy Research in Social Science (IMPRESS) under Indian Council of Social Science Research, Govt. of India (P3271/220).

The authors are grateful to Swarnava Chakraborty for collecting social media dataset as an undergraduate intern student. The authors are also grateful to Sk Arafat Zaman for his work as an Asamiya language translator.

\bibliographystyle{model1-num-names}
\bibliography{sample.bib}

\end{document}